\documentclass[manuscript,trackchanges]{aastex6}

\pdfoutput=1
\usepackage{amssymb}
\usepackage{amsmath}
\usepackage{wrapfig}
\usepackage{rotating}

\newcommand\uiks{$u^*i^\prime K_s$}
\newcommand\ks{$K_s$}
\newcommand\ip{$i^\prime$}
\newcommand\ust{$u^*$}
\newcommand\gp{$g^\prime$}
\newcommand\rp{$r^\prime$}
\newcommand\samename{\vrule height0.4pt depth0.0pt width1.0in \thinspace.}
\newcommand{\rotateonaxis}{\raisebox{0.45ex}{\vrule height 0.025em%
   depth 0pt width 1.5em}\kern-1.05em\scalebox{0.55}[1.1]{\raisebox{%
   -0.6ex}{\rotatebox{60}{$\circlearrowleft$}}}\kern0.55em}

\shorttitle{$N_{\rm GC}$ and $M_\bullet$ in spirals: NGC 4258}

\turnoffedit

\begin{document}

\title{The relation between globular cluster systems and supermassive black holes in spiral galaxies.
 The case study of NGC~4258.
}

\author{Rosa A.\ Gonz\'alez-L\'opezlira\altaffilmark{1,2}, Luis Lomel\'{\i}-N\'u\~nez\altaffilmark{1}, 
Karla \'Alamo-Mart\'{\i}nez\altaffilmark{3}, 
Yasna \'Ordenes-Brice\~no\altaffilmark{3}, Laurent Loinard\altaffilmark{1}, \edit1{Iskren Y.\ Georgiev\altaffilmark{4}}, Roberto P.\ Mu\~noz\altaffilmark{3}, Thomas H.\ Puzia\altaffilmark{3}, 
Gustavo Bruzual A.\altaffilmark{1}, \and Stephen Gwyn\altaffilmark{5}
}
\affil{1 Instituto de Radioastronomia y Astrofisica, UNAM, Campus Morelia,
     Michoacan, Mexico, C.P.\ 58089 }
\affil{2 Argelander Institut f\"ur Astronomie, Universit\"at Bonn, Auf dem H\"ugel 71, D-53121 Bonn, Germany}
\affil{3 Instituto de Astrof\'{\i}sica, Pontificia Universidad Cat\'olica de Chile, Av.\ Vicu\~na Mackenna 4860, 7820436 Macul, Santiago, Chile  }
\affil{4 Max-Planck Institut f\"ur Astronomie, K\"onigstuhl 17, D-69117 Heidelberg, Germany}
\affil{5 Herzberg Institute of Astrophysics, National Research Council of Canada, Victoria, BC V9E 2E7, Canada}

\begin{abstract}
We aim to explore the relationship between globular cluster total
number, $N_{\rm GC}$, and central black hole mass, $M_\bullet$, in
spiral galaxies, and compare it with that recently reported for
ellipticals.  We present results for the Sbc galaxy NGC~4258, from
Canada France Hawaii Telescope data.  Thanks to water masers with 
Keplerian rotation in a circumnuclear disk, 
NGC~4258 has the most precisely measured extragalactic distance and
supermassive black hole mass to date.  The globular cluster (GC)
candidate selection is based on the ($u^*\ -\ i^\prime$) vs.
($i^\prime\ -\ K_s$) diagram, which is a superb tool to distinguish
GCs from foreground stars, background galaxies, and young stellar
clusters, and hence can provide the best number counts of GCs from
photometry alone, virtually free of contamination, even if the galaxy
is not completely edge-on.  The mean optical and optical-near infrared
colors of the clusters are consistent with those of the Milky Way and
M~31, after extinction is taken into account. We directly identify 39
GC candidates; after completeness correction, GC luminosity function
extrapolation and correction for spatial coverage, 
we calculate a
total $N_{\rm GC} = 144\pm31^{+38}_{-36}$ (random and systematic
uncertainties, respectively).  We have thus increased to
6 the sample of spiral galaxies with measurements of both $M_\bullet$
and $N_{\rm GC}$.  NGC~4258 has a specific frequency $S_{\rm N} =                                                     
0.4\pm0.1$ (random uncertainty), and is consistent within 2$\sigma$
with the $N_{\rm GC}$ vs.\ $M_\bullet$ correlation followed by
elliptical galaxies.  The Milky Way continues to be the only spiral
that deviates significantly from the relation.

\end{abstract}


\section{Introduction} \label{sec:intro}

It is virtually certain that all massive galaxies contain central black holes.  In spheroidal
systems, the masses of these black holes, $M_{\bullet}$, correlate with the galaxy bulge
luminosity \citep[the $M_\bullet $--$L_{\rm bulge}$ relation, e.g.,][]{korm93,korm95,mago98, korm01,marc03,gult09}, 
mass \citep[e.g.,][]{dres89,mago98, laor01,mclu02,marc03,hari04},
and stellar velocity dispersion \citep[the $M_\bullet$--$\sigma_*$ relation, e.g.,][]{ferr00,gebh00,trem02,gult09}.
\citet{spit09} also find a relation between black hole mass and
dark matter halo mass, as inferred from globular cluster system total mass.
Such correlations \edit1{could be the result of
central black hole and
spheroid evolution}. Exploring the limits and deviations from
these correlations can help illuminate the different evolutionary
processes that influence the growth of galaxies and their black holes
\citep[e.g.,][]{mcco11}.

{\em It is unclear whether spiral galaxies fall on these relations.}
At least in the case of the $M_\bullet$--$\sigma_*$ and
$M_\bullet$--$M_{\rm bulge}$ relations, although they exist for spirals,
they appear to have larger scatter than for ellipticals.
Also, some barred spirals and pseudobulges can be up to
one order of magnitude offset in the $M_\bullet$--$\sigma_*$ diagram,
with black holes that are too small for the velocity
dispersion of the bulge 
\citep[e.g.,][]{hu08,gree08,gado09,gree10}. 
The scatter could be caused
by measurement uncertainties, but it could also be
intrinsic, e.g., \edit1{if black hole growth is stochastic to 
some extent, depending on inward gas flow and galaxy-to-galaxy differences in gas accretion rates.}  
\citet{grah12a,grah12b} suggests that the
$M_\bullet $--$L_{\rm bulge}$ and $M_\bullet $--$M_{\rm bulge}$ relations might be broken or
curved for galaxies with $M_B > -20.5$ mag, even if they have
real bulges. \citet{lask16} also find hints that a galaxy's
low mass may be more of a determinant than the absence of a classical bulge.

Interestingly, \citet{burk10} and, with more data,
\citet{harr11,harr14} 
have shown that the central black hole mass of
elliptical galaxies increases
almost exactly proportionally to the total number of their
globular clusters, $N_{\rm GC}$. The correlation can be expressed as
$N_{\rm GC} \propto M_\bullet^{1.02\pm0.10}$, spans over 3 orders
of magnitude, and is tighter than
the $M_\bullet-\sigma_*$ relation.

The most likely link between $N_{\rm GC}$ and $M_\bullet$ is the galaxy potential
well/binding energy 
($M_*\sigma_*^2$, where $M_*$ is the bulge stellar mass),
with which both quantities are correlated \citep[e.g.,][]{snyd11,rhod12}.
However, given the extremely disparate scales of
both components, the tightness of the correlation is intriguing and has been intensely explored.
\citet{korm13} review the status of these endeavors.
For example, \citet{harr13} find that
$N_{\rm GC} \propto (R_e \sigma_e)^{1.3}$, 
with $R_e$ the effective radius of the galaxy light profile (bulge+disk) and 
$\sigma_e$ the velocity dispersion at the half-light radius. 
On the other hand, while \citet{sado12} show that the correlation between
$M_\bullet$ and the velocity dispersion of globular cluster {\em systems} as a whole is
tighter for red than for blue GCs \edit1{in a sample of 12 galaxies, \citet{pota13} 
do not find a significantly smaller scatter in the correlation
for either one of the blue and red subsystems in an enlarged sample of 21 galaxies with better $M_\bullet$ measurements. 
Regarding $N_{\rm GC}$, \citet{rhod12}
finds that the number of blue GCs correlates better with $M_\bullet$.
Also, for the few spirals in her sample, \citet{rhod12} gets a better}
correlation of $N_{\rm GC}$ with bulge light than with total light;
however, the correlation of $N_{\rm GC}$ is much tighter with total galaxy
stellar mass than with bulge mass. \edit1{\citet{deso15} perform an exhaustive analysis of the 
$N_{\rm GC}$ vs.\ $M_\bullet$ relation from the point of view of Bayesian statistics; they 
conclude that black hole mass is a good predictor of total number of GCs.}

The $N_{\rm GC}$ vs.\ $M_\bullet$ correlation could be rooted in the initial conditions 
of galaxy formation or in the
process of galaxy assembly, and the scaling relations between different galaxy 
components could give clues about which 
origin is more likely. 

Precisely because of 
the 10 orders of magnitude that separate the spatial scales of 
supermassive black holes (SMBHs) and globular cluster systems, a direct causal link 
has been often quickly dismissed with the argument that bigger galaxies should have
more of everything. 
Nontheless, possible causal relations have been proposed. \citet{harr14} offer a review 
of them,
namely: star (and hence GC) formation driven by AGN jets \citep[e.g.,][]{silk98,fabi12}; BH growth through the cannibalization of
GCs \citep[e.g.,][]{capu01,capu05,capu09,gned14}, especially efficient if all GCs contain intermediate mass BHs
\citep{jala12}. 
Another mechanism for the establishment of scaling relations would be the statistical convergence process of galaxy characteristics 
through hierarchical galaxy formation \citep{peng07,jahn11}.  

Besides the issue of its origin, a well understood $N_{\rm GC}$ -- $M_\bullet$ 
correlation could be useful as a tool to estimate
masses of inactive black holes, that would otherwise be hard to measure.

Before the present work, there were 
only 5 spiral galaxies with precise measurements
of both  $N_{\rm GC}$ and $M_\bullet$:
the Milky Way (MW, Sbc), M~104 (Sa), M~81 (Sab), M~31 (Sb) and NGC~253 (Sc).
All of them, with the notable exception of the MW, fall right 
on the $N_{\rm GC} - M_\bullet$ correlation for
ellipticals. Our Galaxy 
has a black hole that is about 1 order of magnitude
lighter than expected from its $N_{\rm GC}$.

Spirals have both less massive black holes and less rich
GC systems than ellipticals and lenticulars, and hence have 
been studied much less frequently than early type galaxies. 
Concerning in particular GCs, internal extinction and potential
confusion with stars and star clusters in their galactic disks 
have in general further limited studies in spirals to galaxies 
seen edge-on.

In what follows, we present new measurements and analysis of the globular cluster system of the Sbc galaxy NGC~4258 (M~106). Thanks to water masers in a disk orbiting its nucleus, its absolute distance has been derived directly by geometric means \citep{herr99}. NGC~4258 hence has the most precisely measured extragalactic distance and SMBH mass to date, i.e., 
7.60$\pm$0.17$\pm$0.15 Mpc (formal fitting and systematic errors, respectively) and (4.00$\pm$0.09)$\times 10^7  M_\odot$ \citep{hump13}.
In spite of being the archaetypical megamaser galaxy, however, NGC~4258 has a classical bulge, unlike
most other megamaser galaxies \citep{lask14,lask16};
it also falls on the $M_\bullet $--$L_{\rm bulge}$  and  $M_\bullet$--$M_{\rm bulge}$ for ellipticals
\citep{sani11,lask16}.
With an inclination to the line of sight of 67$\degr$ \citep{rc3}, NGC~4258 is not an edge-on galaxy.
Ours is the first application of the ($u^*-i'$) vs.\ ($i'-K_s$) diagram technique \citep[\uiks\ hereafter;][]{muno14}
to a spiral galaxy. We will show that with this procedure one can get the {\it best} total number counts of GCs from photometry alone that are virtually free of contamination from foreground stars and
background galaxies, {\it without the need to 
obtain radial velocity measurements}. In addition, the \uiks\ plot can easily weed out young star clusters
in a spiral disk. We will quantify the efficiency of the \uiks\ diagram to produce a clean sample of GC candidates (GCCs) in a disk galaxy.    
\section{Data} \label{sec:data}

All data for the present work were obtained with the Canada-France-Hawaii-Telescope (CFHT).
The optical images of NGC~4258 are all archival, and were acquired with MegaCam
\citep{boul03}.
MegaCam has 36 2048$\times$4612 CCD42-90 detectors operating at -120$\degr$ C, and 
sensitive from 3700 to 9000 $\AA$; read-out-noise
and gain are, respectively, $<$ 5 e$^-$ pixel$^{-1}$ and $\sim$ 1.62 e$^-$ ADU$^{-1}$.
The detectors are arranged in a 9$\times$4 mosaic with 13$\arcsec$ small gaps between 
CCD columns, and 80$\arcsec$ large gaps between CCD rows. The field-of-view (FOV) of 
MegaCam is 0$\fdg96 \times 0\fdg$94, with a plate scale of 0$\farcs$186 pixel$^{-1}$. 

Images were originally secured through programs 
08BH55, 09AH42, 09AH98, 09BH95 (P.I.\ E.\ Magnier, \ust-band); 09AC04 (P.I.\ R.\ L\"asker, \ust\ and
\ip\ filters); 10AT01 (P.I.\ C.\ Ngeow, \gp, \rp, and \ip\ bands), 
and 11AC08 (P.I.\ G.\ Harris, \gp\ and \ip\ data). 

Before archiving, MegaCam images are ``detrended", i.e., corrected for the instrumental response (bad pixel removal, overscan and bias
corrections, and flat-fielding, plus defringing for the \ip\ and $z^\prime$ bands) with the Elixir software  
\citep{magn04}. Elixir also provides a global astrometric calibration with an accuracy of  
0$\farcs$5 -- 1$\farcs$0, and a photometric calibration with a uniform
zero point whose internal accuracy is better than 1\% for the entire image.

To make the final \ust, \gp, \rp, and \ip\ mosaics of NGC~4258, individual images
were combined with the MegaPipe pipeline \citep{gwyn08}. MegaPipe groups the images
by passband; resamples them to correct the geometric distortion of the MegaCam focal plane;
recalibrates the astrometry to achieve internal and external accuracies of 0$\farcs$04 and
0$\farcs$15, respectively; recalibrates the photometry, mainly to account for modifications made to the Elixir  
pipeline in different epochs, with an output accuracy of 0.03 mag; and finally stacks them. 
  
The \ks-band images of NGC~4258 were acquired on 2013 March 27 UT, through proposal 13AC98 
(P.I.\ R.\ Gonz\'alez-L\'opezlira), with the Wide-field InfraRed Camera 
\citep[WIRCam;][]{puge04}. WIRCam has four 2048$\times$2048 HAWAII2-RG HgCdTe
detectors, cooled cryogenically to $\sim$ 80$\degr$ K, and sensitive at
0.9-2.4 $\mu$m; read-out-noise and gain are, respectively, 
30 e$^-$ pixel$^{-1}$ and 3.7 e$^-$ ADU$^{-1}$. The detectors are organized in a 2$\times$2 mosaic with
45$\arcsec$ interchip gaps. The FOV of the WIRCam is  
$\sim 21^\prime \times 21^\prime$, with a plate scale of 0$\farcs$307 pixel$^{-1}$.

\begin{figure}[ht!]
\plotone{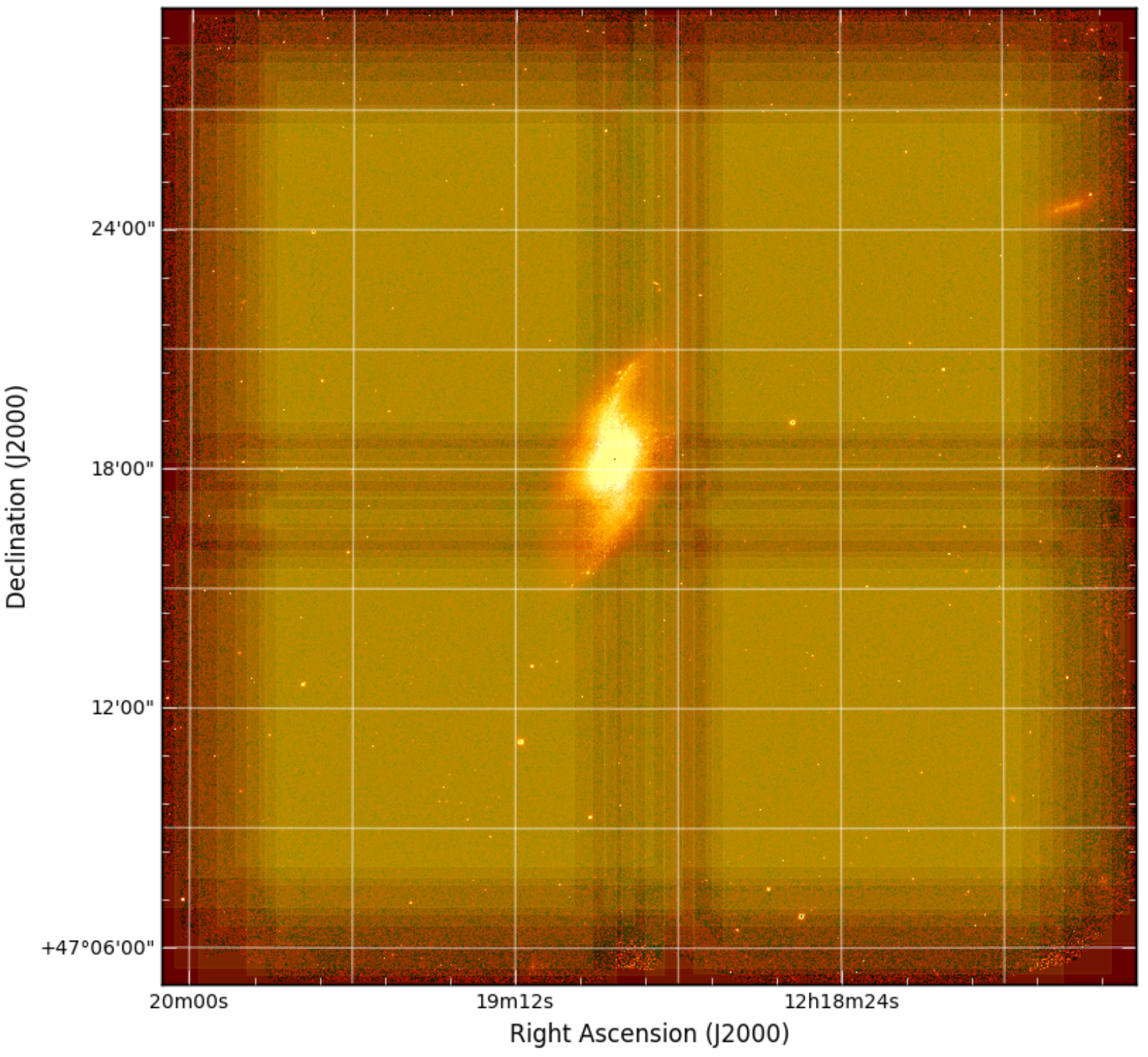}
\caption{NGC~4258 coverage map on our final \ks-band image. Number of overlapping exposures increases with color intensity,
up to 10. {\em This figure was made with the Kapteyn software package} \citep{terl15}.
\label{fig:galcov}}
\end{figure}

We obtained 10$\times$20 s individual \ks\ exposures of NGC~4258. 
The telescope pointing was dithered in an approximately circular pattern with a radius $\sim$ 1$\farcm$6.
Because of the galaxy large diameter compared to WIRCam's FOV
\citep [$R_{25} = 9\farcm3$;][]{rc3}, separate 10$\times$20 s sky frames were taken 2$\fdg$1 away,   
with the same circular pattern, 
using the following target(T)-sky(S) sequence:
STTSSTTSS...TTS. 
Figure~\ref{fig:galcov} displays the sky-coverage map for the galaxy overlaid on
our final \ks-band mosaic; 
number of overlapping exposures increases with color intensity, up to a maximum of 10.

In the case of WIRCam images, detrending is 
performed by the I'iwi pipeline.\footnote{ 
\url{http://www.cfht.hawaii.edu/Instruments/Imaging/WIRCam/IiwiVersion1Doc.html}.}
Detrending includes saturated pixel flagging; non-linearity correction;
bias (reference pixel) and dark current subtraction; flat-fielding, and bad pixel masking.
I'iwi also gives a global astrometric calibration with an accuracy of  
0$\farcs$5 -- 0$\farcs$8; the photometric zero point however, is different for each of the  
WIRCam detectors at the end of the I'iwi processing.

To produce the final \ks-band mosaic of NGC~4258, individual images were sky-subtracted and 
combined with the WIRwolf pipeline \citep{gwyn14}.
Like MegaPipe, WIRwolf resamples
the individual images; recalibrates their astrometry to a typical internal accuracy of 0$\farcs$1; provides a uniform
zero point for the four detectors; and stacks them.  

Both MegaPrime and WIRwolf output images with photometry in the AB system \citep{oke74}, with zero points $zp$ = 30.
Given its smaller FOV and significantly shorter exposure time, all our analysis will be limited by 
the \ks-band image of NGC~4258.

Table~\ref{tab:n4258data} gives a summary of the NGC~4258 observations.

\floattable
\begin{deluxetable}{CRRrlccl}
\tablecaption{NGC~4258 Observation Log\label{tab:n4258data}}
\tablewidth{0pt}
\tablehead{
\colhead{{\rm Filter}} & \colhead{$\lambda_{\rm cen}^a$} & \colhead{ {\rm FWHM}$^b$} & \colhead{Exposure} & \colhead{Camera} &\colhead{Pixel size} &\colhead{ Program }  &\colhead{Date} \\
\colhead{           }  & \colhead{                 }  & \colhead{          } & \colhead{s} &
\colhead{} & \colhead{$\arcsec$} & \colhead{} & \colhead{UT} 
}
\startdata
  u^*        &  3793\ \AA & 654\ \AA &  13360 & Megacam & 0.186  & 08BH55 & 2008 December 22, 23 \\
	     &           &          &        &         &        & 09AC04 & 2009 February 18 \\
	     &           &          &        &         &        & 09AH42 & 2009 March 27, 30\\
	     &           &          &        &         &        & 09AH98 & 2009 April 19    \\
	     &           &          &        &         &        & 09BH95 & 2009 December 11 \\
  g^\prime   &  4872\ \AA & 1434\ \AA &  10400 & Megacam & 0.186  & 10AT01 & 2010 June 11, 12; July 7, 8 \\
	     &           &          &        &         &        & 11AC08 & 2011 March 1 \\
  r^\prime   &  6276\ \AA & 1219\ \AA &   3500 & Megacam & 0.186  & 10AT01 & 2010 June 16; July 8, 12    \\
  i^\prime   &  7615\ \AA & 1571\ \AA &   8080 & Megacam & 0.186  & 09AC04 & 2009 February 26 \\
	     &           &          &        &         &        & 10AT01 & 2010 June 10, 12, 13, 15; July 8,12 \\
  K_s  &  2.15\ \mu{\rm m} & 0.33\ \mu{\rm m} & 200 &  WIRCam  &   0.307 &  13AC98 & 2013 March 27  \\
\enddata
\tablecomments{\ $^a$The central wavelength between the two points defining FWMH 
(\url{http://svo2.cab.inta-csic.es/svo/theory/fps3/index.php?id=CFHT/}). \\
$^b$ {\it Ibid.} 
}
\end{deluxetable}

\section{Detection and photometry}\label{sec:detection}

Source detection and photometric measurements in all the stacked images were
carried out with SExtractor \citep{bert96} and PSFEx 
\citep{bert11}. PSFEx employs point sources detected in a first pass of SExtractor to
build a point-spread function (PSF) model that SExtractor can then apply in a second pass to 
obtain PSF magnitudes of sources.  We used versions 2.8.6 (first pass) and 2.19.4 (second pass) of 
SExtractor, and PSFEx version 3.9.1.

\begin{figure}[ht!]
\plotone{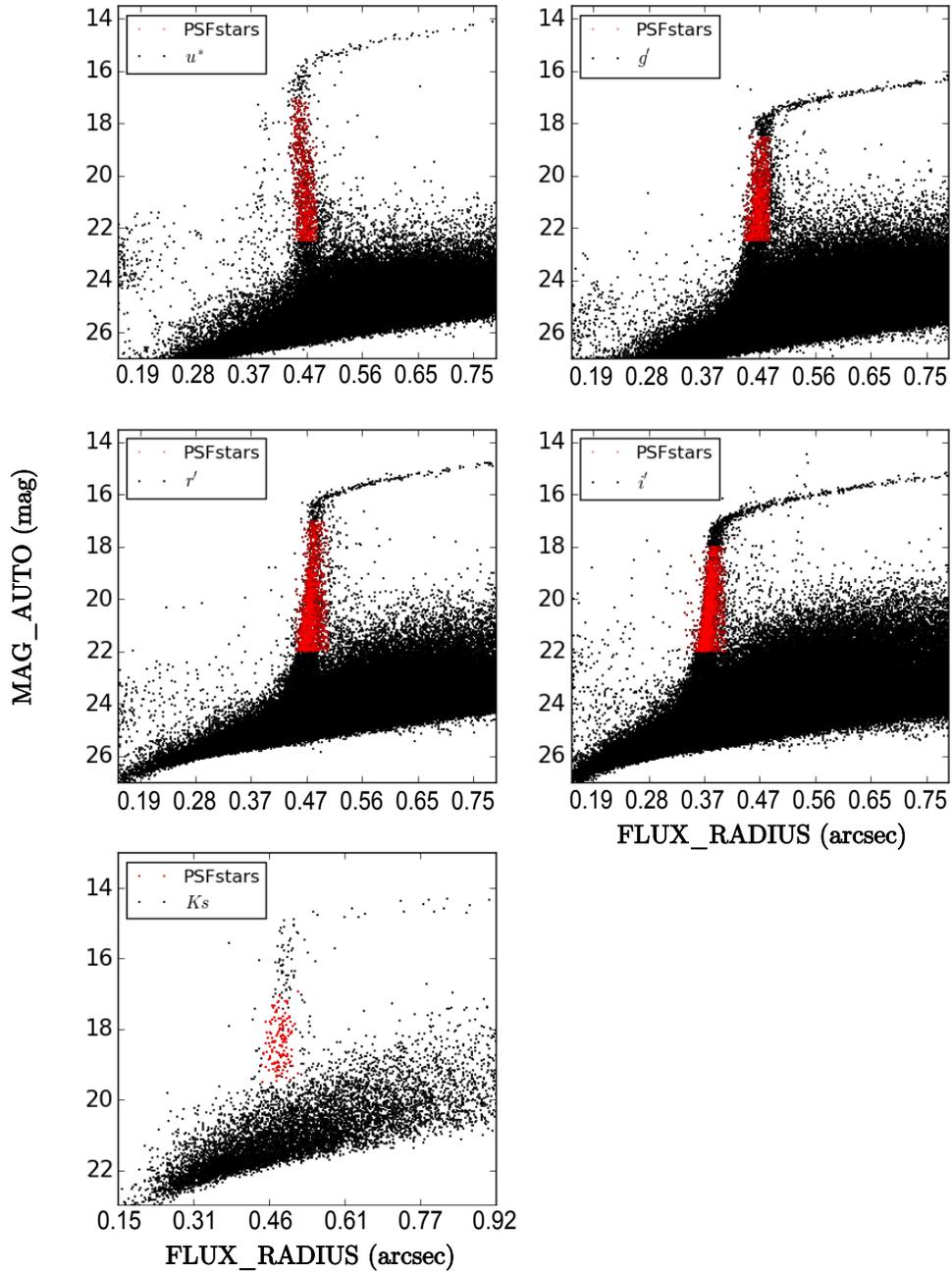}
\caption{PSF star selection. Plots of MAG\_AUTO vs.\ FLUX\_RADIUS for the \ust\ (top left),
\gp\ (top right), \rp\ (middle left), \ip\ (middle right), and \ks\ (bottom left) bands.
Point sources are located in the vertical columns of dots; saturated sources lie in the
plume at roughly constant MAG\_AUTO; the cloud at faint magnitudes contains both extended
and spurious sources, like cosmic rays. Selected PSF stars are shown as red dots. 
\label{fig:psfsel}}
\end{figure}

Although PSFEx can select automatically sources (PSF stars) to build the model PSF, instead we
chose adequate stars manually, 
carefully discarding saturated, extended, and spurious sources.  
The selection is based on the brightness vs.\ compactness parameter space,
as measured by SExtractor parameters MAG\_AUTO and FLUX\_RADIUS,\footnote{
FLUX\_RADIUS estimates the radius of the circle centered on the light barycenter that encloses about 
half of the total flux. 2$\times$FLUX\_RADIUS equals the FWHM for a Gaussian profile, but for 
seeing-limited images 2$\times$FLUX\_RADIUS $\sim$ 1.05 -- 1.1 FWHM; for profiles 
with a significant fraction of flux in the wings, the difference
can be much larger (\url{http://www.astromatic.net/forum/showthread.php?tid=516}). 
MAG\_AUTO is a Kron-like \citep{kron80} 
elliptical aperture magnitude (\url{http://terapix.iap.fr/article.php?id\_article=628}). 
The radius, ellipticity, and position angle of the aperture are defined from the first and second moments of
the object's light profile \citep{bert96}.
} respectively. 
The plot is shown in Figure~\ref{fig:psfsel} for all passbands, with PSF stars 
highlighted as red dots. The selection criteria, as well as the number of 
detected (in both the first, SEx 1, and second, SEx 2, runs of SExtractor) and chosen points at each wavelength, are listed in Table~\ref{tab:psfcrit}. In order for PSFEx to work, one needs to
have VIGNET as an output parameter in a previous run of SExtractor. VIGNET(width,height) specifies the size    
in pixels of the region around each PSF star that will be considered; one also needs to 
instruct SExtractor to perform photometry in a circular aperture with a diameter PHOT\_APERTURES of the same size. 
These size and aperture must be enough to include most of the stellar flux, but 
as small as possible to minimize the risk of contamination from other objects.   
\edit1{The VIGNET (and hence aperture) sizes} are likewise shown in Table~\ref{tab:psfcrit}.

The spatial variations of the PSF were modeled with polynomials of degree 4
for the optical mosaics, and of degree 2 for the smaller \ks\ image.
When running with a PSF model, SExtractor can measure PSF magnitudes (MAG\_PSF) and produce another, very
useful, estimator of shape, called SPREAD\_MODEL. The SPREAD\_MODEL value for each object 
results from the comparison between its best fitting PSF, and the convolution of such PSF with 
an exponential disk with scale length FWHM$_{\rm PSF}$/16., where FWHM$_{\rm PSF}$ is the 
full width at half max of the same PSF \citep{desa12}. 

\floattable
\begin{deluxetable}{CCcccccr}
\tablecaption{PSFEx Parameters\label{tab:psfcrit}}
\tablewidth{0pt}
\tablehead{
\colhead{      }& \colhead{}& \multicolumn2c{FLUX\_RADIUS}& \multicolumn2c{MAG\_AUTO}& \multicolumn2c{$N$ sources}\\
\colhead{Filter}& \colhead{VIGNET}& \colhead{Min}&  \colhead{Max}& \colhead{Min}& \colhead{Max}& \colhead{Total}& \colhead{PSF stars} \\
\colhead {} & \colhead{pixel $\times$ pixel} & \multicolumn2c{pixel} & \multicolumn2c{mag} &\colhead{SEx 1/ SEx 2} & \colhead{}
}
\decimals
\startdata
  u^*        &  20\times 20     &     2.35 &    2.60  &  17.0  &   22.5 &   132183/132276  &  915  \\ 
  g^\prime   &  20\times 20     &     2.35 &    2.60  &  18.5  &   22.5 &   233053/232304  & 1247  \\
  r^\prime   &  20\times 20     &     2.40 &    2.70  &  17.0  &   22.0 &   122867/122867  & 1855  \\
  i^\prime   &  20\times 20     &     1.80 &    2.20  &  18.0  &   22.0 &   218200/218142  & 2173  \\
  K_s  &  15\times 15     &     1.45 &    1.70  &  16.0  &   19.5 &     9276/9272  &  145  \\
\enddata
\end{deluxetable}

\section{Completeness} \label{sec:completeness}

Completeness tests were only performed on the \ks-band data; this is the smallest and shallowest
image and in fact sets our object detection limit. In order to determine the GC detection completeness
as a function of magnitude, $\sim$ 320,000 artificial point sources
were generated based on the PSF model, in the 
interval  17 mag $< m_{Ks} <$ 23 mag, 
and with a uniform, or box-shaped, magnitude distribution.
Due to the WIRCam pixel size, 0$\farcs$307 or $\gtrsim$ 11 pc at the distance of NGC~4258, 
GCCs are unresolved in the \ks-band image.

The artificial objects were added only $\sim$1000 at a time (we call this one simulation), 
in order to prevent creating artificial crowding. 
The \ks-band image of NGC~4258 was divided
into square bins 155 pixels on the side, 
and only one artificial object was added in each bin. 
The positions of the added sources were random, but assigned avoiding
overlapping with other artificial objects or with real sources
identified by the SExtractor segmentation map. 
To reach 320,000 artificial sources for the whole image, we carried out 160 simulation pairs, 
i.e., we produced 320 images, each one with $\sim$ 1000 added sources, but there were
only 160 different sets of positions. 
The artificial objects in a simulation pair had the same positions, but their magnitudes were   
different in each of the two members of the pair.

SExtractor was then run on each one of the 320 simulated images, with the same parameters used for the 
original \ks-band image of NGC~4258, 
and the recovered artificial
sources were identified by cross-matching the positions of all detections with the known input coordinates of
the added objects. 
Aside from non-detections, objects with SExtractor output parameter FLAGS $\neq$ 0 were eliminated.\footnote{Different values of FLAGS indicate various problems with the photometry. For instance, sources with close bright neighbors have FLAGS = 1; blended objects, FLAGS = 2; saturated sources, FLAGS = 4.}
The requirement FLAGS=0 excludes artificial objects falling on top of other artificial or real sources, and hence discards preferentially added objects in crowded regions. The criterion FLAGS=0 was also applied in the selection of true sources (see Section~\ref{sec:uiks}). 

Given that NGC~4258 is a spiral galaxy, and not edge-on, it was of particular interest to
quantify the effect of the disk on source detection. 
Since the detection magnitude limit is affected by object crowding and background brightness level, we 
estimated completeness in four different regions within 1.7 $R_{25}$ of NGC~4258, roughly at the edge of our \ks-band image \citep[$R_{25}= 9\farcm3$ or 20.5 kpc;][]{rc3}.\footnote{ 
Assuming $R_{25}$ = 13.4 kpc for the Milky Way \citep{good98},
$\sim$ 85\% of its GCs lie within 1.7 $R_{25}$ (\url{http://physwww.physics.mcmaster.ca/~harris/mwgc.dat}).
}

We explored an ellipse with semimajor axis = 0.5 $R_{25}$, and three elliptical annuli, respectively, from   
0.5 to 1.0 $R_{25}$, from 1.0 to 1.4 $R_{25}$, and between 1.4 and 1.7 $R_{25}$. All 4 regions have the axis ratio and 
position angle (P.A.) of
the observed disk of NGC~4258, i.e., 0.39 and 150$\degr$, respectively \citep{rc3}.
In order to estimate completeness in these four regions with statistically equivalent samples, we made sure that the added sources were 
inversely proportional to their area.
Hence, the number of simulations considered for each region was different, respectively, 320, 106, 83, and 97.\footnote{The outermost annulus spills over the borders of the image, and hence does not have the area of a complete elliptical ring.} 
On average, $\sim$ 13,300 artificial sources were added to each one of the three annuli. We will discuss the central ellipse separately below. 

In the case of a box-shaped magnitude distribution, the fraction of recovered sources as a function
of magnitude is well described by the Pritchet function \citep[e.g.,][]{mcla94}:  

\begin{equation}
f(m) = \frac{1}{2}  \left[ 1\ -\ \frac{\alpha_{\rm cutoff} (m - m_{\rm lim})}{\sqrt{1 + \alpha_{\rm cutoff}^2(m - m_{\rm lim})^2}}  \right];
\label{eq:pritchet}
\end{equation}

\noindent
$m_{\rm lim}$ is the magnitude at which completeness is 50\%, and
$\alpha_{\rm cutoff}$ determines the steepness of the cutoff.

The fraction of recovered to added sources as a function of output PSF magnitude, in bins
0.25 mag wide, was fit
with equation~\ref{eq:pritchet} separately for each region.
The values of the fit parameters $m_{\rm lim}$ and $\alpha_{\rm cutoff}$ are shown in Table~\ref{tab:prit}.

\floattable
\begin{deluxetable}{ccc}
\tablecaption{Completeness Fit Parameters\label{tab:prit}}
\tablewidth{0pt}
\tablehead{
\colhead{Region} & \colhead {$m_{\rm lim}$} & \colhead{$\alpha_{\rm cutoff}$} \\
\colhead {$R_{25}$} & \colhead{$K_s$ AB mag} & \colhead{} 
}
\decimals
\startdata
0.0 -- 0.5 & 21.59 $\pm$ 0.01$^a$   & 5.57 $\pm$ 0.48$^a$ \\
\edit1{0.5 -- 1.0} & 21.74 $ \pm$ 0.01  & 14.63 $\pm$ 115.34 \\
1.0 -- 1.4 & 21.72 $ \pm$ 0.01  & 8.90 $\pm$ 2.54 \\
1.4 -- 1.7 & 21.59 $ \pm$ 0.01  & 3.87 $\pm$ 0.25 \\
\enddata
\tablecomments{\ $^a$Errors calculated by Monte Carlo resampling, with dispersion estimated from
the rms of the fit.
}
\end{deluxetable}

\begin{figure}[ht!]
\begin{tabular}{lll}
\hspace*{-0.8cm}\includegraphics[scale=0.25]{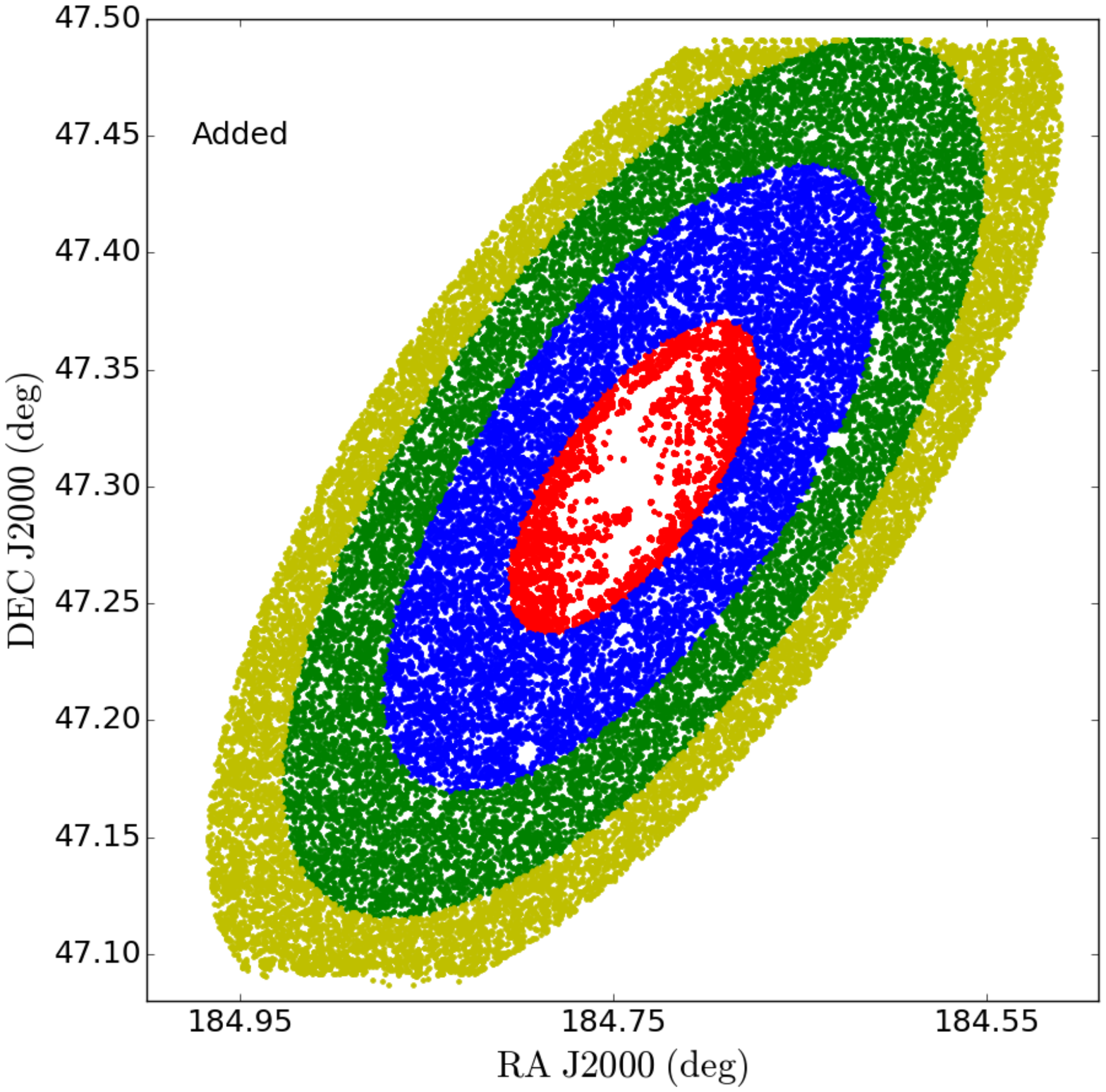}
&
\hspace*{-0.5cm}\includegraphics[scale=0.25]{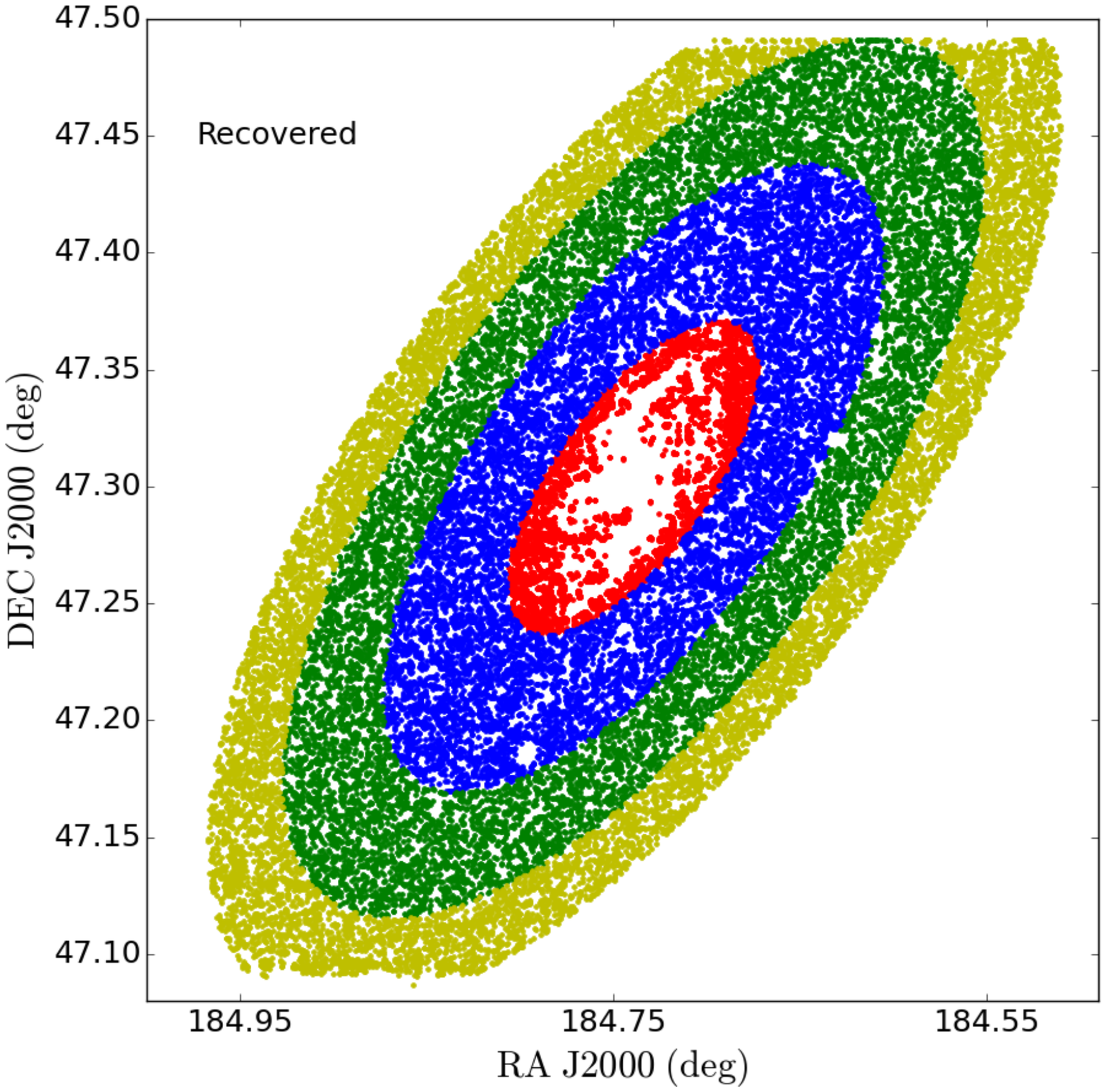}
&
\hspace*{-0.2cm}\includegraphics[scale=0.25]{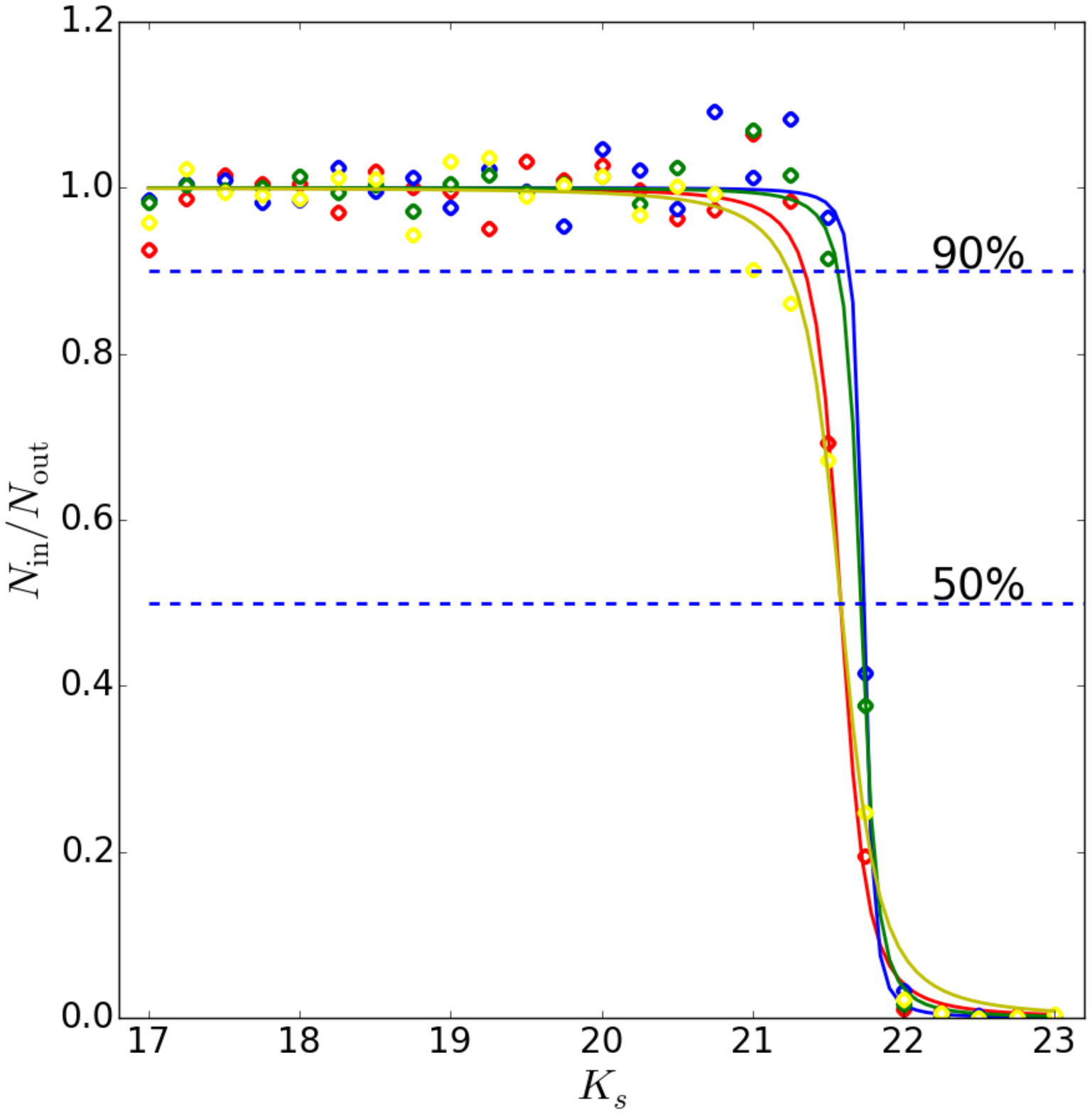}
\end{tabular}
\caption{Completeness tests. 
{\it Left:} added sources. 
{\it Center:} recovered sources.   
{\it Right:} fits ({\it solid lines}) to recovered fractions 
({\it dots}) with eq.~\ref{eq:pritchet}. Colors ({\it red, blue, green, yellow}) refer to centermost ellipse, and inner, middle, and external
annuli, respectively. {\it Blue dotted lines} indicate the 90\% and 50\% completeness values.  
\label{fig:cmpltnss}}
\end{figure}

In Figure~\ref{fig:cmpltnss} we display the added and recovered sources, in the
left and center panels, respectively. The right panel shows
the fractions of recovered to added objects vs.\ \ks\ mag,
as well as the fits to them (solid lines) with eq.\ref{eq:pritchet}. 

There are some results of this completeness test that stand out: 
(1) the similarity between the detection magnitude limit in the \ks\ band data and 
the $m_{\rm lim}$ values in all the annuli beyond 0.5 $R_{25}$; 
(2) the brighter $m_{\rm lim}$ of the ring with the 
largest galactocentric distance (1.4 to 1.7 $R_{25}$, yellow points), compared with those of 
the two intermediate annuli (0.5 to 1.4 $R_{25}$, blue and green), and 
(3) the fact that there 
is a region close to the center where the condition to avoid overlap with real sources prevented the
addition of artificial objects.

The similarity between the $m_{\rm lim}$ values and the source detection limit beyond 0.5 $R_{25}$ is due to the lack of 
a large low surface brightness envelope around the galaxy. In ellipticals, an envelope of completely unresolved
stars results in a radial gradient of S/N that is reflected in a $m_{\rm lim}$ that likewise changes with
radius.

\begin{figure}[ht!]
\begin{tabular}{lll}
\hspace*{-0.8cm}\includegraphics[scale=0.25]{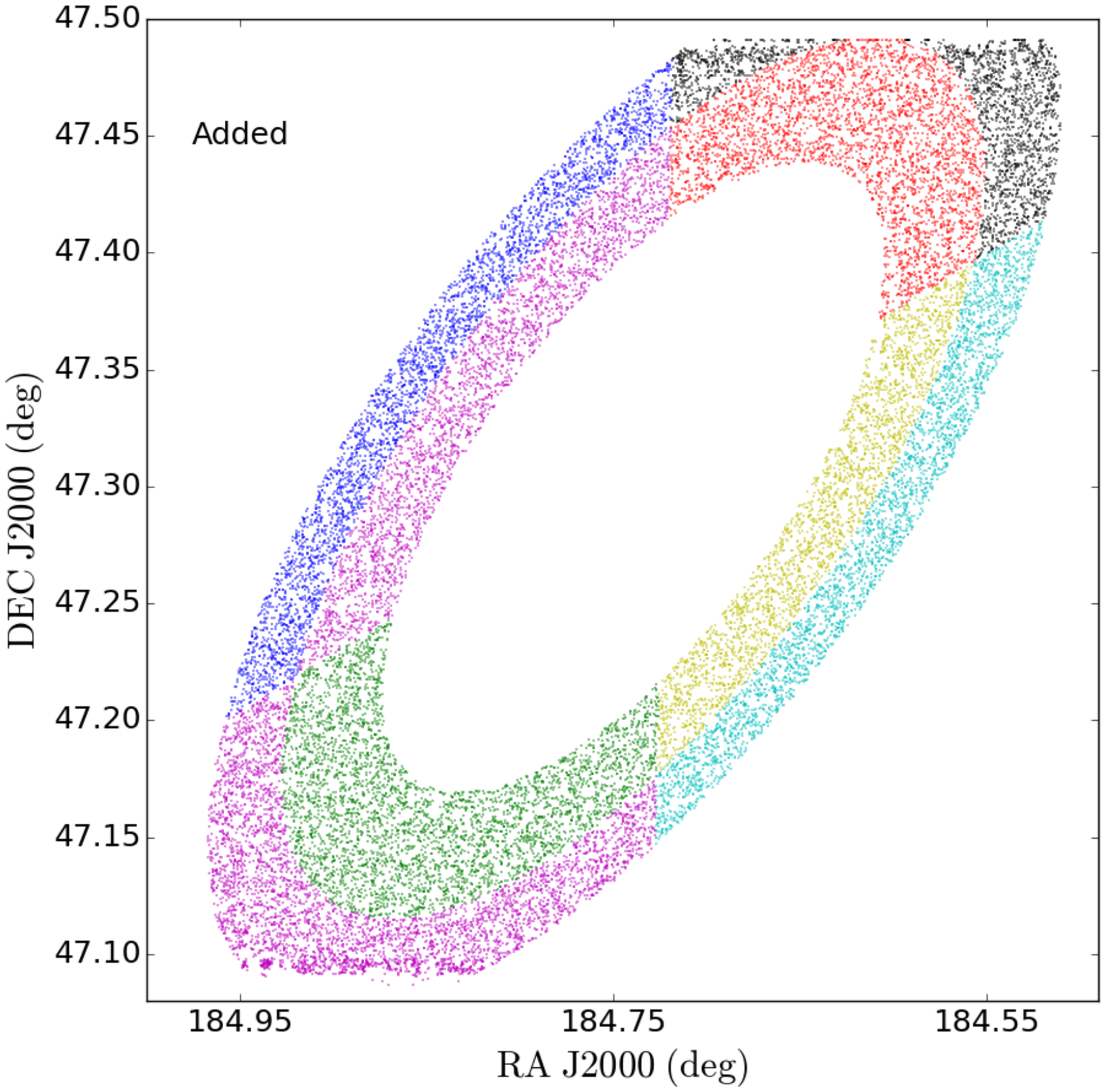}
&
\hspace*{-0.5cm}\includegraphics[scale=0.25]{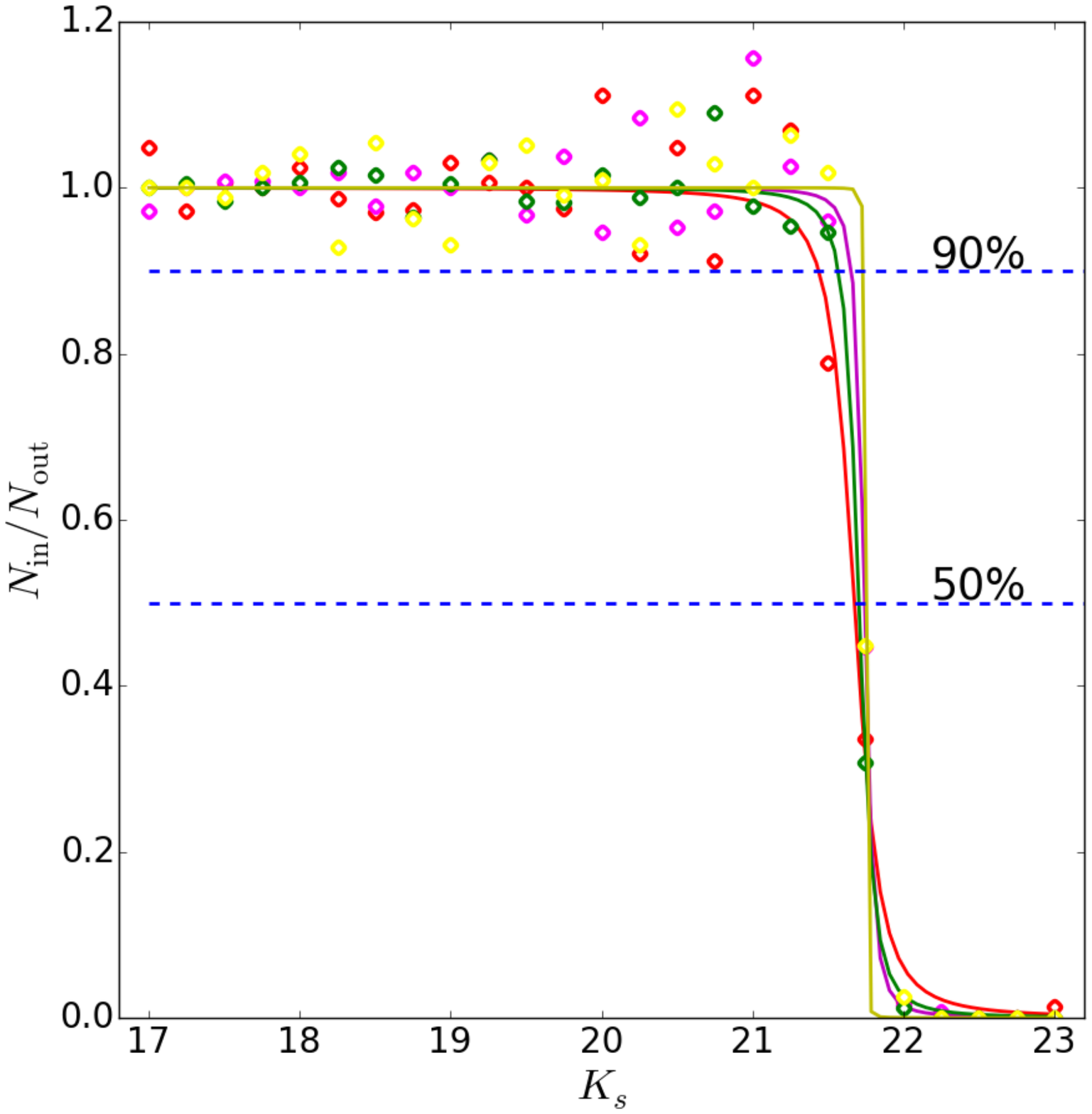}
&
\hspace*{-0.2cm}\includegraphics[scale=0.25]{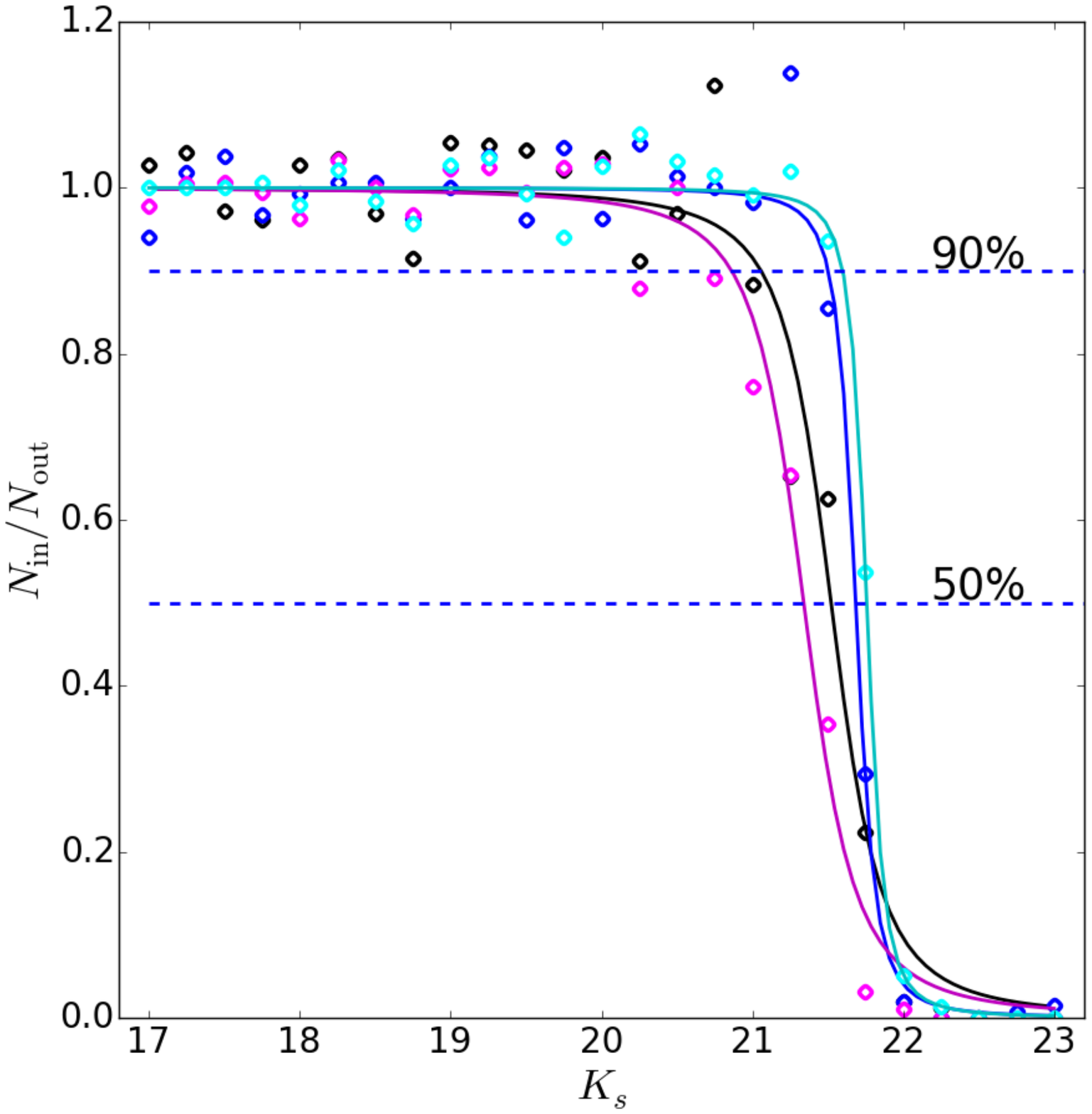}
\end{tabular}
\caption{Completeness tests for middle and external annuli by sections. 
{\it Left:} Added sources to rings between 1.0 and 1.4 $R_{25}$ (counterclockwise, from top, 
{\it red/north, magenta/east, green/south,} and {\it yellow/west}), and from 1.4 to 1.7 $R_{25}$ ({\it black/north, blue/east, magenta/south, cyan/west}). 
{\it Center:} fits ({\it solid lines}) to recovered fractions ({\it dots}) in middle annulus with eq.~\ref{eq:pritchet}. 
{\it Right:} fits to recovered fractions in external ring. 
\label{fig:anillo_secciones}}
\end{figure}

Thus, compared to the usual case of ellipticals, the brighter $m_{\rm lim}$ in the annulus furthest away
from the galaxy is counterintuitive.
However, the brighter $m_{\rm lim}$ of the annulus between 1.4 and 1.7 $R_{25}$ is a consequence of it being too
close to the top and bottom edges of the image, where fewer frames contribute to the final mosaic and hence the
signal-to-noise ratio is slightly smaller. We confirmed this by dividing the two most external rings in four sections, 
as shown in the left panel of Figure~\ref{fig:anillo_secciones}. 
The central and 
right panels of Figure~\ref{fig:anillo_secciones} show the fit with eq.~\ref{eq:pritchet} to the fraction of recovered to added sources vs.\  
\ks\ magnitude for the four sections, respectively, of the inner and outermost rings; the parameters of the fits
are listed in Table~\ref{tab:anillo_secciones}.

\floattable
\begin{deluxetable}{cccc}
\tablecaption{Completeness Fit Parameters of Middle and External Annuli by Section\label{tab:anillo_secciones}}
\tablewidth{0pt}
\tablehead{
\colhead{Radius}& \colhead{Section} & \colhead {$m_{\rm lim}$} & \colhead{$\alpha_{\rm cutoff}$} \\
\colhead {$R_{25}$}& Color & \colhead{$K_s$ AB mag} & \colhead{} 
}
\decimals
\startdata
1.0 -- 1.4 & Red/north & 21.68 $\pm$ 0.02$^a$   & 5.66 $\pm$ 1.34$^a$ \\
	   & Magenta/east & 21.743 $ \pm$ 0.004  & 15.8 $\pm$ 296.1 \\
	   & Green/south & 21.71 $ \pm$ 0.01  & 9.9 $\pm$ 30.1 \\
	   & Yellow/west & 21.7493 $ \pm$ 0.0008  & 146.6 $\pm$ 139.1 \\
1.4 -- 1.7 & Black/north & 21.52 $\pm$ 0.04   & 2.87 $\pm$ 0.47 \\
	   & Blue/east & 21.68 $ \pm$ 0.02  & 7.40 $\pm$ 20.08 \\
	   & Magenta/south & 21.34 $ \pm$ 0.03  & 2.79 $\pm$ 0.31 \\
	   & Cyan/west & 21.759 $ \pm$ 0.005  & 8.39 $\pm$ 8.27\\
\enddata
\tablecomments{\ $^a$Errors calculated by Monte Carlo resampling, with dispersion estimated from
the rms of the fit.
}
\end{deluxetable}

Whereas the completeness is basically the same for all four sections of the ring between 1.0 and 1.4 $R_{25}$,
for the outermost annulus there is a significant difference between the top/north and bottom/south sections, on the one hand,
and the left/east and right/west sections, on the other. This is consistent with our interpretation that the source
detection limit in the ring between 1.4 and 1.7 $R_{25}$ is brighter because fewer sub-integrations contribute to the 
final mosaic at the edges. 

On the other hand, the inability to add sources in the brightest areas of the bulge and arms implies that 
they are strongly affected by confusion, due to partially resolved stars and star clusters. 
Indeed, we are not reporting the detection of any real sources there 
(see Figure~\ref{fig:spatial}), precisely because no objects have been found in those areas that are not blended or close to
bright neighbors. We are aware that we have very likely missed GCCs in the brightest regions
close to the center. 
On this same issue, it is truly remarkable that the galactic disk between $\sim$ 0.5 and 1 $R_{25}$ is not having a very significant effect, neither on the addition of artificial sources nor on their recovery, compared with regions farther away from the center of the galaxy. An inspection of Figure~\ref{fig:cmpltnss} reveals that confusion should not be hampering our GCC detection beyond $\sim 0.4 R_{25}$. 
Given the very small completeness corrections that we have derived beyond 0.5 $R_{25}$ (and that we will duly apply in Section~\ref{subsec:gclf}), the most important result of the completeness simulations is precisely this realization. 

\section{The \uiks\ diagram}\label{sec:uiks}

Color-color diagrams are routinely used as efficient tools for the classification and 
selection of sources in astrophysical surveys. In particular, for GCs, examples include
(\rp\ - \ip) vs.\ (\gp\ - \ip) with CFHT data \citep{pota15}; (F438W - F606W) vs.\ (F606W - F814W) 
with Hubble Space Telescope ({\sl HST}) filters \citep{fedo11}, and ($U\ - B$) vs.\ ($V\ - I$) with Very Large 
Telescope (VLT) images \citep{geor06}. 

Recently, \citet{muno14} have shown that the combination of optical and near infrared (NIR) data
into the (\ust - \ip) vs.\ (\ip - \ks) color-color diagram provides the most powerful 
photometric only method for the selection of a clean sample of GC candidates.
The \uiks\ diagram in fact makes use of the whole spectral range between the ultraviolet (UV) 
and NIR atmospheric cutoffs, at 3200 $\AA$ and 2.2 $\mu$m, respectively. Thus, it samples
simultaneously the main-sequence turnoff, and the red giant and horizontal branches of the stellar populations in GCs.
In the \uiks\ plane, the nearly simple stellar populations (SSPs) of GCs occupy a region that is well separated 
from, respectively, the loci of the composite stellar populations of background galaxies, and of foreground
stars in our own Galaxy. 
As demonstrated also by \cite{muno14}, the separation of these 3 distinct regions
(i.e., GCs, galaxies, and MW stars) from the point of view of stellar populations is completely consistent
with the shape parameters of the objects that inhabit them. In the case of Virgo data acquired with the
CFHT\footnote{Next Generation Virgo Survey \citep[NGVS;][]{ferra12} and NGVS-IR \citep{muno14}.}, 
sources in the galaxy region are clearly extended, those in the star band are
pointlike, and \edit1{many} GCCs are marginally resolved. 

\begin{figure}[ht!]
\plotone{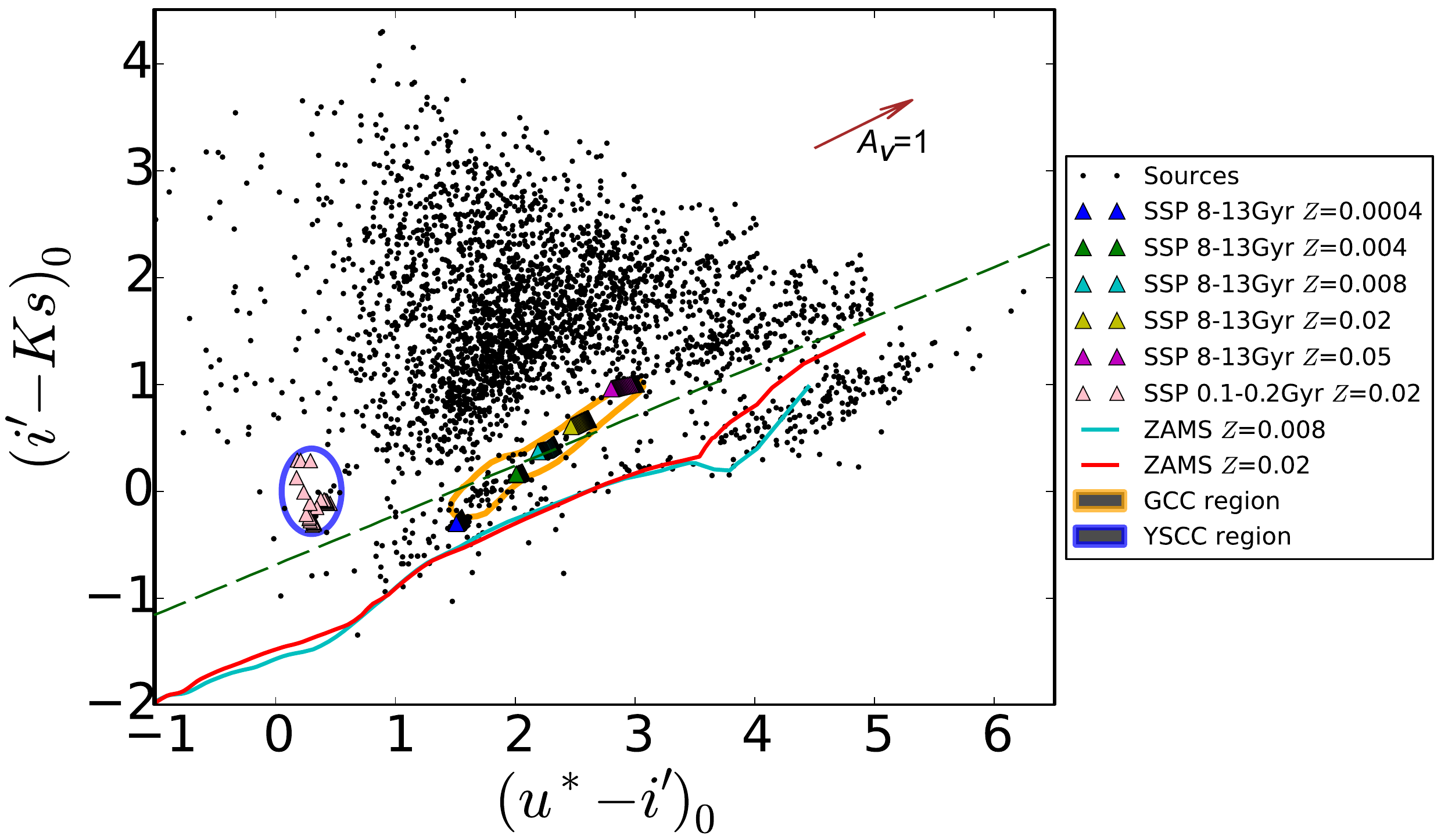}
\caption{\uiks\ color--color diagram of NGC~4258. {\it Black dots:} sources with MAGERR\_PSF $<$ 0.2 mag in the \ip-band.  
Clusters of triangles inside the orange ellipse are age sequences, from 8 to 12 Gyr, of old SSPs with a single metallicity, i.e.,
{\it blue:} $Z=$ 0.0004; {\it dark green:} $Z=$ 0.004; {\it cyan:} $Z=$ 0.008; {\it olive:} $Z=$ 0.02; 
{\it purple:} $Z=$ 0.05. Data points inside the orange ellipse are GCCs. {\it Pink triangles} inside the blue ellipse are an age sequence, from 10$^7$ to 10$^8$ yr, for an 
SSPs with solar metallicity; data points there are YSCCs.  The {\it cyan} and {\it red} lines sketch the loci of ZAMSs with $Z=$ 0.008 and $Z=$ 0.02, respectively. 
\label{fig:uiks}}
\end{figure}

Figure~\ref{fig:uiks} presents the \uiks\ plot for our NGC~4258 data (small solid black dots).
Sources were selected by cross-matching detections by RA, DEC in our \ust, \ip, and \ks\ catalogs, with a tolerance of 
1$^{\prime\prime}$.  The data were corrected for Galactic extinction with the \citet{schlaf11} values given for the 
Sloan Digital Sky Survey (SDSS) filters by the 
NASA Extragalactic Database:\footnote{The NASA/IPA Extragalactic Database (NED) is operated by the Jet
Propulsion Laboratory, California Institute of Technology, under contract with the National Aeronautics and 
Space Administration.} $A_u = 0.069$; $A_g = 0.054$; $A_r = 0.037$; $A_i = 0.028$; $A_{Ks} = 0.005$.
Only data with SExtractor FLAGS$=$0 in all filters, and an error MAGERR\_PSF $<$ 0.2 mag in MAG\_PSF in the \ip-band were included.
Also, since we are interested in GCs, whose luminosity function (GCLF) is practically universal, we
only kept those objects within $\pm\ 3\sigma$ of the expected LF turnover (LFTO) magnitude in 
every filter; \edit1{we assumed $\sigma = 1.2$ mag which, according to \citet{jord07}, is appropriate given the absolute
total dereddened $B$ mag of NGC~4258 \citep[-20.87 mag;][]{rc3}. Values of the turnover in the optical were derived 
by combining the absolute TO magnitude in the $g$ band, $M^0_g = -7.2$ mag, for the same galaxy luminosity \citep{jord07},} with 
the (AB) colors given in the MegaCam filter system by \citet[][hereafter BC03]{bruz03} models for an SSP with $Z = 0.0004$, an age of 12 Gyr, and
a Chabrier initial mass function \citep[IMF;][]{chab03}. \edit1{The TO in the $V$-band would be $M^0_V = -7.4$ mag.} 
The TO magnitude in the \ks-band was taken from 
\citet{wang14}.\footnote{\citet{wang14} determine a TO magnitude $K_{s0} = 14.534^{+0.142}_{-0.146}$ mag for the M~31 GC system, with a Vega zero point. We apply to this value the distance modulus of Andromeda, and the transformation between 
Vega and WIRCam AB magnitudes (see Section~\ref{sec:colordist}).
} 
TO absolute magnitudes, observed TO magnitudes at the distance of NGC~4258, and magnitude ranges of the GCLF are shown in Table~\ref{tab:TOdat}.  

\floattable
\begin{deluxetable}{cccc}
\tablecaption{Turnover magnitudes and GCLF ranges\label{tab:TOdat}}
\tablewidth{20cm}
\tablehead{
\colhead{Filter} & \colhead {$M^0$} & \colhead{$m^0_{\rm TO}$} & \colhead{$m^0_{\rm TO} \pm\ 3\sigma$} \\
\colhead {} & \colhead{mag} & \colhead{mag} & \colhead{mag} 
}
\decimals
\startdata
\ust &\edit1{  -6.34} &\edit1{  23.06}  & \edit1{$19.46 < m^0_{u^*} < 26.66$} \\
\gp &\edit1{  -7.20} &\edit1{  22.20}  & \edit1{$18.60 < m^0_{g^\prime} < 25.8$} \\
\rp &\edit1{  -7.69} &\edit1{  21.71}  & \edit1{$18.11 < m^0_{r^\prime} < 25.31$} \\
\ip &\edit1{  -7.92} &\edit1{  21.48}  & \edit1{$17.88 < m^0_{i^\prime} < 25.08$} \\
\ks &  -8.1 &  21.3  &\edit1 {$17.7 < m^0_{Ks} < 24.9$} \\
\enddata
\end{deluxetable}

In Figure~\ref{fig:uiks}, the green long-dashed line 
separates the cloud of background 
galaxies from the band of foreground stars in the MW:
objects above the line are mostly extended, sources below it are preferentially point sources,
as we will show in the next subsection.
\edit1{The orange contour highlights the region that contains GCCs.
This region was traced based on the area with the highest density of spectroscopically confirmed
GCs in the \uiks\ diagram of M~87 \citep{muno14,powa16}; there are more than 2000 such sources
in the GC system of M~87, and
hence the GCC locus is clearly delimited.   
Inside the region in our plot for NGC~4258, we indicate 
the location of BC03 model SSPs.} Each cluster of triangles 
represents an age sequence with a single metallicity, from \edit1{$Z =$ 0.0004} to $Z =$ 0.05. 
Within each age sequence, from bluer to redder colors, models go from 8 to 12 Gyr. 
For illustration purposes, 
zero age main sequences (ZAMS) 
with $Z = 0.008$ and $Z = 0.02$ are shown; 
they sketch the locus of field stars in the MW. 

Because we are interested in the GC systems of nearby spiral galaxies that are not completely edge-on, it was important to
investigate the locus in the \uiks\ plane of young stellar cluster candidates (YSCCs).  
The blue ellipse has been defined  
around BC03 models of SSPs with solar metallicity and ages, from blue to red in (\ip\ -\ks), between 10$^7$ and 10$^8$ yr 
(pink triangles). They are reasonably far away from models of old stellar populations of all metallicities.
Moreover, we would like to emphasize the direction of
the reddening vector in this parameter space. Although extinction could make a GC look older or more metal rich if detected through the disk of 
its parent galaxy, reddening would not be able to make a YSC in the disk of the galaxy look like an old GC.  
This is a property of the \uiks\ diagram that makes it an ideal tool to study GC systems in 
spiral galaxies.
 
\subsection{Shape parameters} \label{subsec:shape}

\begin{sidewaysfigure}[ht]
\plotone{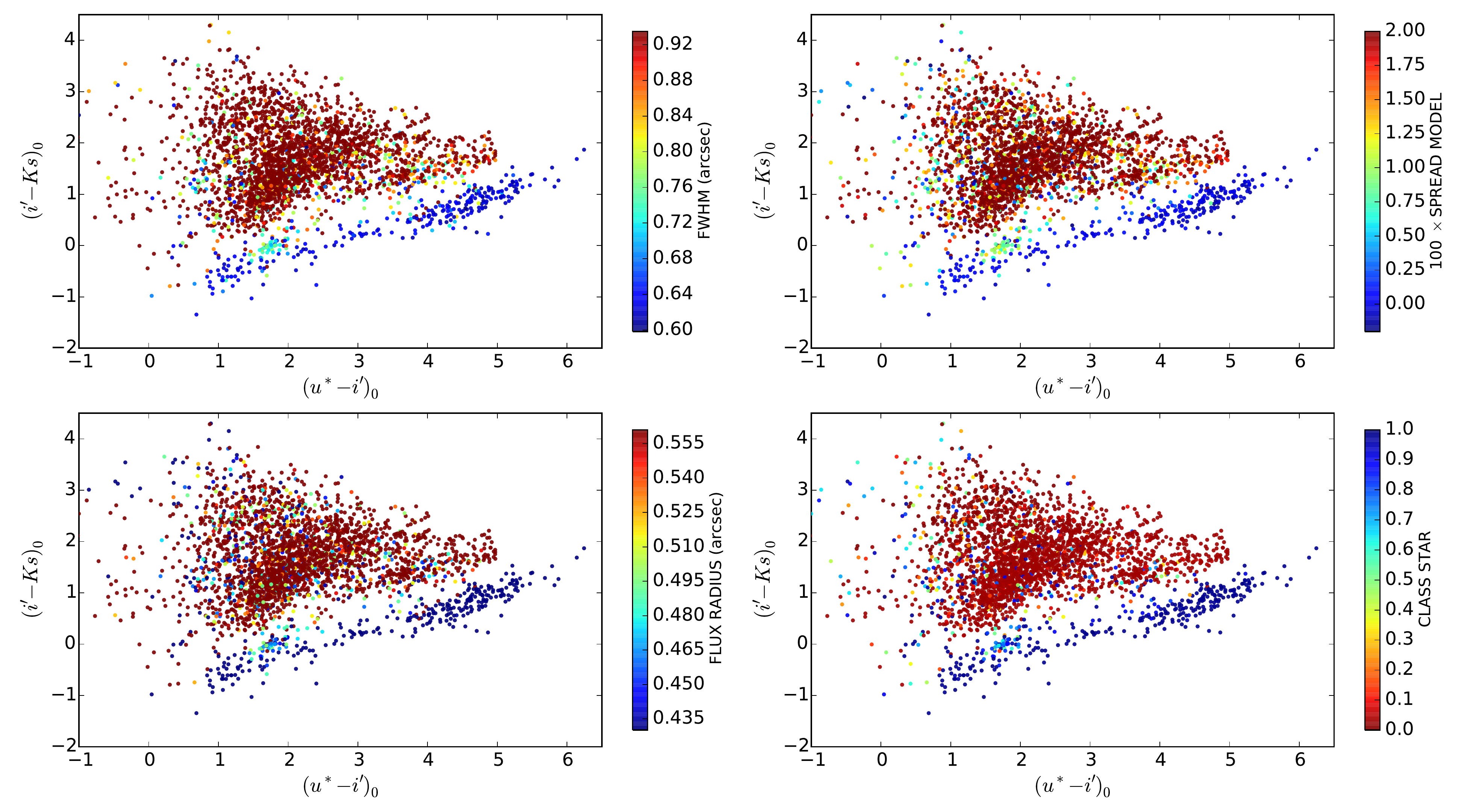}
\caption{Morphology variations across the \uiks\ diagram of NGC~4258. Four different indicators of shape in the \ip-band, 
with values coded as indicated by color bars:
FWHM ({\it top left}); SPREAD\_MODEL ({\it top right}); FLUX\_RADIUS ({\it bottom left}); CLASS\_STAR ({\it bottom right}). 
Marginally resolved sources sport shades of cyan in all indeces.
\label{fig:uiks_shapes}}
\end{sidewaysfigure}

Regarding the shape of the sources on the \uiks\ plot, Figure~\ref{fig:uiks_shapes} shows four different gauges of
compactness in the \ip-band: FWHM (top left); SPREAD\_MODEL (top right); 
FLUX\_RADIUS (bottom left); 
CLASS\_STAR\footnote{
The SExtractor dimensionless index CLASS\_STAR is assigned by a neural network, 
based on the comparison between the 
source and the PSF. It goes from 0 for extended objects to 1 for a perfect point source 
\citep{bert96,hamm10}.} (bottom right).
Their values are coded as indicated by the color bars, and all of them show a correlation between locus in the color-color diagram and light concentration. Objects in the background galaxy cloud (mostly reddish-brown)
are more extended than those in the Galactic star band (mostly blue), and \edit1{many} GCCs at the distance of NGC~4258 appear marginally resolved (cyan). Typical GC have effective radii between one and 20 pc \citep[e.g.]{vand95,jord05,puzi14}; one MegaCam pixel = 6.9 pc at the adopted distance to NGC~4258, and the PSF FWHM of the
\ip\ data is 0$\farcs$60 or 22 pc.    
In Figure~\ref{fig:uiks_shapes} one can see that point sources and marginally resolved 
objects have FWHM $\la 0\farcs8$; SPREAD\_MODEL $<$ 0.015; 
FLUX\_RADIUS $\la 0\farcs5$; \edit1{CLASS\_STAR $\ga$0.4}. Although not discrete, the transition value of SPREAD\_MODEL between point sources and extended objects lies between
$\sim$ 0.003 and 0.005, consistent with those reported by, respectively, \citet{desa12} and \citet{annu13}. 
Separation by compactness is most efficient for FWHM and SPREAD\_MODEL, 
i.e., these parameters provide the best contrast between \edit1{point sources, marginally resolved, and extended sources}. 

\subsection{Other diagrams}\label{subsec:otherccds}

\begin{sidewaysfigure}[ht]
\begin{tabular}{lll}
\includegraphics[scale=0.28]{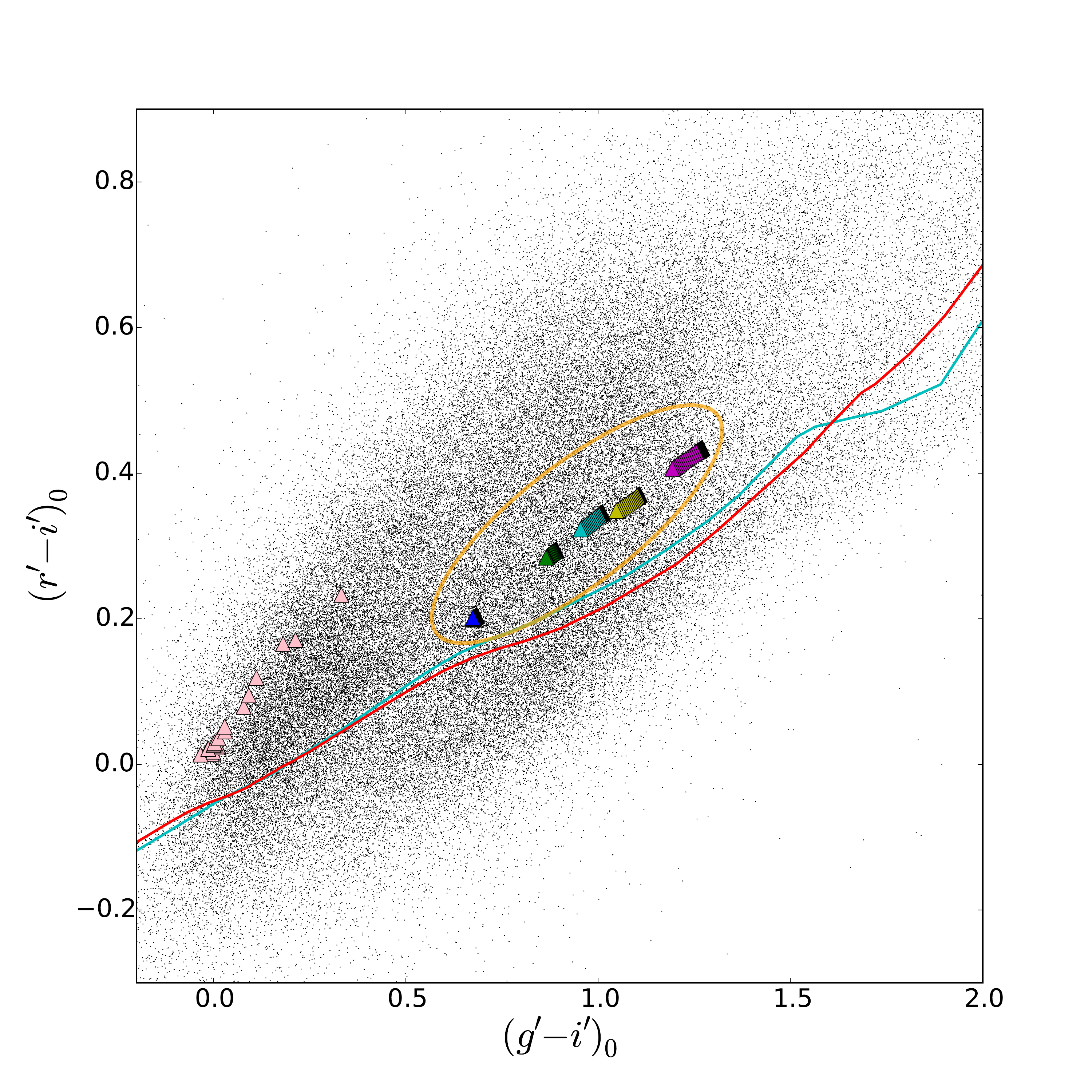}
&
\includegraphics[scale=0.28]{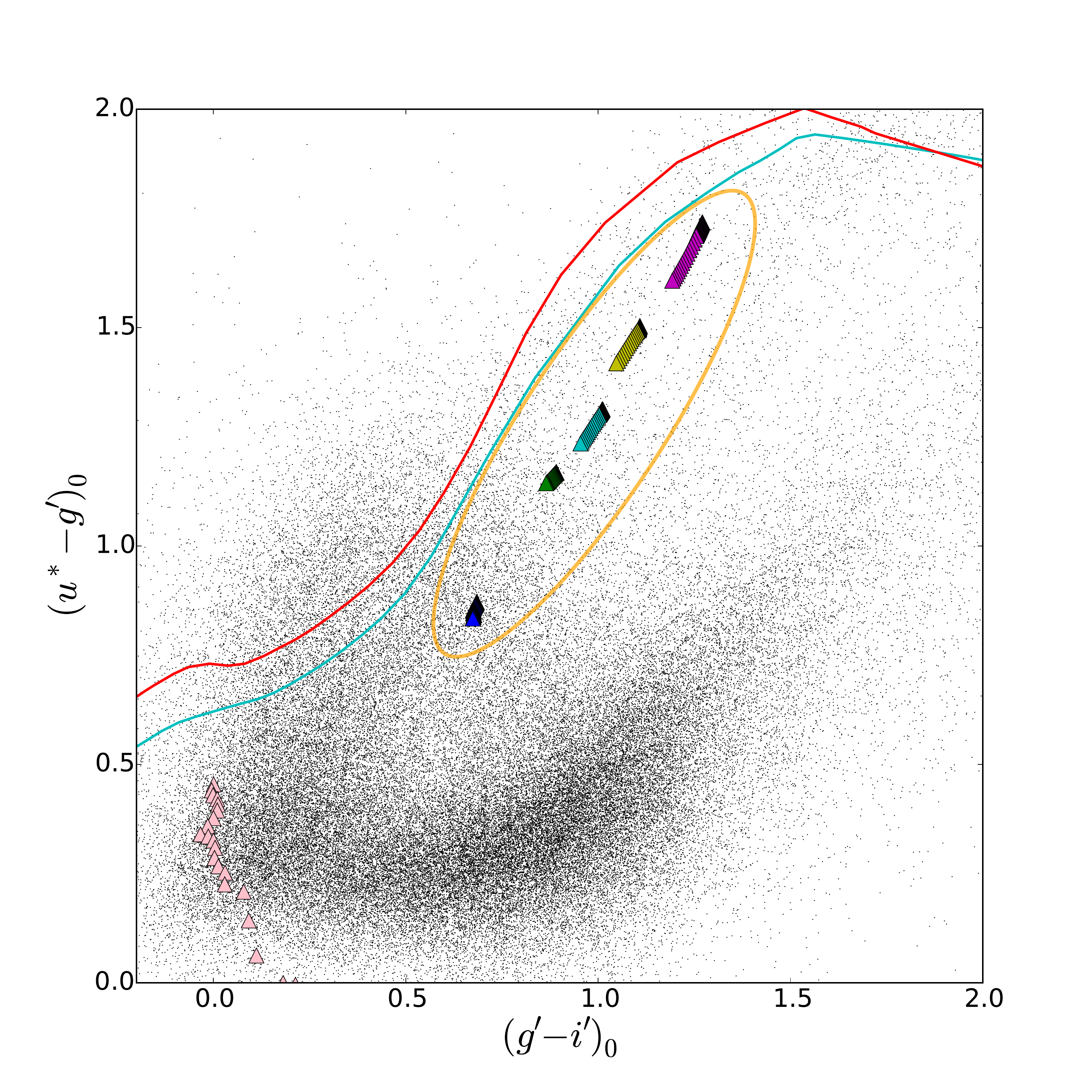}
&
\includegraphics[scale=0.281]{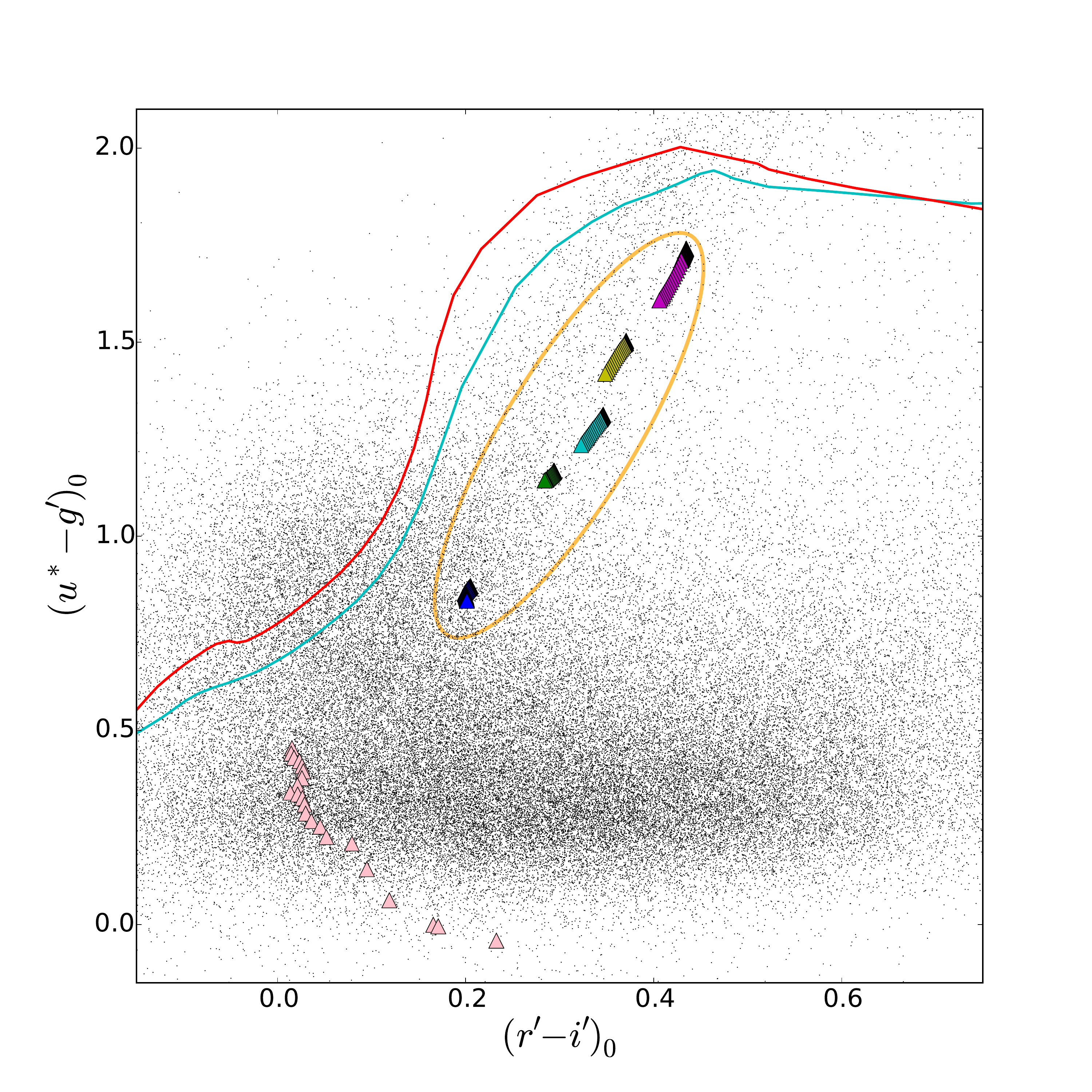}
\end{tabular}
\caption{Alternative color-color diagrams. {\it Left:} (\rp\ - \ip) vs.\ (\gp\ - \ip); {\it middle:} (\ust\ - \gp) vs.\ (\gp\ - \ip); {\it right:} (\ust\ - \gp) vs.\ (\rp\ - \ip). Symbols for model SSPs and ZAMS as in Figure~\ref{fig:uiks}.
\label{fig:otherdiag}}
\end{sidewaysfigure}

For comparison, we show in Figure~\ref{fig:otherdiag} 
color-color diagrams for the sources in the
NGC~4258 field, with other combinations of CFHT filters. These 
are analogous to those used in works by \citet{geor06}, \citet{fedo11}, and \citet{pota15}. The three 
diagrams have many more points than our \uiks\ diagram, since all available archival optical data are much deeper
than our \ks\ image.  

\begin{figure}[ht!]
\plotone{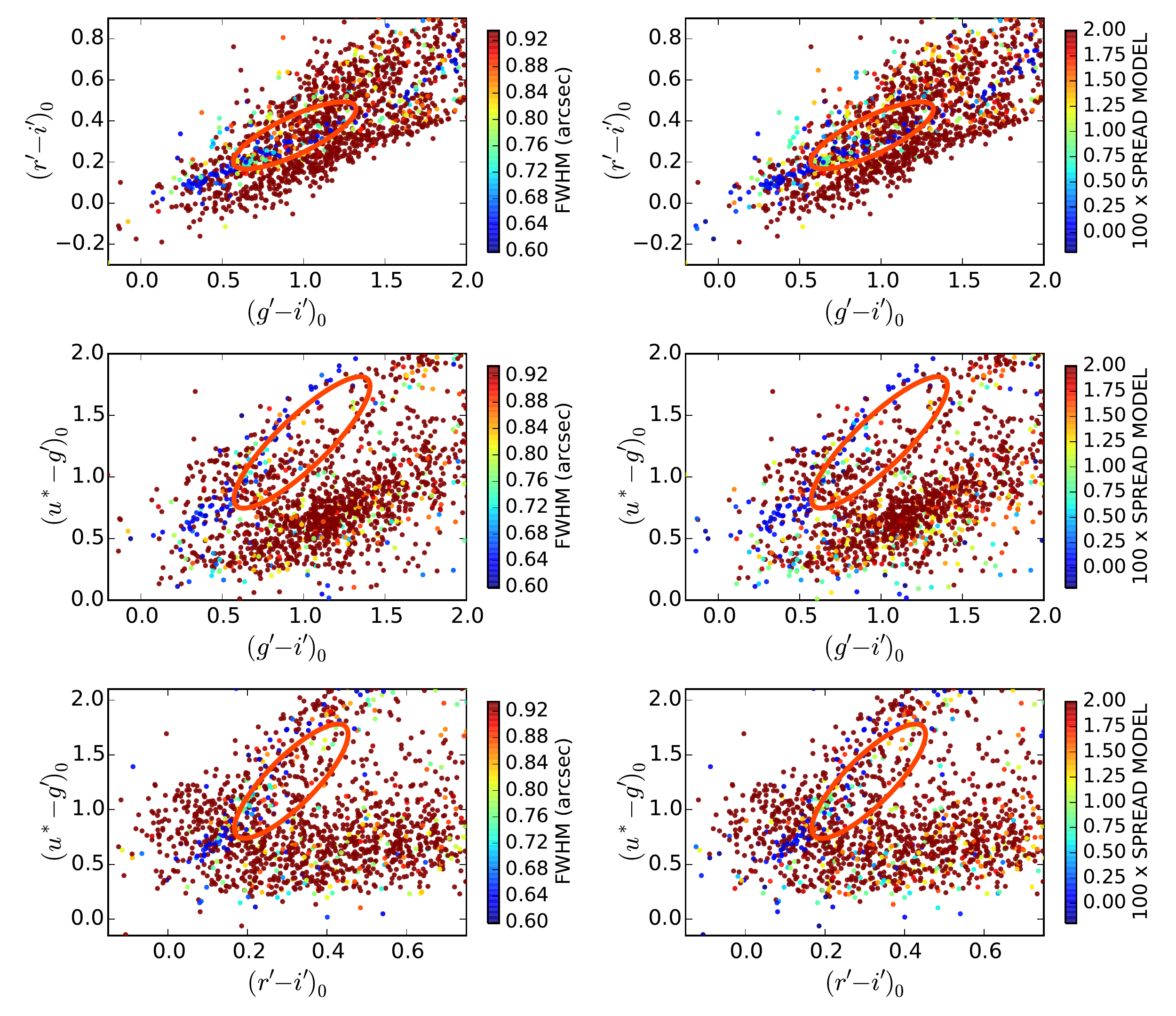}
\caption{Morphology variations in the \ip-band across alternative color-color diagrams of NGC~4258. 
{\it Top:} (\rp\ - \ip) vs.\ (\gp\ - \ip); {\it middle: } (\ust\ - \gp) vs.\ (\gp\ - \ip); 
{\it bottom:} (\ust\ - \gp) vs.\ (\rp\ - \ip).
{\it Left column:} FWHM; {\it right column:} SPREAD\_MODEL. \edit1{Orange} 
ellipses highlight loci of GCCs; values of shape estimators coded as in Figure~\ref{fig:uiks_shapes}.  
\label{fig:otherdiagshapes}}
\end{figure}

The plots display models of young and old SSPs, as well as ZAMS, with the same ages and metallicities as in Figure~\ref{fig:uiks}. With the aid of the old SSP models, we have outlined with orange ellipses the expected loci of GCCs.
In all cases, the contamination by background galaxies and foreground Galactic stars inside the ellipse is much more severe than with the \uiks\ diagram. 
The mixture of source types in the ellipse is quite evident in Figure~\ref{fig:otherdiagshapes}, where the shape estimators FWHM and SPREAD\_MODEL in the \ip-band are \edit1{color-coded as previously, only for the same objects already included in Figures~\ref{fig:uiks} and~\ref{fig:uiks_shapes}} . 
Both outside and within the ellipses, all ranges of light concentration can be found, rendering it questionable to select GCCs based on compactness and color-color diagram locus alone.

\section{GCC final sample} \label{sec:finalsample}

So far, with the aim of determining a sample of GC candidates for NGC~4258 we have imposed several restrictions: 
objects must have MAGERR\_AUTO $< 0.2$ mag in the \ip-band;  
lie within $\pm\ 3\sigma$ of the expected GCLF turnover magnitude in
every filter, assuming \edit1{$\sigma = 1.2$ mag}; and be inside the selection region outlined in orange in 
the \uiks\ color-color diagram (Figure~\ref{fig:uiks}). 

In order to select the most likely candidates, we also rely on the shape parameters explored in 
the previous sections. 
We find that the FWHM and SPREAD\_MODEL parameters calculated by SExtractor are particularly efficient
in separating \edit1{compact and marginally resolved sources in the selection region from more extended 
ones (see Figure~\ref{fig:uiks_shapes}). 
Hence, we keep sources with SPREAD\_MODEL $\leqslant$ 0.017, and FWHM $\leqslant 0\farcs84$ (less than 4.5 MegaCam pixels)} in the \ip-band. 

Given the distance to NGC~4258 (7.6 Mpc) and the angular resolution of the MegaCam data
(one MegaCam pixel $= 0\farcs186 \simeq$ 6.9 pc), we have likewise been able to calculate 
the half-light radii, $r_e$, in the \ip-band (FWHM $= 0\farcs6 \simeq$ 22 pc), of all the sources 
inside the selection \edit1{region (outlined in} orange in Figure~\ref{fig:uiks}) 
\edit1{following the procedure described in \citet{geor14}. 
To this end, we used the program {\sc ishape} \citep{lars99} in the BAOLAB\footnote{\url{http://baolab.astroduo.org}.} software package.
{\sc ishape} measures the size of a compact source by comparing its observed light profile with a suite of  model
clusters, generated by convolving different analytical profiles with the image PSF. 
For data with S/N $\ga$ 30, {\sc ishape} can measure $r_{\rm eff}$ reliably down to $\sim$ 0.1 the 
PSF FWHM, or $\sim$ 2.2 pc at the distance of NGC~4258. Objects smaller than this are effectively unresolved
\citep{lars99,harr09}. 
We constructed a spatially variable PSF from 172 isolated stars,  
and we fitted all the sources with King profiles
\citep{king62,king66} with fixed (KINGx) concentration indices ($C \equiv r_{\rm tidal}/r_{\rm core}$) of 30 and 100, and  
with the concentration index left as a free parameter (KINGn) as well. For each object, the model providing the 
fit to the data with the smallest $\chi^2$ residuals was then used to derive the effective radius. This was done by
applying the conversion factors between $r_{\rm eff}$ and FWHM given in the {\sc ishape} manual. 
For each source, Table \ref{tab:gccshapes} lists its measured $r_e$, the type and $C$ index of the best-fitting King profile,
the SNR of the data, and the $\chi^2$ of the fit. 
}

\edit1{All objects left after the cuts in SPREAD\_MODEL and FWHM have $r_e \leqslant$ 5.9 pc.}    
Finally, we eliminate \edit1{four additional objects. One} 
has colors, other than (\ust\ - \ip) and (\ip - \ks),
that are redder than the reddest BC03 model SSPS defining GCC selection regions in alternative color-color
diagrams (see Section~\ref{subsec:otherccds} and Figure~\ref{fig:otherdiag}); \edit1{it is} likely affected by
dust. \edit1{For three other ones, the $r_e$ fit did not converge; one
seems to be the nucleus of the dwarf galaxy SDSS J121909.07+470523.1, located to the south of NGC 4258.
} 

\begin{figure}[ht!]
\plotone{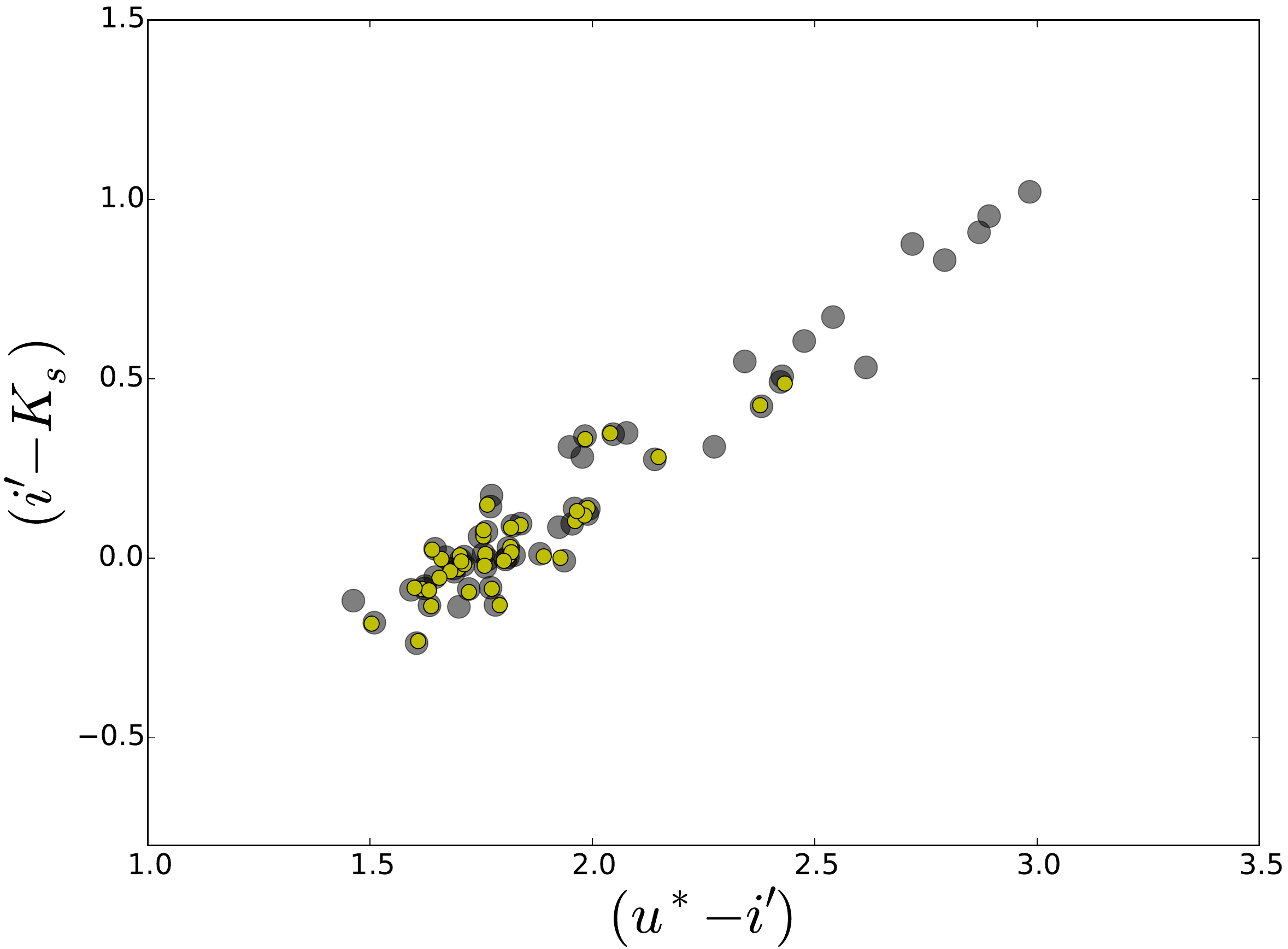}
\caption{Location of the GCC final sample in the \uiks\ color-color diagram. 
Objects in the final selection are shown as {\it yellow dots}. {\it Gray dots:} all \edit1{58 objects in the
orange delimited selection region} in Figure~\ref{fig:uiks}. 
\label{fig:ama_uiks}}
\end{figure}

Our definitive sample comprises \edit1{39} objects, all confirmed by visual inspection. Their locations in the
\uiks\ diagram are shown by the yellow points in Figure~\ref{fig:ama_uiks}; 
the rest of the objects in the selection region 
are displayed in gray. 

\begin{figure}[ht!]
\plotone{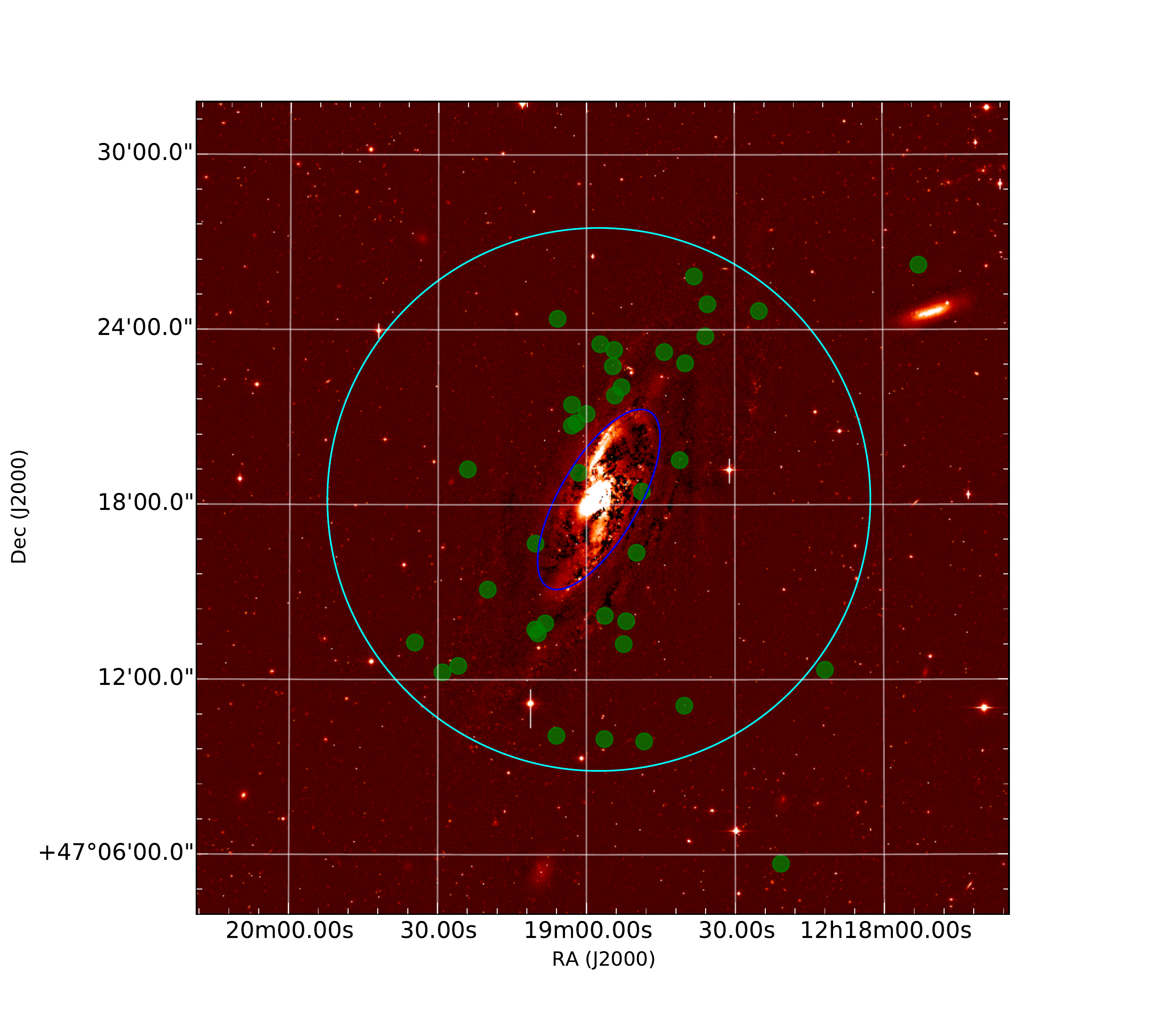}
\caption{Spatial distribution of the GCC sample in NGC~4258. 
The GCCs are shown as {\it green filled circles}. 
{\it Dark blue ellipse:} boundary of area where source confusion is
highest, with major axis $= 0.37 R_{25}$;  
{\it solid cyan line:} $R_{25}$. 
\label{fig:spatial}}
\end{figure}

The spatial distribution is shown in Figure~\ref{fig:spatial}.
The members of the sample are displayed as green circles on the \ip-band 
 image of NGC~4258. The dark blue ellipse outlines the region within 0.37 $R_{25}$, where 
source confusion is highest. The cyan circle indicates $R_{25}$.
\edit1 {Slightly over} 90\% of the GCCs lie within $R_{25}$. We note that \edit1 {one object} outside $R_{25}$, to the northwest
of NGC~4258, \edit1{is} projected close to its companion NGC~4248, an irregular non-magellanic galaxy \citep{rc3} with
roughly the same $B$ luminosity as the Small Magellanic Cloud. 
Colors and concentration parameters of the GCCs in the sample are given, respectively, in Table~\ref{tab:gcccolors} and 
Table~\ref{tab:gccshapes}.\footnote{Errors in the colors are only random, and do not consider correlations between bands.} 

\begin{sidewaystable}[ht]
 \caption{Colors of Globular Cluster Candidates}
  \begin{scriptsize}
 \begin{center}
\hspace*{-1.2cm} \begin{minipage}{280mm}
\vspace*{-0.6cm}
  \begin{tabular}{@{}ccccccccccccccc@{}}
\hline
\hline
\vspace*{-0.3cm}
&&&&&&&&&&&&&&\\
Name &   RA J2000 & DEC J2000  & $g^\prime_0$ &  $\Delta g^\prime_0$ & $K_{s,0}$ & $\Delta K_{s,0}$ & ($u^\ast\ - i^\prime$)$_0$ & $\Delta$ ($u^\ast\ - i^\prime$)$_0$ & ($g^\prime\ - r^\prime$)$_0$ & $\Delta$ ($g^\prime\ - r^\prime$)$_0$ & ($g^\prime\ - i^\prime$)$_0$ & $\Delta$ ($g^\prime\ -i^\prime$)$_0$ & ($i^\prime\ - K_s$)$_0$ & $\Delta$ ($i^\prime\ - K_s$)$_0$  \\
     &    deg     &   deg      &    mag      &      mag           &    mag   &     mag          &          mag              &            mag                     &             mag            &             mag                      &     mag                    &       mag                           &           mag          &             mag                                 \\
\hline
GLL J121752+472613 & 184.4696 & 47.4370 & 21.686 & 0.006 & 21.08 & 0.16 & 1.690 & 0.009 & 0.443 & 0.009 & 0.646 & 0.007 & -0.04 & 0.16 \\
GLL J121811+471220 & 184.5494 & 47.2056 & 19.909 & 0.002 & 19.25 & 0.02 & 1.710 & 0.003 & 0.470 & 0.003 & 0.675 & 0.003 & -0.02 & 0.02 \\
GLL J121820+470541 & 184.5867 & 47.0948 & 19.165 & 0.002 & 18.14 & 0.02 & 1.771 & 0.002 & 0.582 & 0.002 & 0.883 & 0.002 & 0.14 & 0.02 \\
GLL J121825+472438 & 184.6045 & 47.4107 & 20.455 & 0.003 & 19.16 & 0.02 & 1.983 & 0.005 & 0.640 & 0.004 & 0.956 & 0.004 & 0.34 & 0.02 \\
GLL J121835+472452 & 184.6479 & 47.4146 & 19.809 & 0.002 & 19.15 & 0.02 & 1.688 & 0.004 & 0.488 & 0.003 & 0.692 & 0.003 & -0.03 & 0.02 \\
GLL J121835+472346 & 184.6497 & 47.3963 & 21.410 & 0.006 & 20.65 & 0.08 & 1.712 & 0.009 & 0.505 & 0.008 & 0.752 & 0.007 & 0.00 & 0.08 \\
GLL J121838+472549 & 184.6593 & 47.4305 & 21.320 & 0.006 & 20.54 & 0.07 & 1.882 & 0.009 & 0.556 & 0.008 & 0.766 & 0.007 & 0.01 & 0.07 \\
GLL J121840+472251 & 184.6668 & 47.3809 & 22.075 & 0.007 & 21.11 & 0.12 & 1.992 & 0.013 & 0.591 & 0.011 & 0.830 & 0.009 & 0.14 & 0.12 \\
GLL J121840+471106 & 184.6677 & 47.1852 & 21.102 & 0.004 & 20.44 & 0.06 & 1.805 & 0.007 & 0.449 & 0.007 & 0.664 & 0.006 & -0.00 & 0.06 \\
GLL J121841+471931 & 184.6715 & 47.3255 & 20.653 & 0.003 & 19.94 & 0.04 & 1.746 & 0.005 & 0.467 & 0.005 & 0.656 & 0.004 & 0.06 & 0.04 \\
\hline
\vspace*{-0.8cm}
\end{tabular}
\end{minipage}
\end{center}
\hspace*{-1.15cm}{{\sc Note}---Table~\ref{tab:gcccolors} is published in its entirety in the machine-readable format.
      A portion is shown here for guidance regarding its form and content.
}
\label{tab:gcccolors}
\end{scriptsize}
\end{sidewaystable}

\begin{sidewaystable}[ht]
 \caption{\ip-Band Shape Parameters of Globular Cluster Candidates}
  \begin{scriptsize}
 \begin{center}
\hspace*{-1.2cm} \begin{minipage}{280mm}
\vspace*{-0.6cm}
  \begin{tabular}{@{}cccccccccrrc@{}}
\hline
\hline
\vspace*{-0.3cm}
&&&&&&&&&&&\\
 Name  &     RA J2000  &   DEC J2000   &    100 $\times$(SPREAD\_MODEL)$_{i^\prime}$  &   FWHM$_{i^\prime}$  &  CLASS\_STAR$_{i^\prime}$ & FLUX\_RADIUS$_{i^\prime}$  &   $r_{\rm e, i^\prime}$  &   Shape  &    $C$ index   &     SNR     &     $\chi^2$ \\
       &      deg      &     deg       &                                             &     arcsec          &                         &      arcsec               &           pc            &          &            &             &                                  \\
%
GLL J121752+472613 & 184.4696 & 47.4370 & 0.88 & 0.72 & 0.90 & 0.43 & 5.27$^{+3.97}_{-4.13}$ & KINGn  & 11.4 & 33.60 & 0.24 \\
GLL J121811+471220 & 184.5494 & 47.2056 & 0.84 & 0.72 & 0.79 & 0.46 & $<$ 2.2                & KINGn  & 289.2 & 168.20 & 2.83 \\
GLL J121820+470541 & 184.5867 & 47.0948 & 0.05 & 0.62 & 0.99 & 0.39 & 1.35$^{+4.80}_{-1.32}$ & KINGn  & 0.0 & 337.30 & 0.67 \\
GLL J121825+472438 & 184.6045 & 47.4107 & 0.92 & 0.72 & 0.80 & 0.49 & $<$ 2.2                & KINGn  & 635.0 & 142.20 & 2.84 \\
GLL J121835+472452 & 184.6479 & 47.4146 & 1.20 & 0.76 & 0.62 & 0.50 & 2.31$^{+0.41}_{-1.70}$ & KINGn  & 86.2 & 191.90 & 6.50 \\
GLL J121835+472346 & 184.6497 & 47.3963 & 0.82 & 0.71 & 0.56 & 0.44 & 2.64$^{+0.19}_{-0.20}$ & KINGx  & 30.0 & 46.90 & 0.37 \\
GLL J121838+472549 & 184.6593 & 47.4305 & 0.78 & 0.72 & 0.95 & 0.44 & 2.78$^{+0.24}_{-0.13}$ & KINGx  & 30.0 & 51.80 & 0.47 \\
GLL J121840+472251 & 184.6668 & 47.3809 & 0.77 & 0.73 & 0.98 & 0.42 & 5.24$^{+1.81}_{-4.34}$ & KINGn  & 12.0 & 28.80 & 0.27 \\
GLL J121840+471106 & 184.6677 & 47.1852 & 0.88 & 0.70 & 0.82 & 0.47 & $<$ 2.2                & KINGn  & 720.9 & 63.50 & 0.56 \\
GLL J121841+471931 & 184.6715 & 47.3255 & 0.90 & 0.72 & 0.62 & 0.59 & $<$ 2.2                & KINGx  & 100.0 & 81.20 & 1.06 \\
\hline
\vspace*{-0.8cm}
\end{tabular}
\end{minipage}
\end{center}
\hspace*{-1.15cm}{{\sc Note}---Table~\ref{tab:gccshapes} is published in its entirety in the machine-readable format.
      A portion is shown here for guidance regarding its form and content.
}
\label{tab:gccshapes}
\end{scriptsize}
\end{sidewaystable}

\subsection{Decontamination} \label{sec:decont}

In order to estimate the possible number of contaminants (foreground stars and background
galaxies) in our sample, we use as control field CFHT observations within the Extended Groth Strip (EGS). 
The EGS is one of the deep fields that have been observed repeatedly in
the last two decades in all the regions of the electromagnetic spectrum
accessible with current technology. The original motivation to observe these fields, of which the first
one was the famous Hubble Deep Field \citep[HDF,][]{will96}, is the
study of the distant universe. Therefore, they are preferably at high Galactic latitudes,
in order to reduce the contribution from sources in the Milky Way, and they avoid lines of sight with
known nearby galaxies.

The EGS is a 1$\fdg$1 by 0$\fdg$15 region located at RA = 14h19m17.84s, DEC= +52d49m26.49s (J2000), in the constellation
Ursa Major. 
Observations of the EGS are coordinated by the
All-Wavelength EGS
International Survey \citep[AEGIS;][]{davi07} project.
In particular, the EGS is one of the four deep fields observed by CFHT within its
Legacy Survey \citep[CFHTLS;][]{gwyn12}, and the one closest to
NGC~4258 in angular distance (19$\fdg$9). The D3 field of the CFHTLS
consists of observations 
within the EGS centered at RA = 14h19m27.00s, DEC= +52d40m56s,
1 square degree in the bands $u^*$, $g^\prime$, $r^\prime$, $i^\prime$, and $z^\prime$, and 0.4165 square degree 
in $J$, $H$, and $K_s$. The Galactic extinction values in this direction are
$A_u = 0.037$, $A_i = 0.015$ and $A_K = 0.003$ \citep{schlaf11}.
All final mosaics are available in the CFHT archive,\footnote{
\url{http://www.cadc-ccda.hia-iha.nrccnrc.gc.ca/en/cfht/}.}
and have a pixel scale of 0$\farcs$186. The exposure times of the EGS D3 CFHTLS mosaics at \ust, \ip, and \ks\ are, respectively, 19800 s, 64440 s, and 17500 s.

\begin{figure}[ht!]
\hspace*{0.5cm}\includegraphics[scale=0.45]{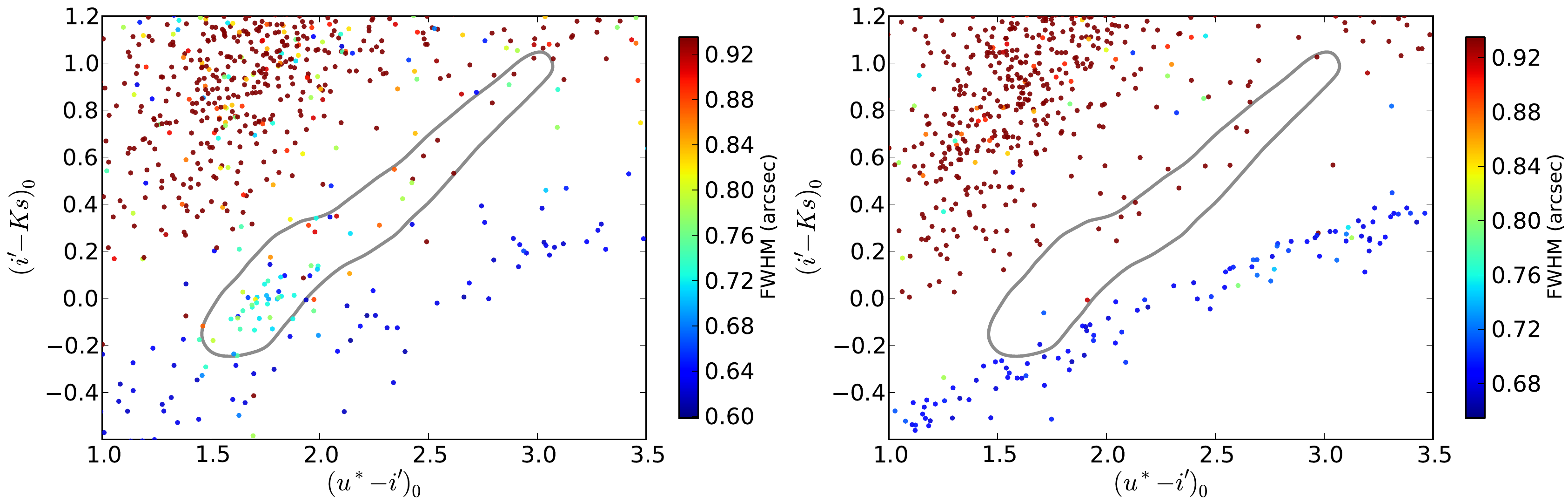}
\hspace*{0.5cm}\includegraphics[scale=0.45]{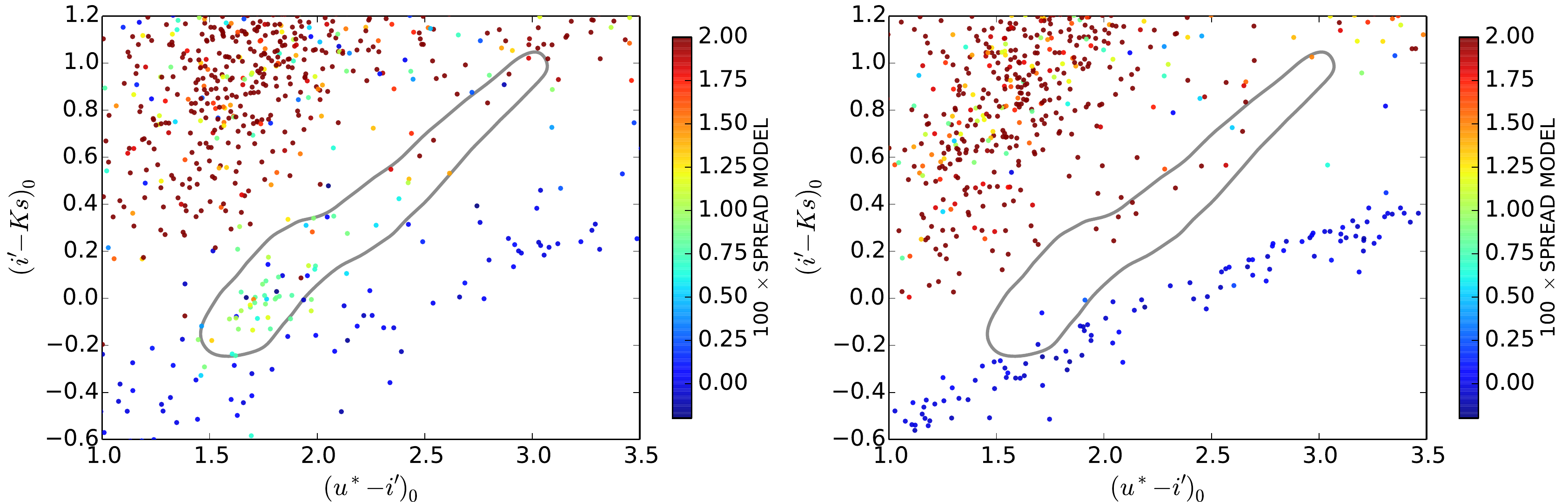}
\hspace*{0.5cm}\includegraphics[scale=0.45]{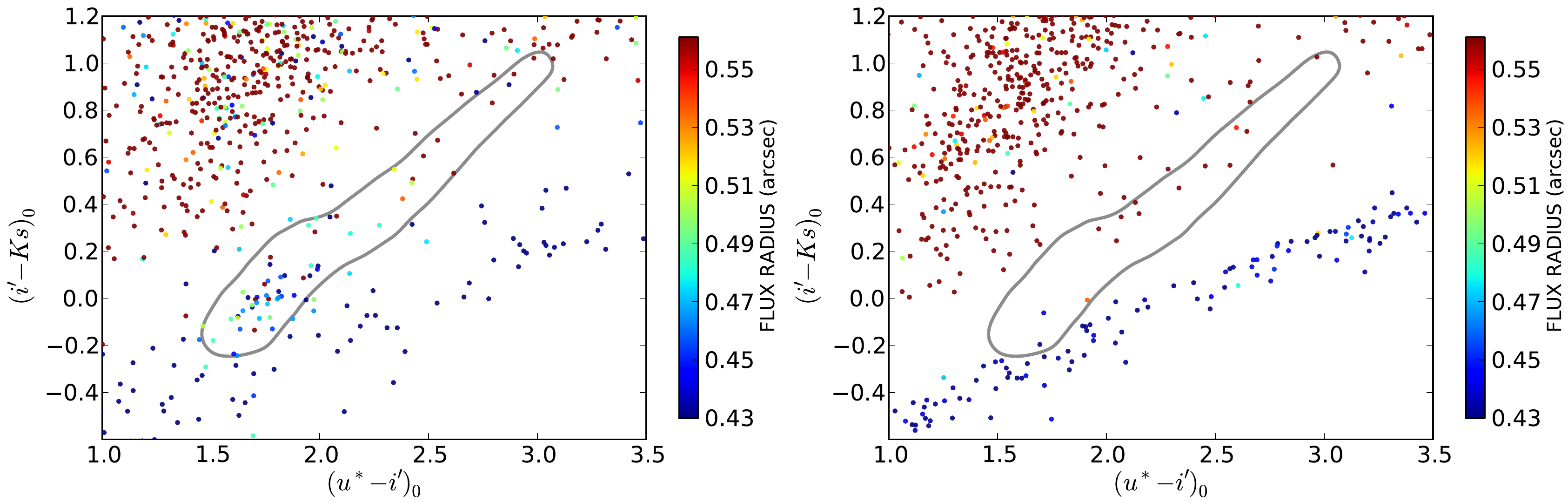}
\hspace*{0.5cm}\includegraphics[scale=0.45]{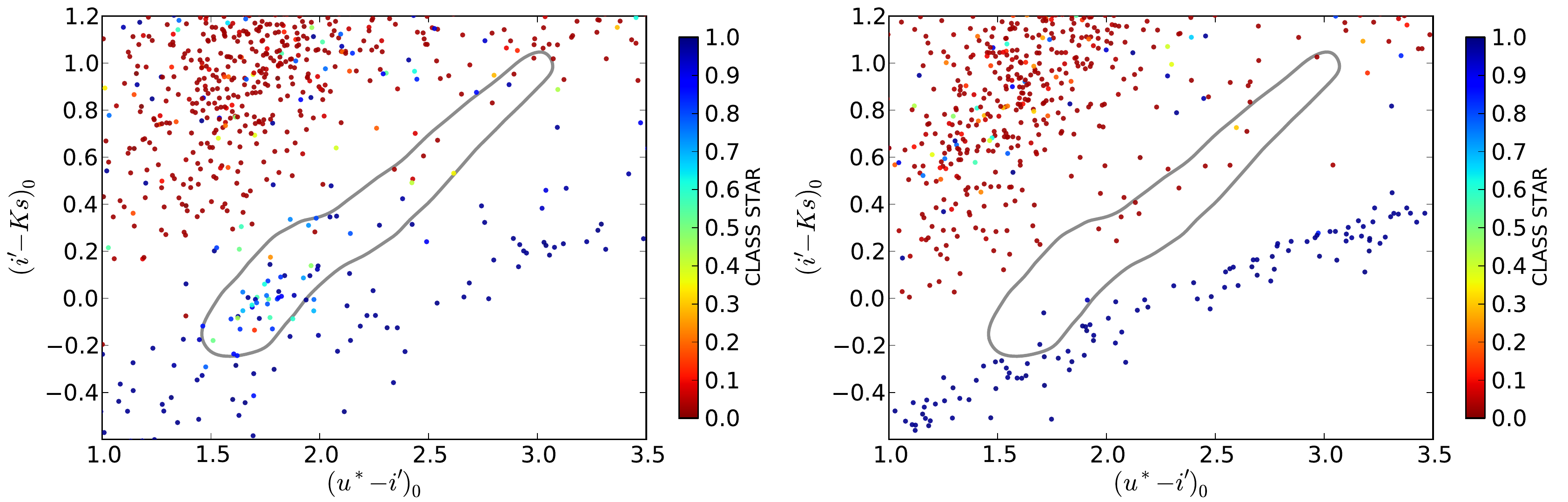}
\caption{NGC~4258 ({\it left column}) and Extended Groth Strip ({\it right column}). 
Comparison of structural parameters in the GCC selection
region. From {\it top} to {\it bottom:} FWHM; 100$\times$SPREAD\_MODEL; FLUX\_RADIUS; CLASS\_STAR. 
\edit1{The gray contour outlines the GCC selection region.}
GCCs appear with shades of \edit1 {blue and} cyan in all compactness estimators.
Since the average FWHM is slightly larger for the strip mosaic than for NGC~4258 (0$\farcs$67 vs.\ 0$\farcs$60), we have 
increased accordingly
the lower value of the FWHM range covered by the respective colormap. 
This way, point sources in the Groth Strip will show approximately
with the same color as those in the NGC~4258 field.
\label{fig:visavis}}
\end{figure}

We remind the reader that all our analysis is limited by the $K_s$-band image of NGC~4258, given
its FOV and shallowness. Figure~\ref{fig:visavis} presents the comparison between the shape parameters of 
the objects in the \uiks\ diagrams of NGC~4258 and the Groth Strip, respectively. To produce this diagram,
we only include sources in an area of the EGS equal to the effective FOV of the \ks-band mosaic of 
NGC~4258 (612.1 arcmin$^2$, excluding unexposed borders), that are brighter at \ks\ than the limiting magnitude
of the spiral galaxy data (21.7 mag). \edit1{The selection region in the \uiks\ color-color diagram is highlighted with the 
gray contour. The scarcity of unresolved (blue) and marginally resolved (cyan) sources
within the contour in} the EGS is remarkable. 
If we only used one parameter at a time, 
\edit1{FLUX\_RADIUS and FWHM would classify two --the same--  objects 
as GCCs; CLASS\_STAR would classify a third one, and SPREAD\_MODEL would 
classify yet a fourth, redder one, as a GCC in the EGS field.  
Hence, the combination of 
FWHM and SPREAD\_MODEL that we employed for our selection of GCCs in NGC~4258 likely 
allowed two contaminants to be classified as GCCs.}

\edit1{We adopt 2 contaminants and hence 37} as our final number of GC candidates.
Incidentally, this estimate of contaminating objects 
coincides with the results of the very recent work by \citet{powa16}. They have very carefully examined,
also using structural parameters,
the \uiks\ diagram of spectroscopically confirmed GCs in the M~87 field produced by \citet{muno14}; 
Powalka et al.\  concluded, conservatively, that the contamination is at most 5\%. 
Both M~87 and NGC~4258 are at high Galactic 
latitudes (respectively, at $b =74\fdg5$ and $b = 68\fdg8 $) and, indeed, the Besan\c con model\footnote{ 
\url{http://model.obs-besancon.fr/}
} \citep{robi03}
predicts 8\% fewer stars in the direction of NGC~4258 relative to the M~87 line of sight.

\section{Properties of the GC system of NGC~4258} \label{sec:properties}

\subsection{Color distribution} \label{sec:colordist}

\begin{sidewaysfigure}[ht]
\begin{tabular}{llllll}
\hspace*{-1.2cm}\includegraphics[scale=0.20]{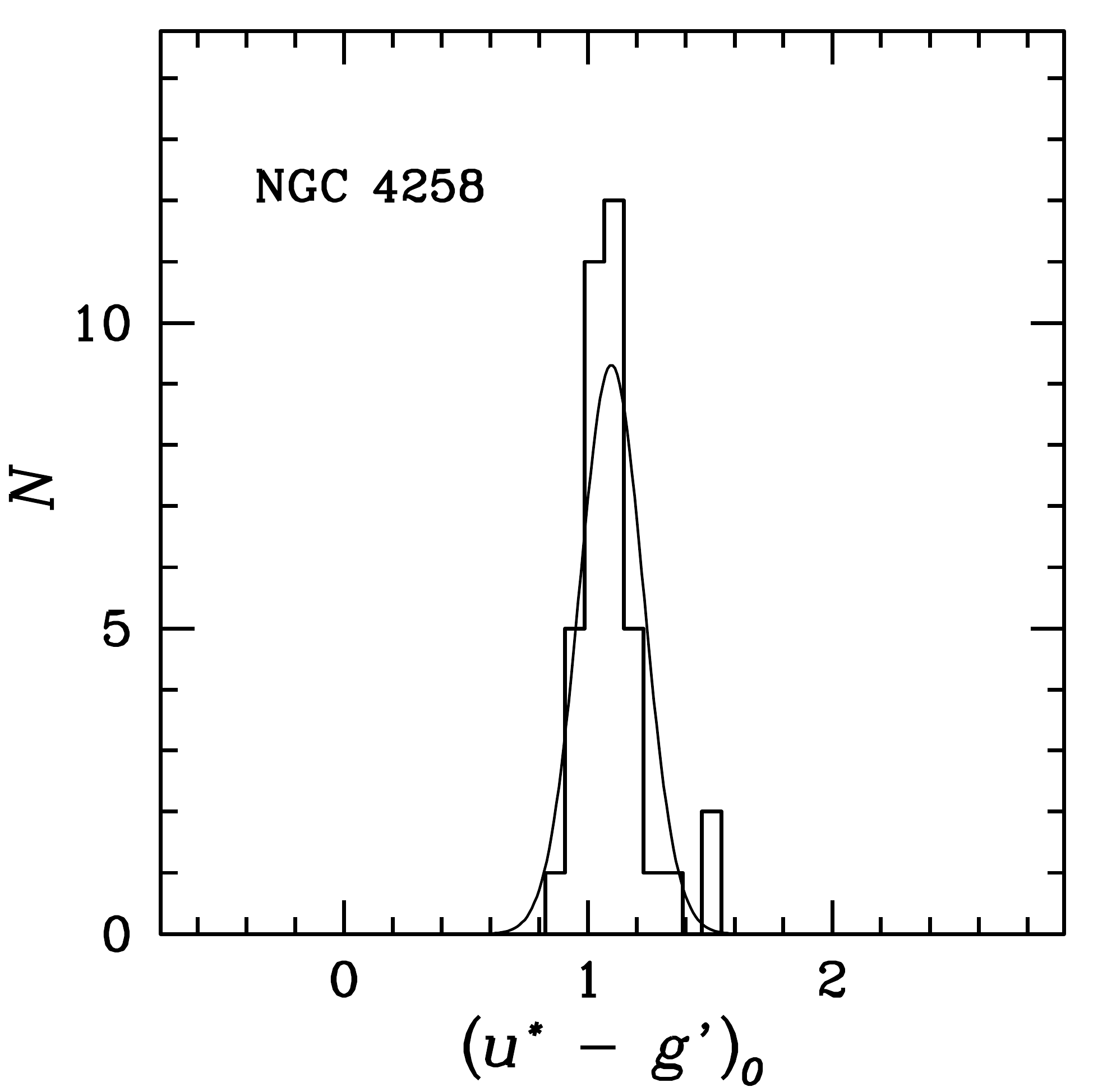}
&
\includegraphics[scale=0.20]{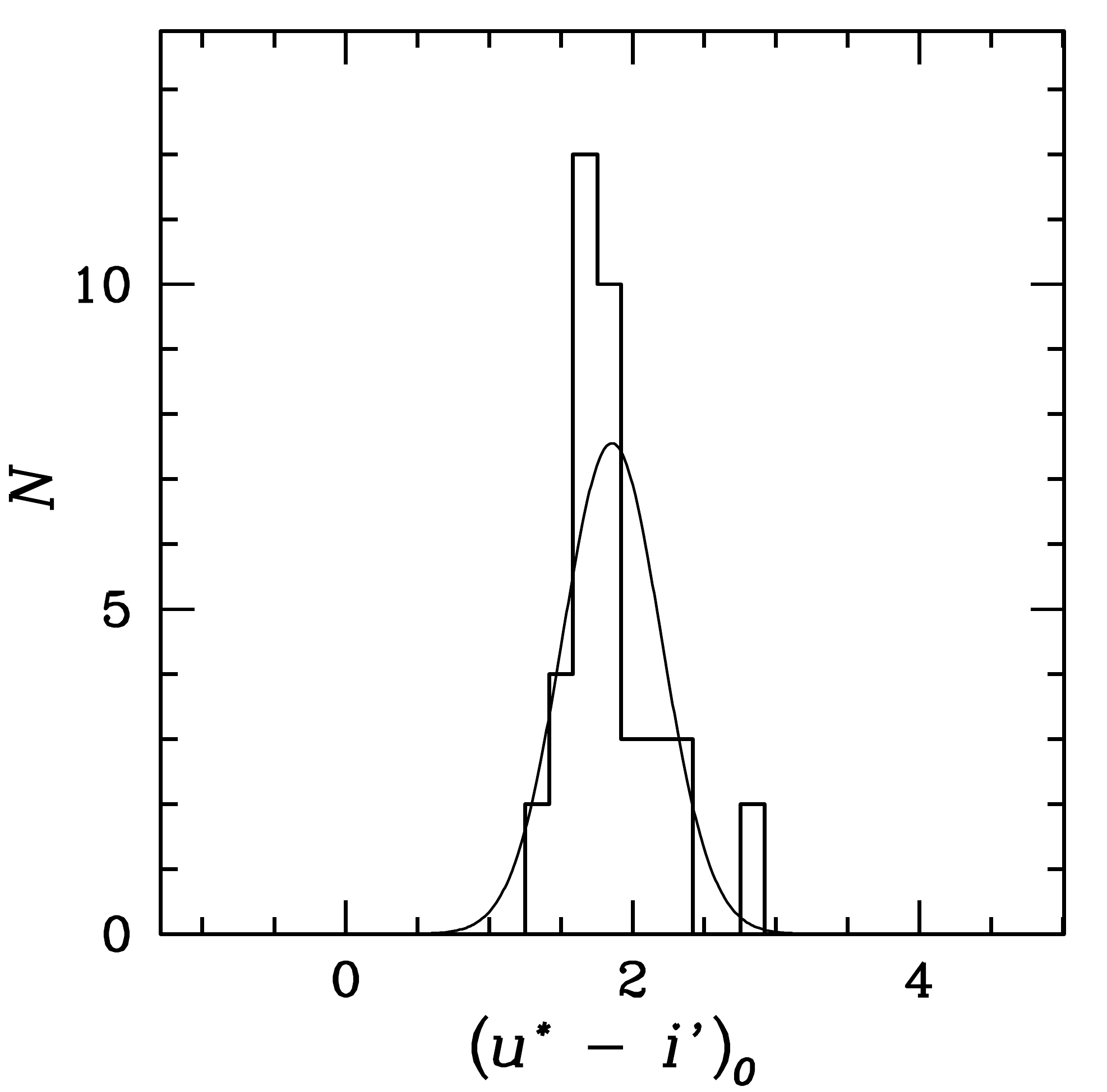}
&
\includegraphics[scale=0.20]{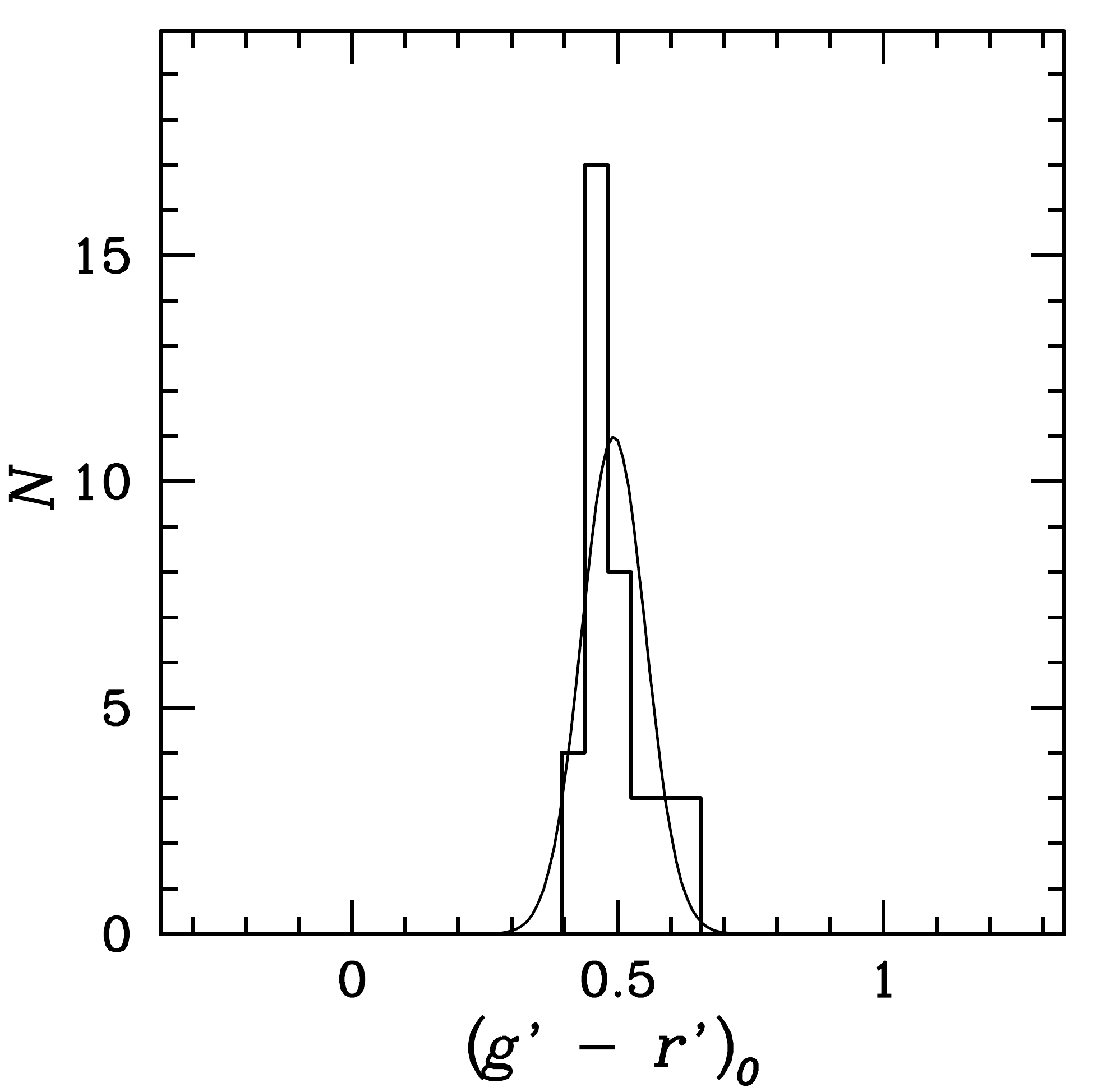}
&
\includegraphics[scale=0.20]{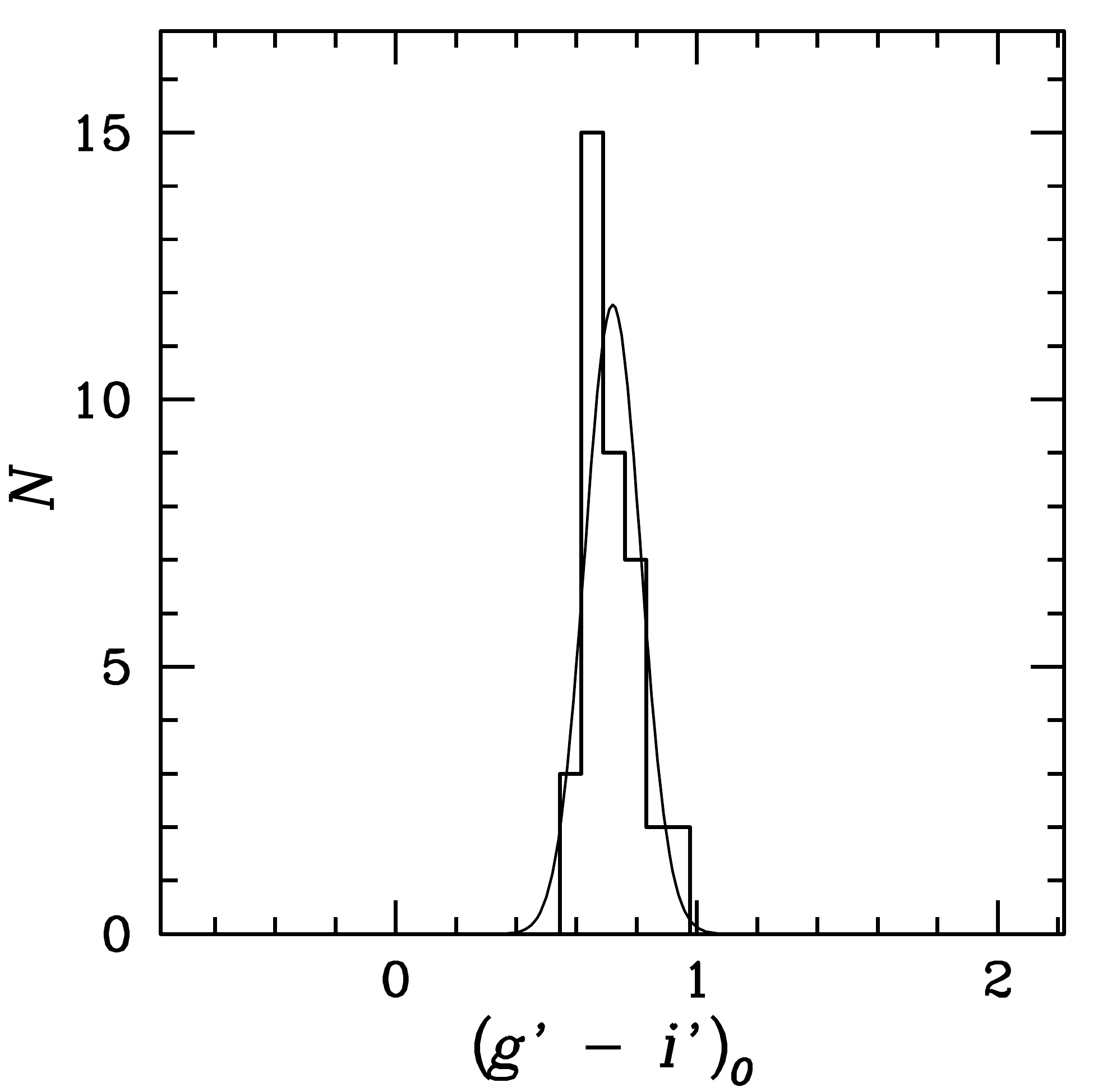}
& 
\includegraphics[scale=0.20]{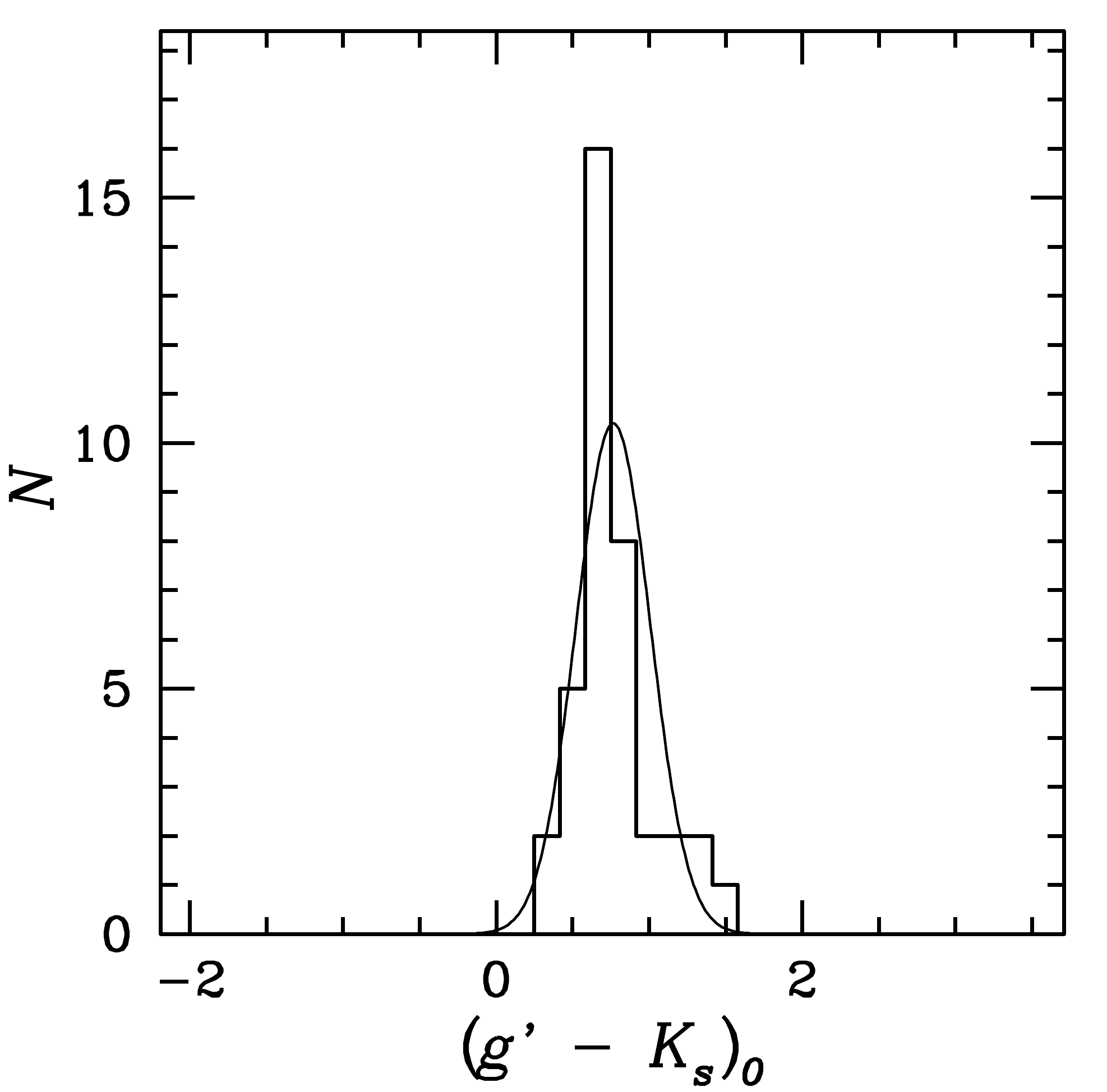}
&
\includegraphics[scale=0.20]{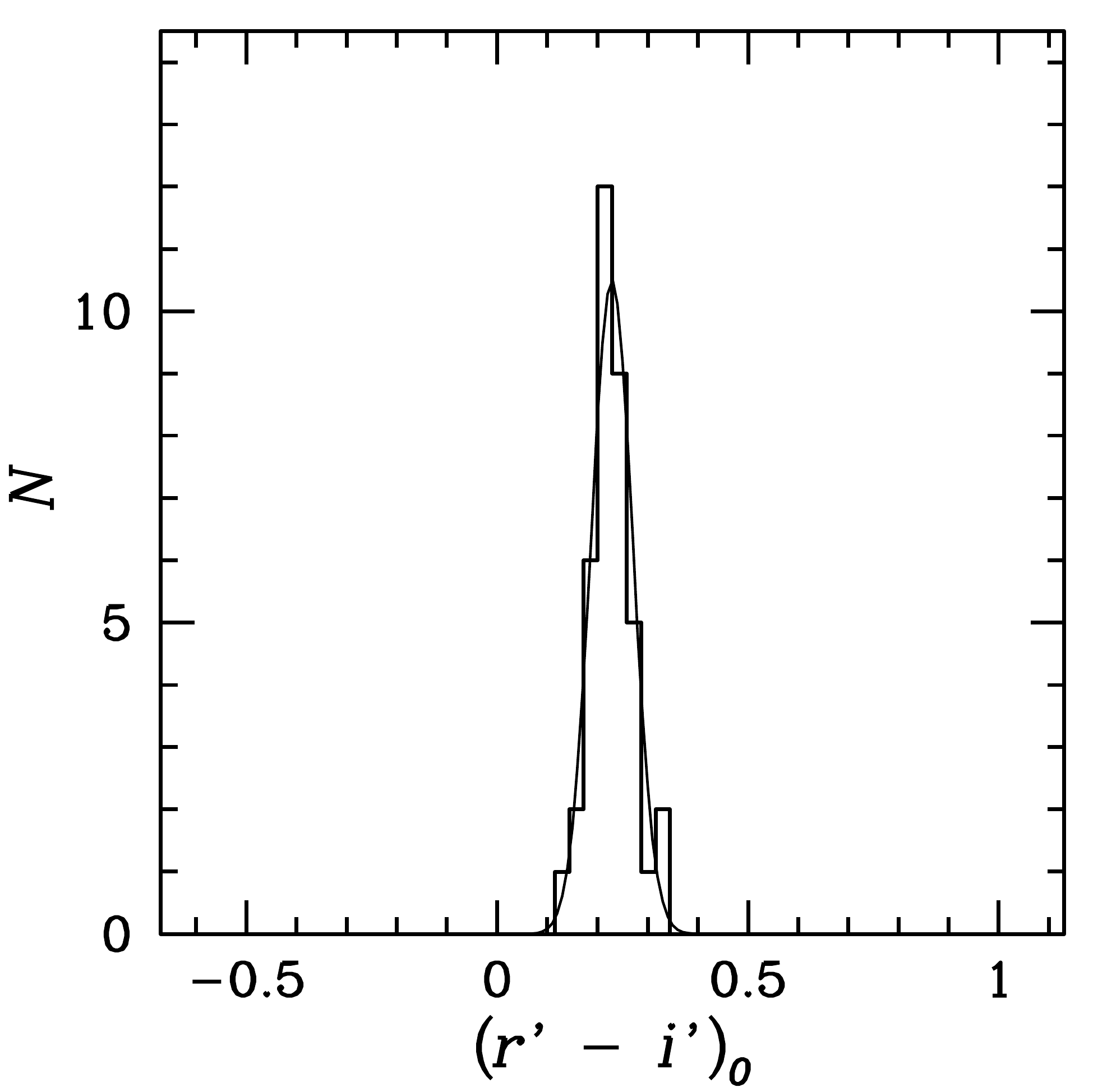}
\end{tabular}
\begin{tabular}{lllll}
\hspace*{-1.2cm}\includegraphics[scale=0.20]{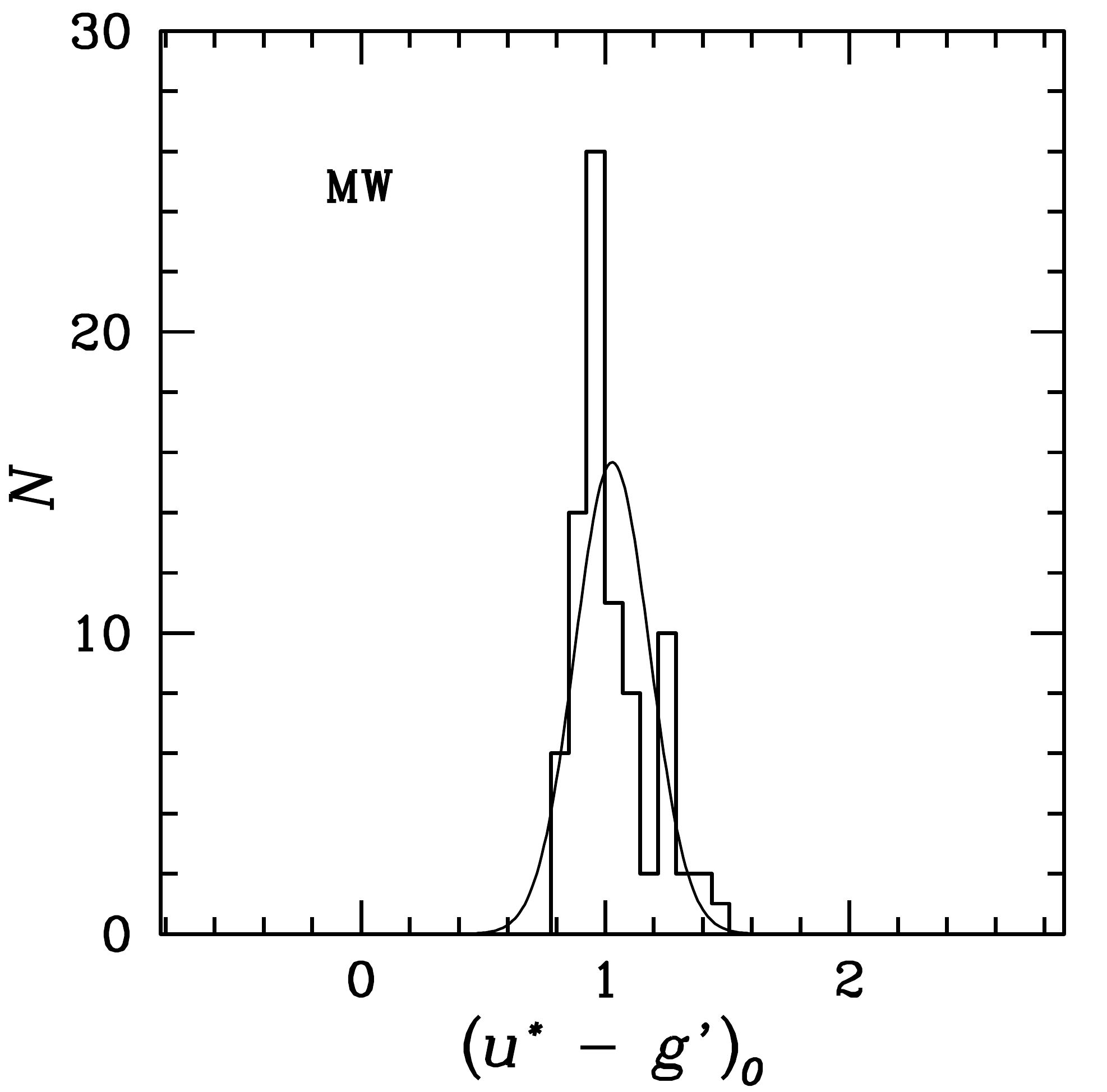}
&
\includegraphics[scale=0.20]{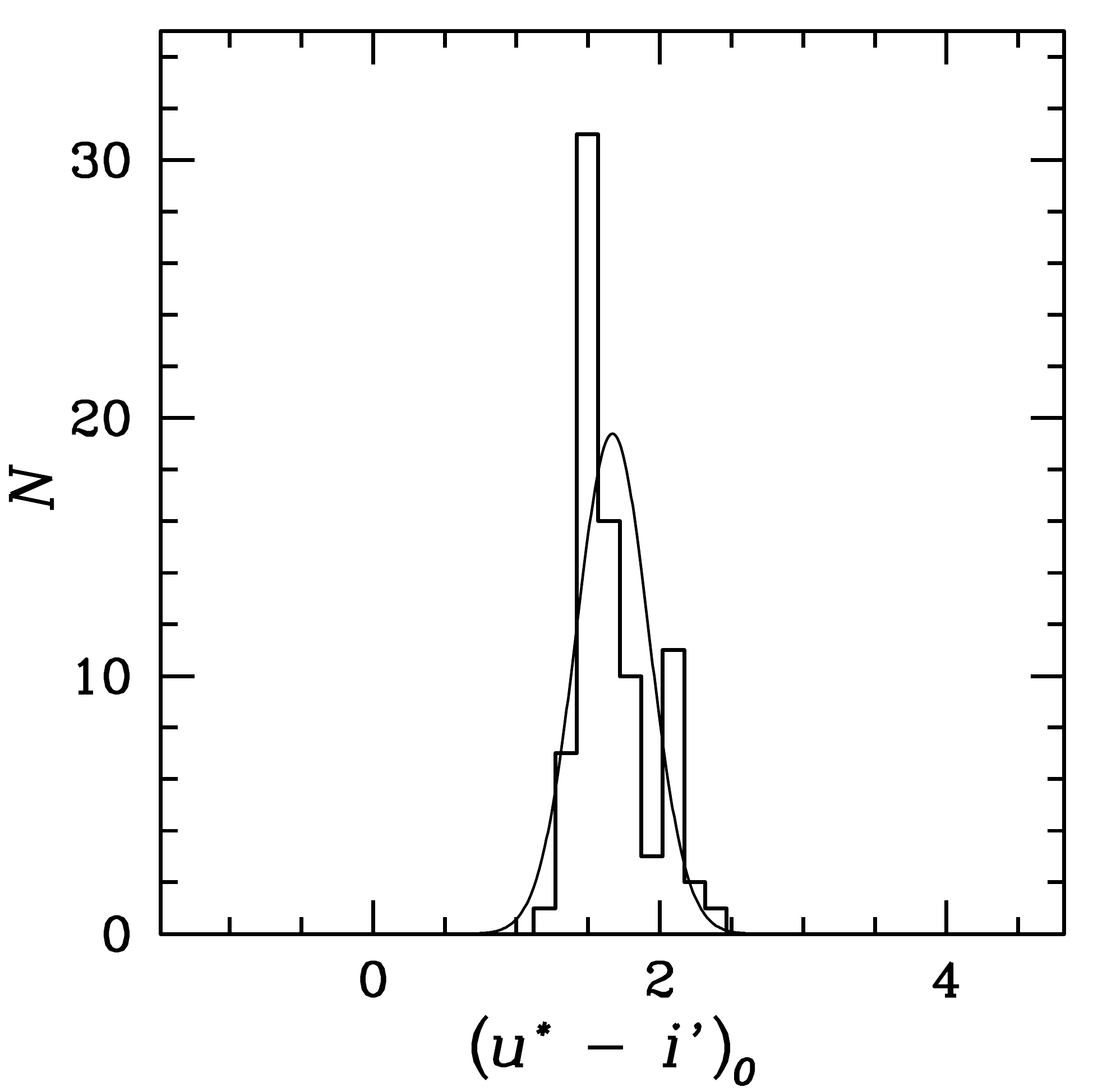}
&
\includegraphics[scale=0.20]{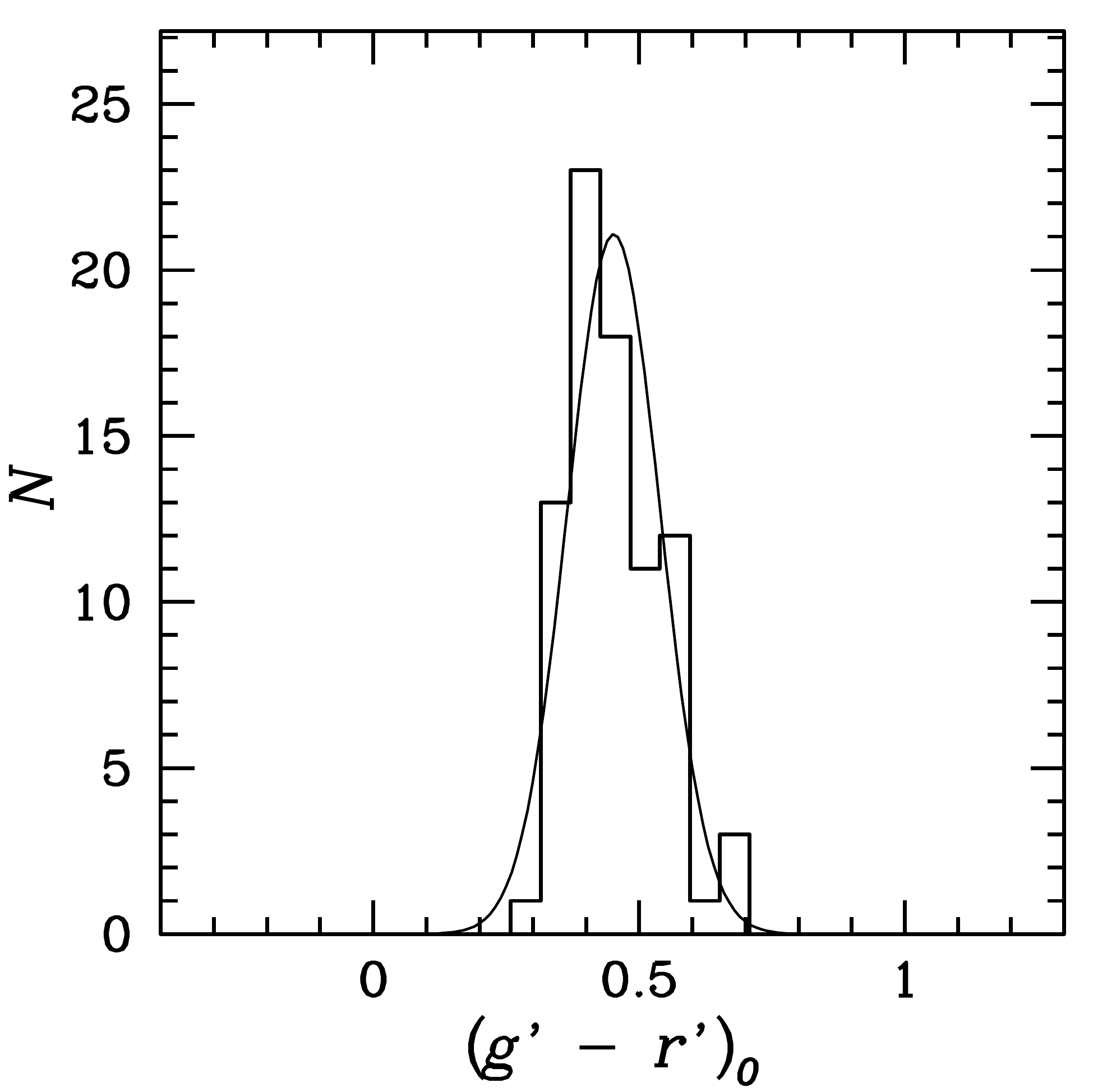}
&
\includegraphics[scale=0.20]{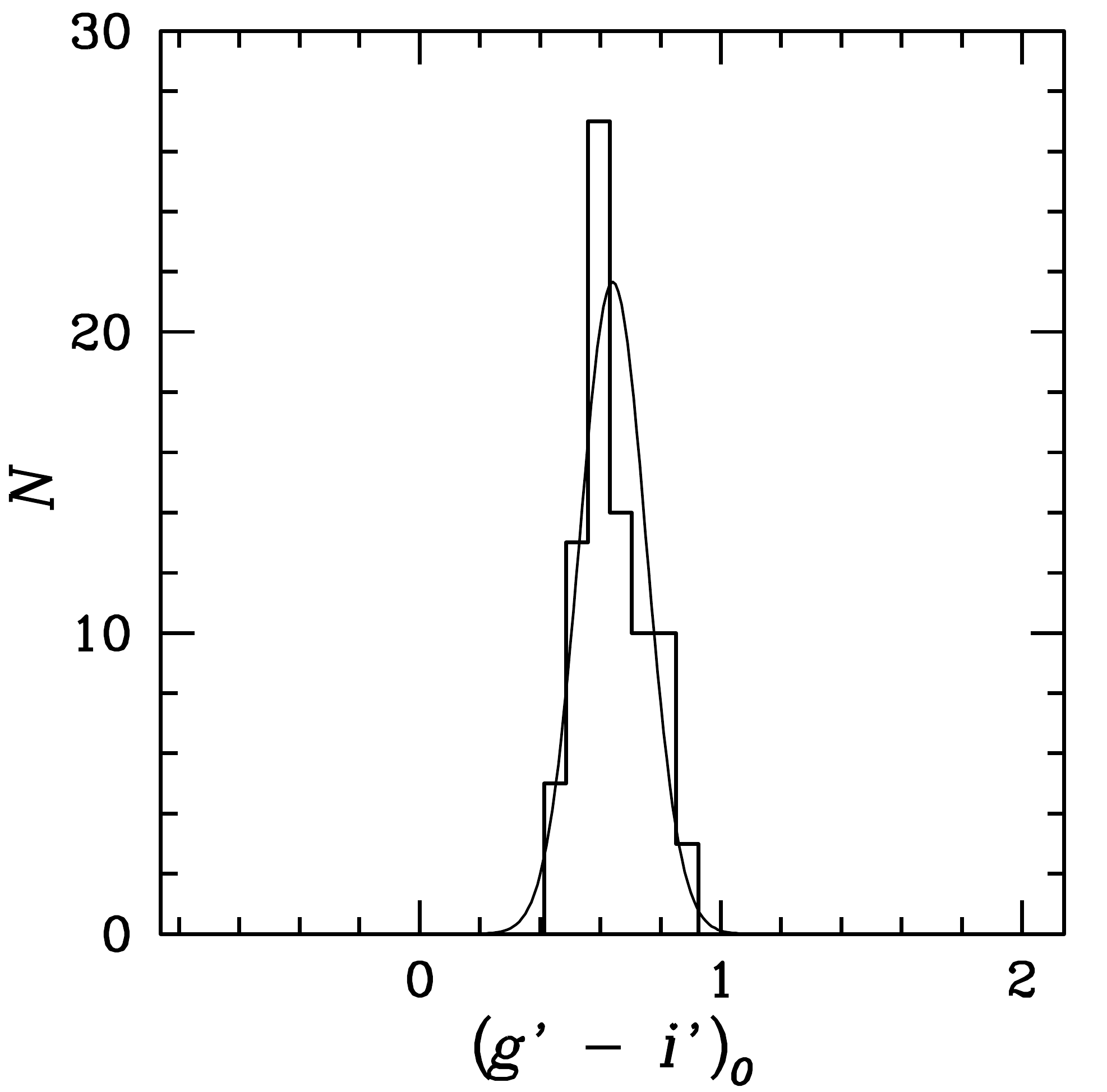}
& 
\hspace*{4.35cm}\includegraphics[scale=0.20]{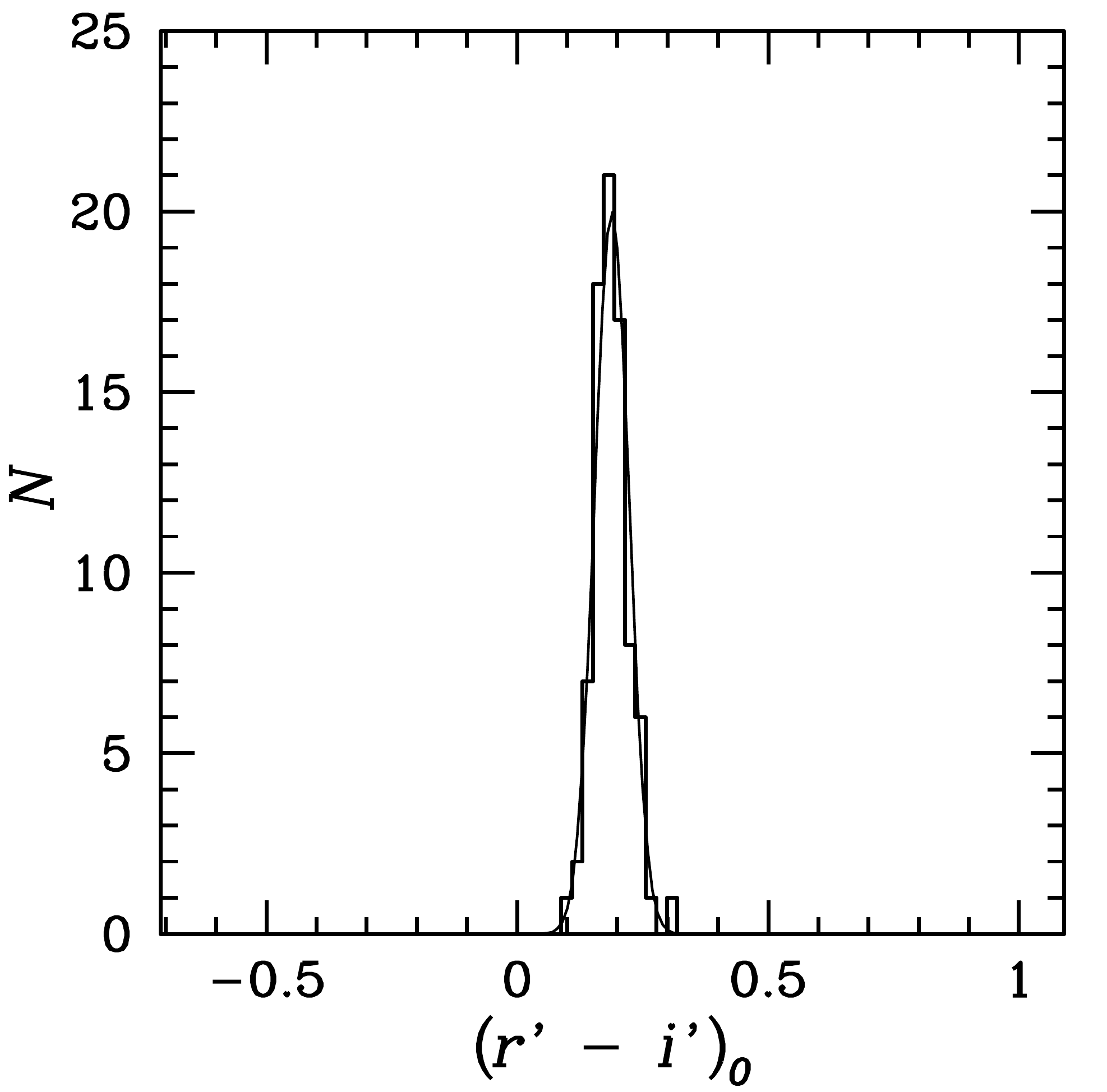}
\end{tabular}
\begin{tabular}{llllll}
\hspace*{-1.2cm}\includegraphics[scale=0.20]{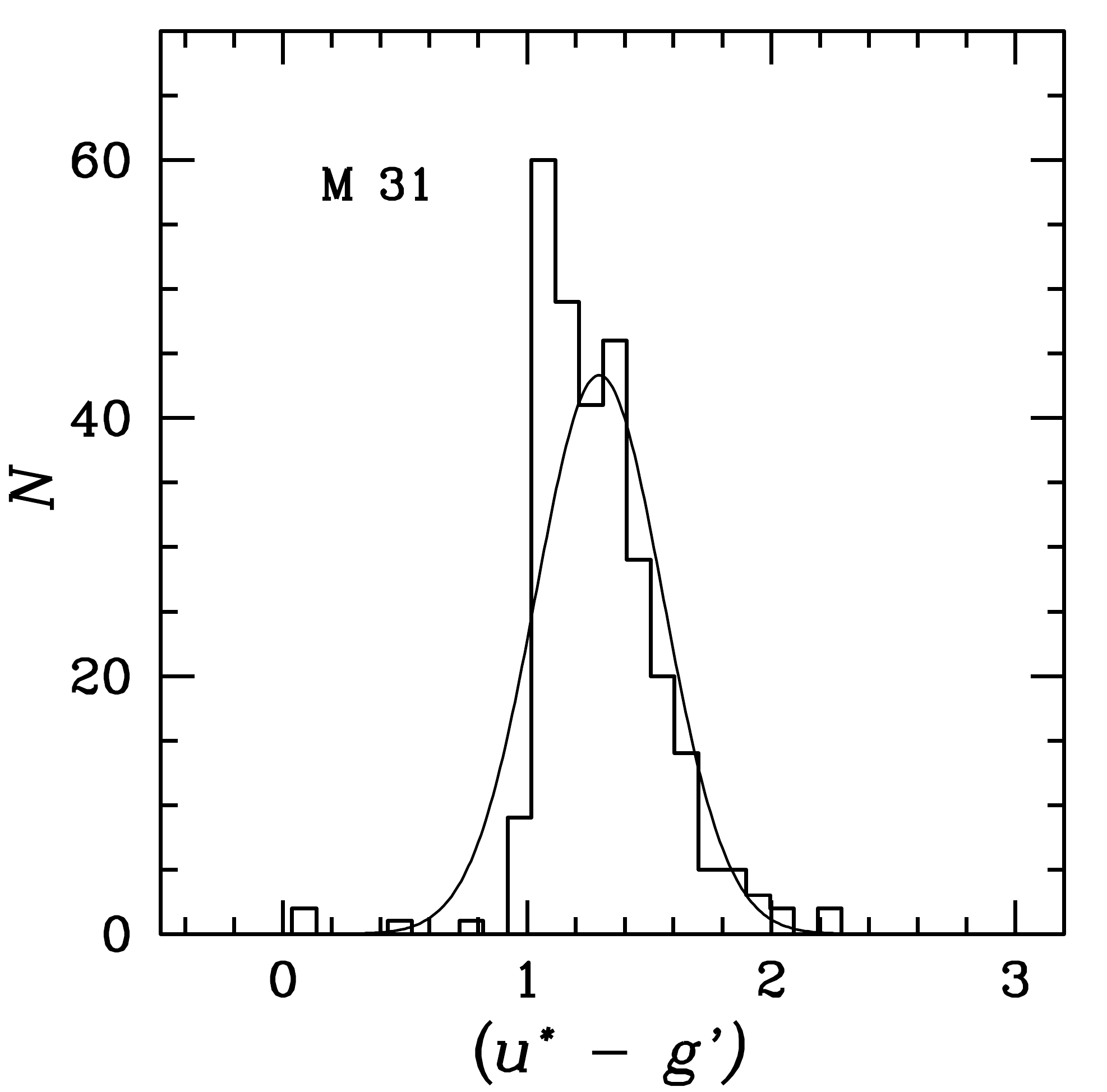}
&
\includegraphics[scale=0.20]{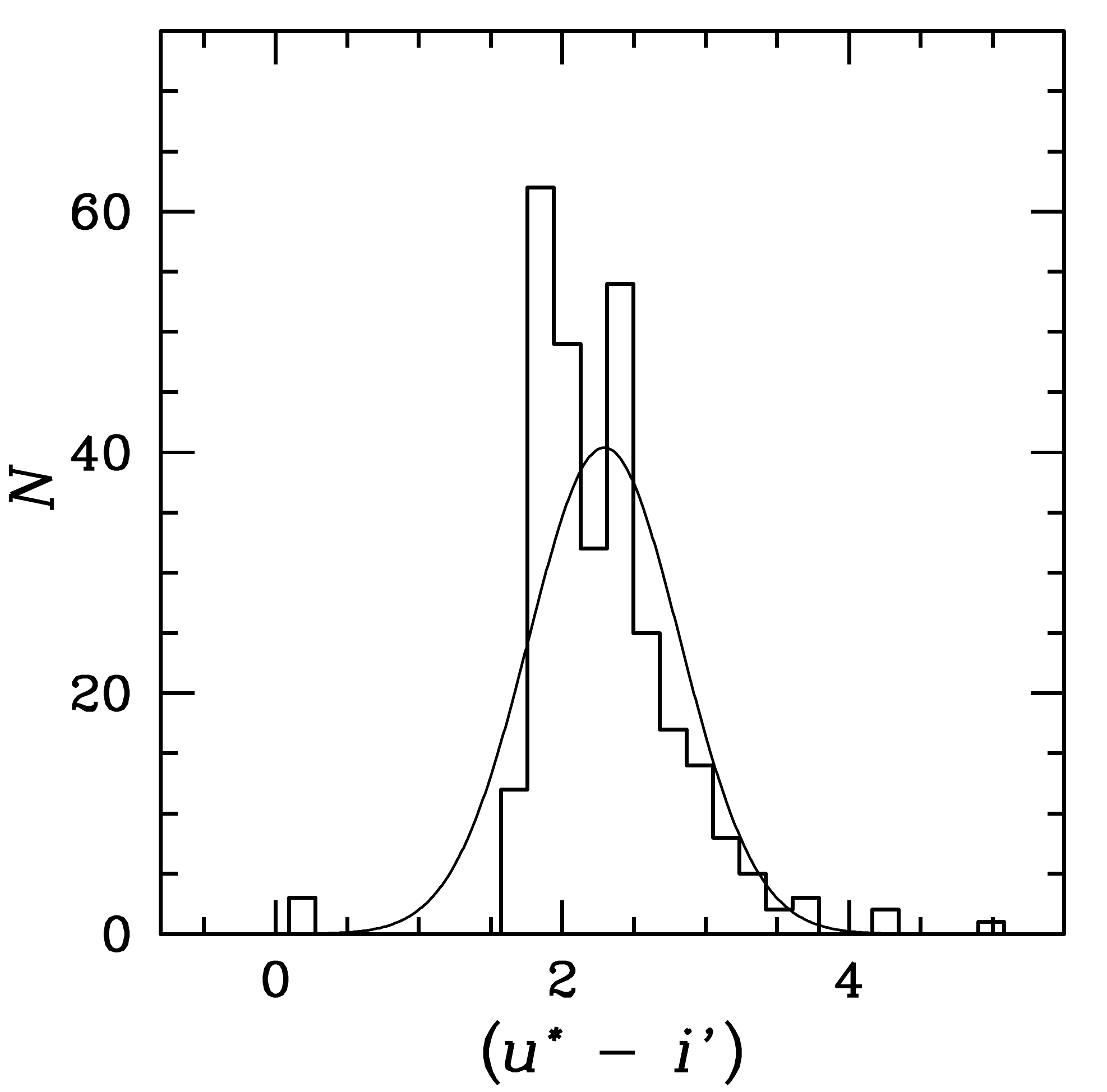}
&
\includegraphics[scale=0.20]{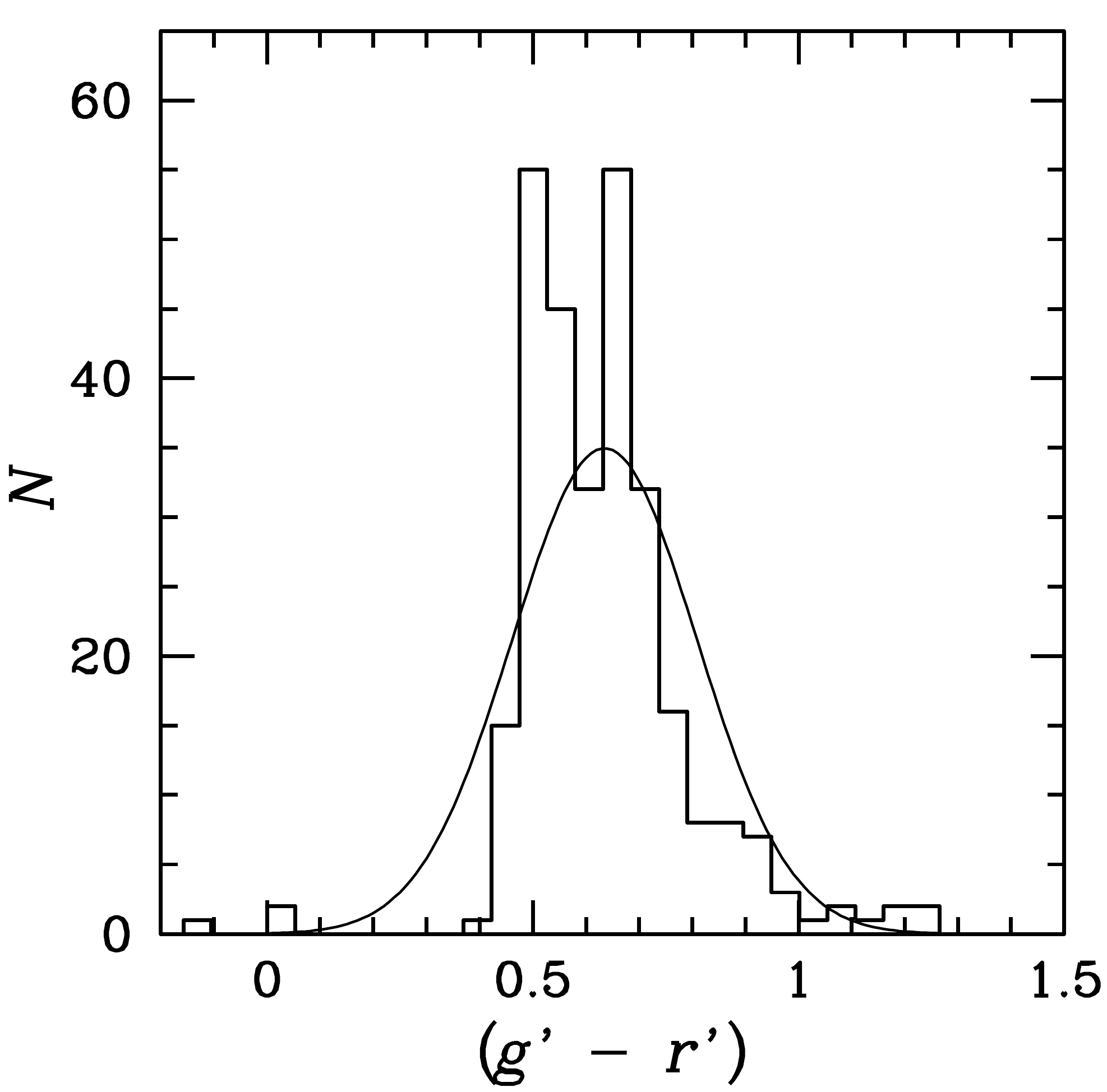}
&
\includegraphics[scale=0.20]{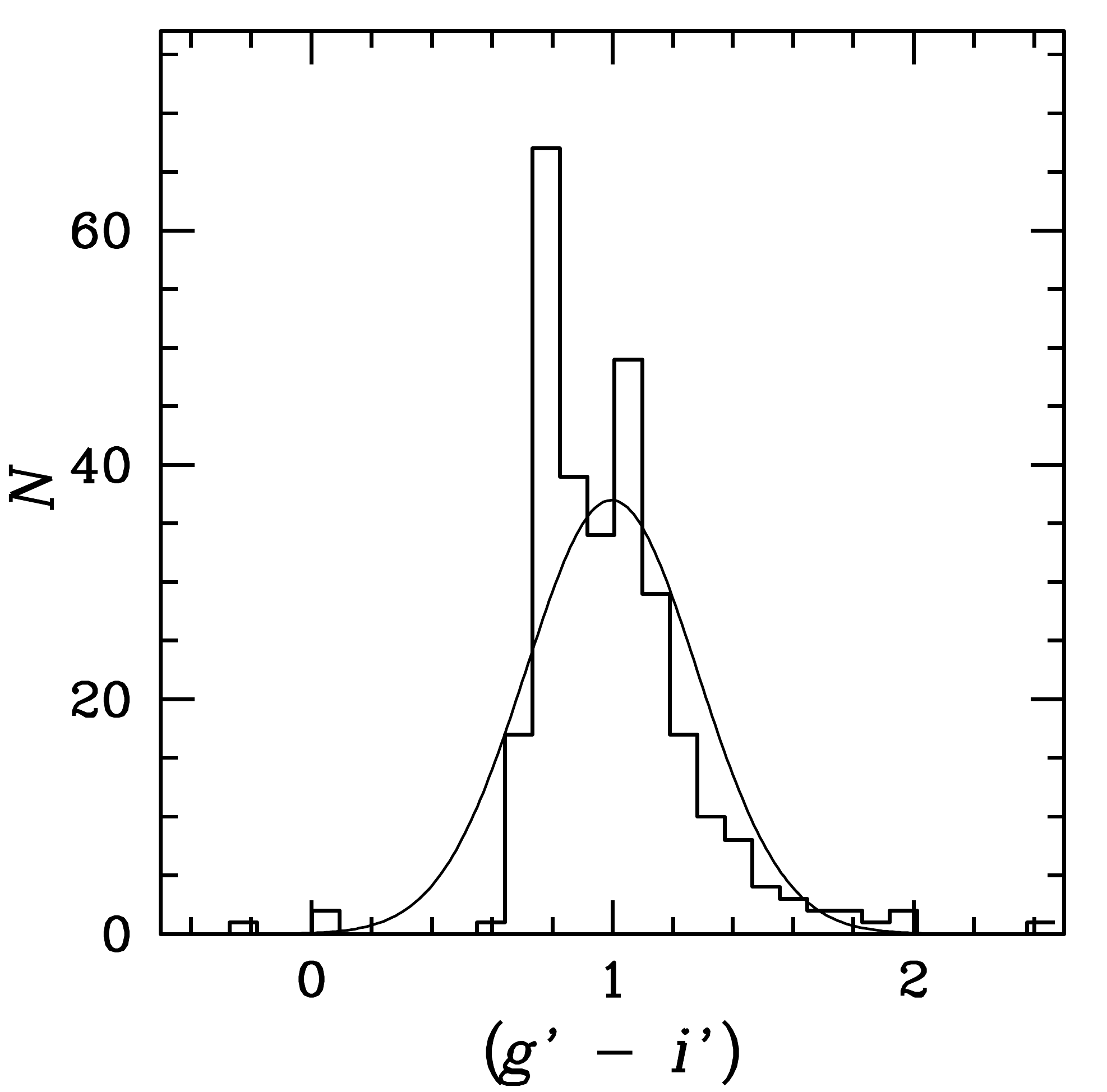}
& 
\includegraphics[scale=0.20]{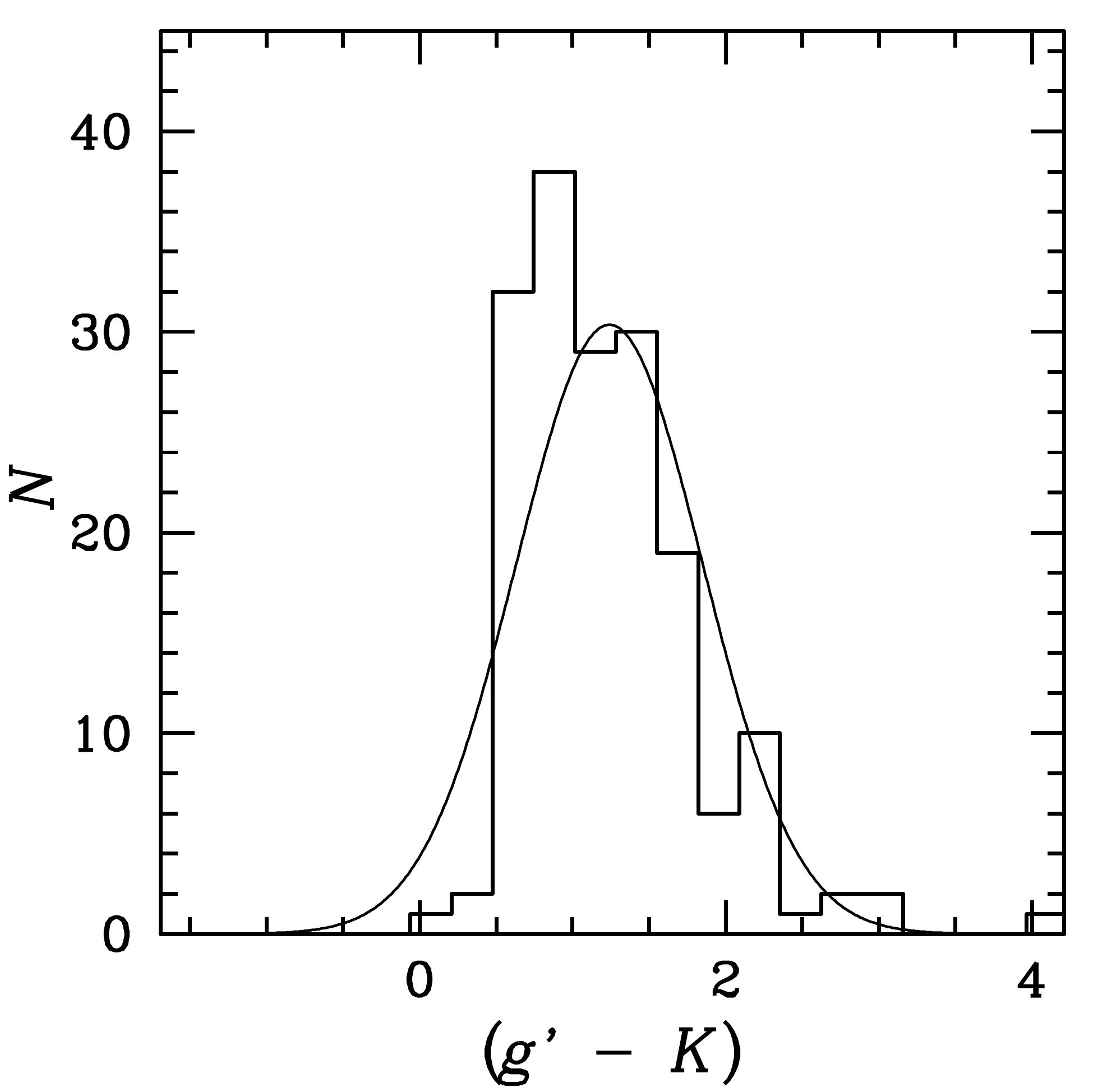}
&
\includegraphics[scale=0.20]{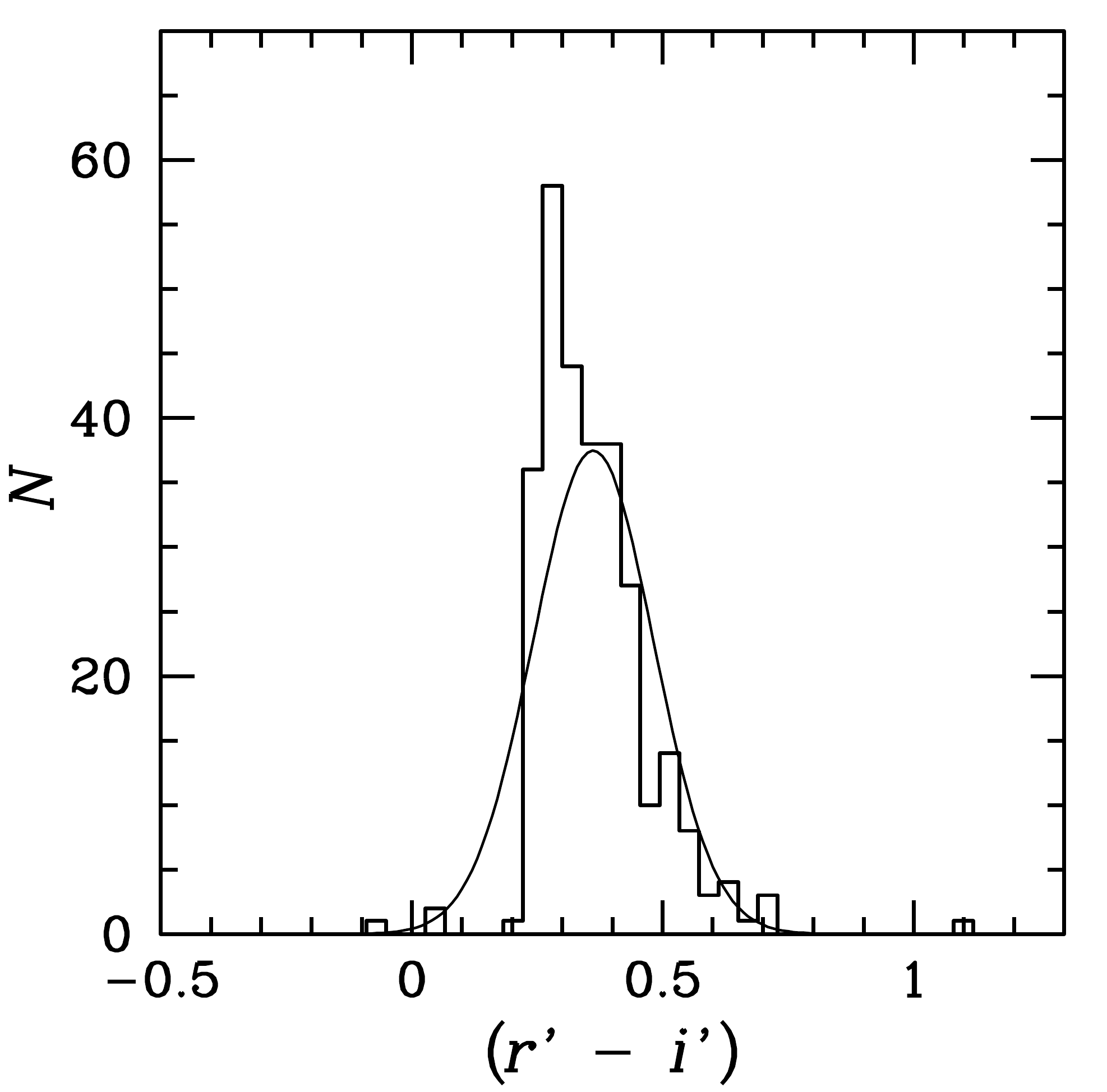}
\end{tabular}

\caption{Color distributions of GC systems. {\it Top:} NGC~4258 GCCs (this work); {\it middle:} MW globulars \citep{harr96}; {\it bottom:} 
M~31 old clusters \citep{peac10}. From left to right: ($u^\ast\ - g^\prime$), ($u^\ast\ - i^\prime$), ($g^\prime - r^\prime$), ($g^\prime - K_s$),
($r^\prime - i^\prime$). For M~31, near-IR filter is actually $K$, although calibrated with the \ks-band data of the 2MASS survey.   
\label{fig:colordist}}
\end{sidewaysfigure}

The distributions of our final sample of GCCs in the colors (\ust\ - \gp), (\ust\ - \ip), 
(\gp\ - \rp), (\gp - \ip), (\rp - \ip), and (\gp - \ks) are shown in the top panels of
Figure~\ref{fig:colordist}. The colors of the individual candidates have been corrected for foreground
extinction in the Milky Way, as mentioned before (Section~\ref{sec:uiks}). The means and dispersions of Gaussian 
fits to the color distributions, \edit1{performed with the algorithm GMM \citep{mura10},} are presented in Table~\ref{tab:colordist}. 

We now compare the colors of the GCCs of NGC~4258 with those of the globulars of our Galaxy
and M~31. For the Milky Way, we use the compilation 
by \citet{harr96}.
This catalog comprises to date information on 147 objects, but only 82 have photometry on all
$UBVRI$ bands, which we need to obtain their colors in the MegaCam system. 
We first transform the values in the catalog to the SDSS (AB) system with the equations in
\citet{jest05} for stars with $R_c\ - I_c < 1.15$, and then to MegaCam \ust \gp \rp \ip 
$z^\prime$ magnitudes with the relations given in the MegaPipe webpage.\footnote{ 
\url{http://www.cadc-ccda.hia-iha.nrc-cnrc.gc.ca/en/megapipe/docs/filt.html}.}
We have dereddened individually each of the 82 MW GCs 
using the $E(B\ - V)$ values in \citet{harr96}, assuming the 
relative extinctions for the Landolt filter set derived by \citet{schleg98} from the $R_V = 3.1$ laws of 
\citet[][UV and IR]{card89}, and \citet[][optical]{odon94}. 

Our transformation equations are given below, where names without superscripts refer to the SDSS system,
and those with superscripts allude to the MegaCam filters:

\begin{align}
u\ - g & =  1.28 \times [U\ - B - 3.1\times E(B\ - V) \times (1.664 - 1.321)] + 1.13 \\
g\ - r & =  1.02 \times [B\ - V - 3.1\times E(B\ - V) \times (1.321 - 1.020)] - 0.22 \\
g\ - i & =  0.92 \times [V\ - R - 3.1 \times E(B\ - V) \times (0.819 - 0.594)] - 0.20 \\
u^\ast - g' & =  0.759 \times (u\ - g) + 0.153 \times (g\ - r) \\
u^\ast - i' & = 0.759 \times (u\ - g) + 1.003 \times (r\ - i) \\
g'\ - r' & = 0.871 \times (g\ - r) \\
g'\ - i' & = 0.847 \times (g\ - r) + 1.003 \times (r\ - i) \\
r'\ - i' & = 1.003 \times (r\ - i) - 0.024 \times (g\ - r).
\end{align}

(\ust\ - \gp), (\ust\ - \ip),
(\gp\ - \rp), (\gp - \ip), and (\rp - \ip) color distributions of the MW globulars are shown in the middle panels of Figure~\ref{fig:colordist};
means and dispersions of Gaussian fits to them are also presented in Table~\ref{tab:colordist}.

In the case of M~31, we take the sample of {\it old} clusters with SDSS photometry in \citet{peac10}. There are 289 objects with 
$u$, $g$, $r$, and $i$ data, as well as 173 GCs that have also been observed in the $K$ filter with the
Wide Field Camera (WFCAM) mounted on the United Kingdom Infrared Telescope (UKIRT). Whereas the 
optical magnitudes are given in the AB system, the infrared $K$ values are quoted in Vega magnitudes. 
To transform the optical colors to the MegaCam system, we use eqs.~(6)--(10). For (\gp\ - \ks), we apply:

\begin{equation}
g'\ - K_s = g\ - 0.153 \times (g\ - r) - K - 1.834, 
\end{equation}

\noindent
where the constant 1.834 is introduced to convert $K$ to the AB system used by WIRwolf
(\url{http://www.cfht.hawaii.edu/Instruments/Imaging/WIRCam/dietWIRCam.html}).
We note that \citet{peac10} have calibrated their $K$ magnitudes with Two Micron 
All Sky Survey \citep[2MASS;][]{skru06} \ks-band data, 
but the two filters are different. The \ks\ filter cuts off at 2.3 $\mu$m in order to 
reduce the contribution from the thermal background; its minimum, maximum, and central
wavelengths are, respectively,  $\lambda_{\rm min} = 1.95\mu$m,
$\lambda_{\rm max} = 2.36\mu$m, $\lambda_{\rm cen} = 2.16 \mu$m, vs.\  $\lambda_{\rm min} = 1.98\mu$m,
$\lambda_{\rm max} = 2.47\mu$m, $\lambda_{\rm cen} = 2.23 \mu$m for the $K$ filter 
(\url{http://svo2.cab.inta-csic.es/svo/theory/fps3/}).

The color distributions of the M~31 old clusters are shown in the bottom panels of Figure~\ref{fig:colordist}, and the parameters
of the Gaussian fits are listed in Table~\ref{tab:colordist}. 

The colors reported in \citet{peac10} have not been corrected for extinction, 
but we can estimate the reddening for the whole system as we describe below. 
\citet{barm00} have published $UBVRI$ colors derived consistently for 
more than 400 M~31 clusters. Although these authors do not provide extinction corrections for their individual objects, they do find a global average reddening $E(B\ - V) = 0.22$. 
Using again the reddening law tabulations in \citet{schleg98}, 
we find $E(u^\ast\ - g')$ = 0.25; $E(u^\ast\ - i')$ = 0.57; $E(g'\ - r')$ = 0.19; $E(g'\ - i')$ = 0.32; $E(g'\ - K_s)$ = 0.68; 
$E(r'\ - i')$ = 0.13. Even without considering errors, if we correct the mean MegaCam colors of the M~31 GC system for these excesses, they are virtually identical to the colors
of the MW system, as can be seen in Table~\ref{tab:colordist}. 
The redder colors of the M~31 can then be fully explained by extinction, of which 
about one third would be due to the relatively uncertain foreground of the MW 
[$E(B\ -V) = 0.062$; \citealt{schleg98}], and the rest would be internal to M~31. 
The colors of the NGC~4258 system are slightly redder than those of the MW globulars. This small difference could also be partially due to
unaccounted for internal extinction in NGC~4258.

Since in the NGC~4258 data we are 100\% complete for objects brighter than the GC LFTO, we also derive colors only for the globulars
in the MW and M~31 systems with $M_V \leq -7.5$ mag; 
for the \citeauthor{peac10} M~31 sample, we use the transformation for the SDSS filters $V = g - 0.59 \times (g\ - r) - 0.01$ 
\citep{jest05}, and assume a distance modulus for Andromeda of 24.45 (mean and median from NED).  The colors of these subsets
are likewise tabulated in Table~\ref{tab:colordist}, and 
are virtually identical to the colors of the full systems.

\begin{sidewaystable}[ht]
 \caption{Parameters of Gaussian Fits to Color Distributions}
 \begin{center}
 \begin{minipage}{280mm}
  \begin{tiny}
  \begin{tabular}{@{}lcccccccccccccccccccccccc@{}}
\hline
\hline
System & \multicolumn{2}{c}{$(u^\ast - g')$} & \multicolumn{2}{c}{$(u^\ast - g')_0$} & \multicolumn{2}{c}{$(u^\ast - i')$} & \multicolumn{2}{c}{$(u^\ast - i')_0$} & 
                   \multicolumn{2}{c}{$(g' - r')$} & \multicolumn{2}{c}{$(g' - r')_0$} & \multicolumn{2}{c}{$(g' - i')$} & \multicolumn{2}{c}{$(g' - i')_0$} & 
                   \multicolumn{2}{c}{$(g' - K_s^a)$} & \multicolumn{2}{c}{$(g' - K_s^a)_0$} & \multicolumn{2}{c}{$(r' - i')$} & \multicolumn{2}{c}{$(r' - i')_0$} \\
 & $\mu$ & $\sigma$ & $\mu$ & $\sigma$ & $\mu$ & $\sigma$ & $\mu$ & $\sigma$ &
           $\mu$ & $\sigma$ & $\mu$ & $\sigma$ & $\mu$ & $\sigma$ & $\mu$ & $\sigma$ &
           $\mu$ & $\sigma$ & $\mu$ & $\sigma$ & $\mu$ & $\sigma$ & $\mu$ & $\sigma$  \\
\hline
NGC~4258               &      &      &\edit1{ 1.10} &\edit1{ 0.13} &      &      &\edit1{ 1.86} &\edit1{ 0.34} &      &      &\edit1{ 0.49} & 0.06 &      &      &\edit1{ 0.72} & 0.09 &      &      &\edit1{ 0.76} &\edit1{  0.24 } &      &      &\edit1{ 0.23} &\edit1{ 0.04} \\
Milky Way              &      &      & 1.03 & 0.15 &      &      & 1.67 & 0.25 &      &      & 0.45 & 0.09 &      &      & 0.64 & 0.11 &      &      &      &        &      &      & 0.19 & 0.03 \\
Milky Way $M_V < -7.5$ &      &      & 1.02 & 0.15 &      &      & 1.65 & 0.25 &      &      & 0.45 & 0.09 &      &      & 0.64 & 0.11 &      &      &      &        &      &      & 0.18 & 0.03 \\
M~31                   & 1.30 & 0.26 & 1.05 &      & 2.29 & 0.53 & 1.72 &      & 0.64 & 0.17 & 0.45 &      & 1.00 & 0.28 & 0.68 &      & 1.24 & 0.61 & 0.56 &        & 0.36 & 0.12 & 0.23 &      \\
M~31 $M_V < -7.5$      & 1.30 & 0.26 &      &      & 2.29 & 0.52 &      &      & 0.64 & 0.17 &      &      & 1.00 & 0.28 &      &      & 1.23 & 0.59 &      &        & 0.36 & 0.11 &      &      \\
\hline
\vspace*{-0.5cm}
\end{tabular}
\end{tiny}
\end{minipage}
\end{center}
$^a$ For M~31, near-IR filter is actually $K$, although calibrated with the \ks-band data of the 2MASS survey.
Colors of MW clusters are extinction-corrected; colors of objects in NGC~4258 are corrected for foreground Galactic extinction.
For old clusters in M~31, we give uncorrected colors, and colors corrected by the average of both foreground extinction in the
MW and reddening internal to M~31. 
\label{tab:colordist}
\end{sidewaystable}

This brief analysis shows that, like the GCs of M~31 and our Galaxy, the GCCs in NGC~4258 do not have a bimodal color distribution, and that
the colors of the three systems are remarkably consistent, once extinction has been taken into account. 

\subsection{Luminosity function}\label{subsec:gclf}

\begin{figure}[ht!]
\plotone{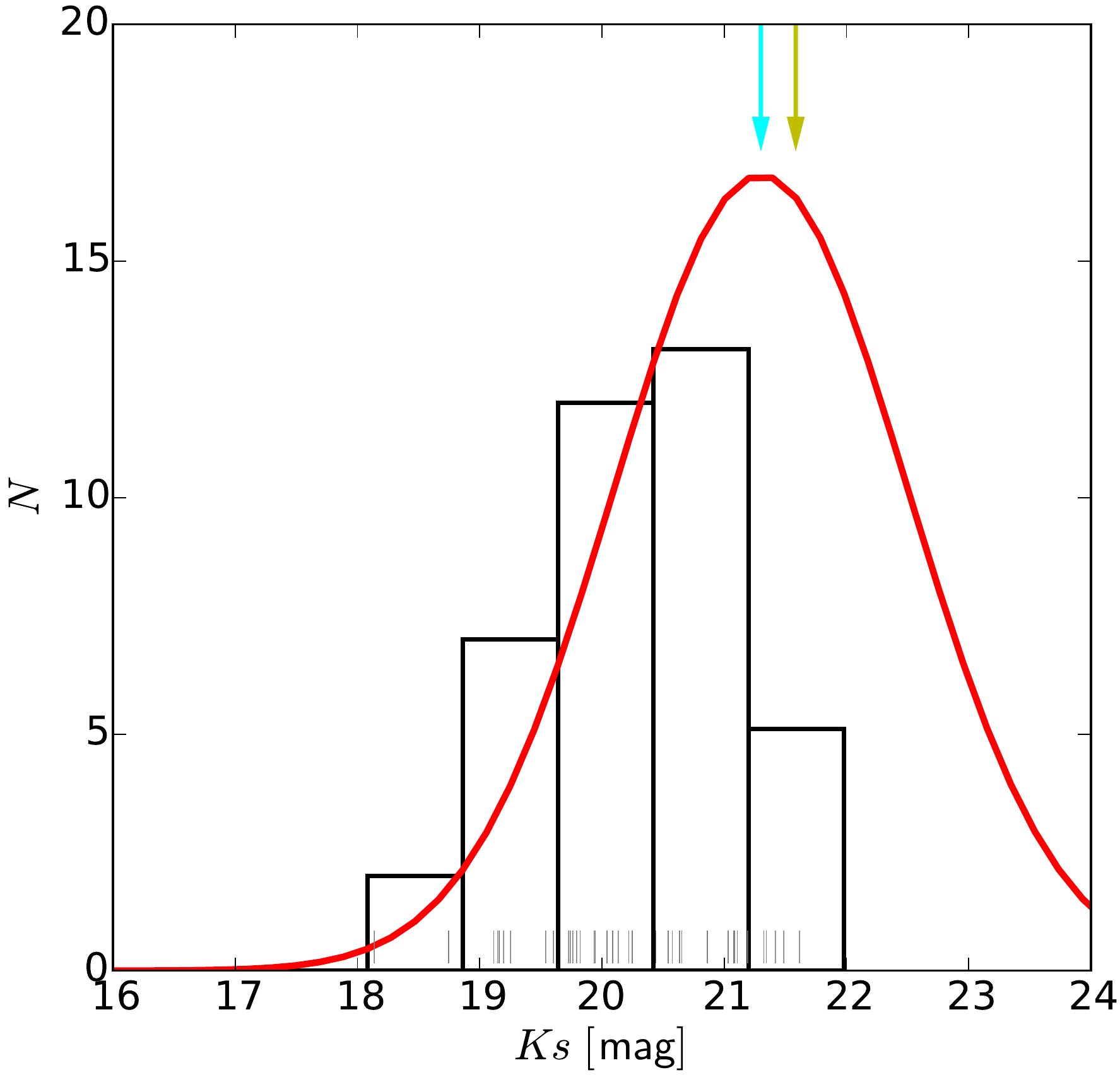}
\caption{\ks-band luminosity function of the GCCs in NGC~4258. 
{\it Histogram:} numbers of detected objects, corrected for incompleteness using 
parameters derived for concentric regions (see Section~\ref{sec:completeness}). 
\edit1{{\it Thin bars:} individual object \ks\ magnitudes.}
{\it Cyan arrow:} \ks\ TO (21.3 mag); {\it olive arrow:} 21.58 mag, for which 
completeness is 50\%. 
{\it Solid red line:} Gaussian GCLF derived from fit to the corrected histogram. 
\label{fig:gclf}}
\end{figure}

The \ks-band luminosity function for the final sample of \edit1{39} GCCs is shown in Figure~\ref{fig:gclf}.
The histogram includes
the corrections for incompleteness derived for the four different elliptical/annular regions 
considered in Section~\ref{sec:completeness} and listed in Table~\ref{tab:prit}. With these corrections, the number of 
sources increases to \edit1{40}. 
The solid red line is a Gaussian fit to the histogram,
with mean $\mu$ at the turnover magnitude $m^0_{\rm TO} = 21.3$, and $\sigma = \edit1{1.2}$ mag.  
Extrapolating over the GCLF yields \edit1{65} objects.

We would like to stress here that we do not use the extrapolation over the luminosity function
to derive the total number of globular clusters in the system. We perform the completeness and 
GCLF corrections implicitly, as we explain in the next section.

\section{Total number of globular clusters} \label{sec:totalnum}

\subsection{Correction for incomplete spatial coverage}\label{subsec:spatcov}

Usually, mostly in elliptical galaxies, the total number of globular clusters is derived from the 
observed sources by (1) applying the completeness correction, (2) extrapolating over the  
LF to account for clusters below the 
detection limit, and (3) accounting for incomplete spatial coverage, both due to the image FOV and 
to occultation/obscuration by the galaxy itself.
In particular, in the case of NGC 4258, not being an
edge-on spiral, it is essential to estimate the number of clusters that are lost to crowding 
in the brightest regions of the bulge and arms (see Section~\ref{sec:completeness}).
Both the small number of clusters detected, and the lack of constraints in the centermost regions of 
the galaxy due to confusion prevent us
from deriving this last correction from the fit to the radial distribution of GCCs customarily performed in
elliptical galaxies.

Conversely, we adapt the procedure introduced by \citet{kiss99}, based on a comparison with the
GC system of the MW. With this method, the correction for incomplete spatial coverage implicitly takes care 
of the completeness correction and the extrapolation over the LF.  
\citet{kiss99} and \citet{goud03} have discussed in detail the assumptions involved and their implications, 
as well as how the results are consistent with those from a fit to the radial number density, when 
both can be applied to the same object.

Summarizing, one 
estimates the number of GCs that one would detect
if the MW were at the distance and orientation of the galaxy of interest, and were
observed with the same instrument and to the same depth. The total number of globular clusters 
would then be:

\begin{equation}
N_{\rm GC} = N_{\rm GC} {\rm (Milky Way)} \frac{N_{\rm obs}}{N_{\rm FOV}}.
\label{eq:kissler}
\end{equation}

\noindent
$N_{\rm GC}$(Milky Way) is the total number of clusters in the MW, including objects invisible 
to us behind the bulge; $N_{\rm obs}$ is the number of clusters observed in the
target galaxy, in our case \edit1{37} after correcting for 
contamination, and $N_{\rm FOV}$ is the number of objects recovered in the artificial observation of the MW. 
We take $N_{\rm GC}$ (Milky Way) = \edit1{160$\pm$10 \citep{harr14}}. 

In order to estimate $N_{\rm FOV}$  we use, again, the catalog by \citet{harr96}; 
it provides $X,Y,Z$ coordinates for 144 MW GCs.
This coordinate system's origin is at the position of the Sun; the $X$ axis points towards the Galactic center,
the $Y$ axis in the direction of Galactic rotation, $Z$ is perpendicular to the
Galactic plane. 
We define a new coordinate system $X^{\prime}YZ$ with origin in the Galactic center, 
assuming a Galactocentric distance for the Sun R${_0}$ = 8.34 kpc \citep{reid14}.  
The main difference with
previous applications \citep[e.g.,][]{kiss99,goud03} is that NGC 4258 is not edge-on. Hence,
we have to first rotate the MW GC system in 3-D 67$\degr$ with respect to the $X^\prime$ or $Y$ axes, 
then project it on the plane of the sky; the rotated pairs ($X^\prime_{{\rm rot},X^\prime}, Y_{{\rm rot},X^\prime}$), 
($X^\prime_{{\rm rot},Y}, Y_{{\rm rot},Y}$), where the subscript indicates the Galactic axis of 3-D rotation, are directly the projected 
coordinates.
The easiest way to place the projected system on the 
WIRCam square FOV is to then apply a  
rotation around the line of sight to P.A. $= 150\degr$.  
We mask out an ellipse centered on the galaxy, with semi-major axis = 3$\farcm$5, or 0.37 $R_{25}$, axis ratio = cos(67$\degr$), and the same P.A. of $= 150\degr$ (default mask). This is the area most affected by confusion (see Section~\ref{sec:completeness} and 
Figure~\ref{fig:cmpltnss}, leftmost panel).   
As for the  limiting magnitude, we set \edit1{$V = 22.4$} mag,
combining the magnitude limit at \ks\ with the difference between the \ks\ and $V$ bands TO magnitudes; 
we first correct the $V$ mag of each cluster for the Galactic extinction given 
in the \citet{harr96} catalog, and afterwards apply the foreground Galactic reddening in the direction of
NGC~4258.

\begin{figure}
\plottwo{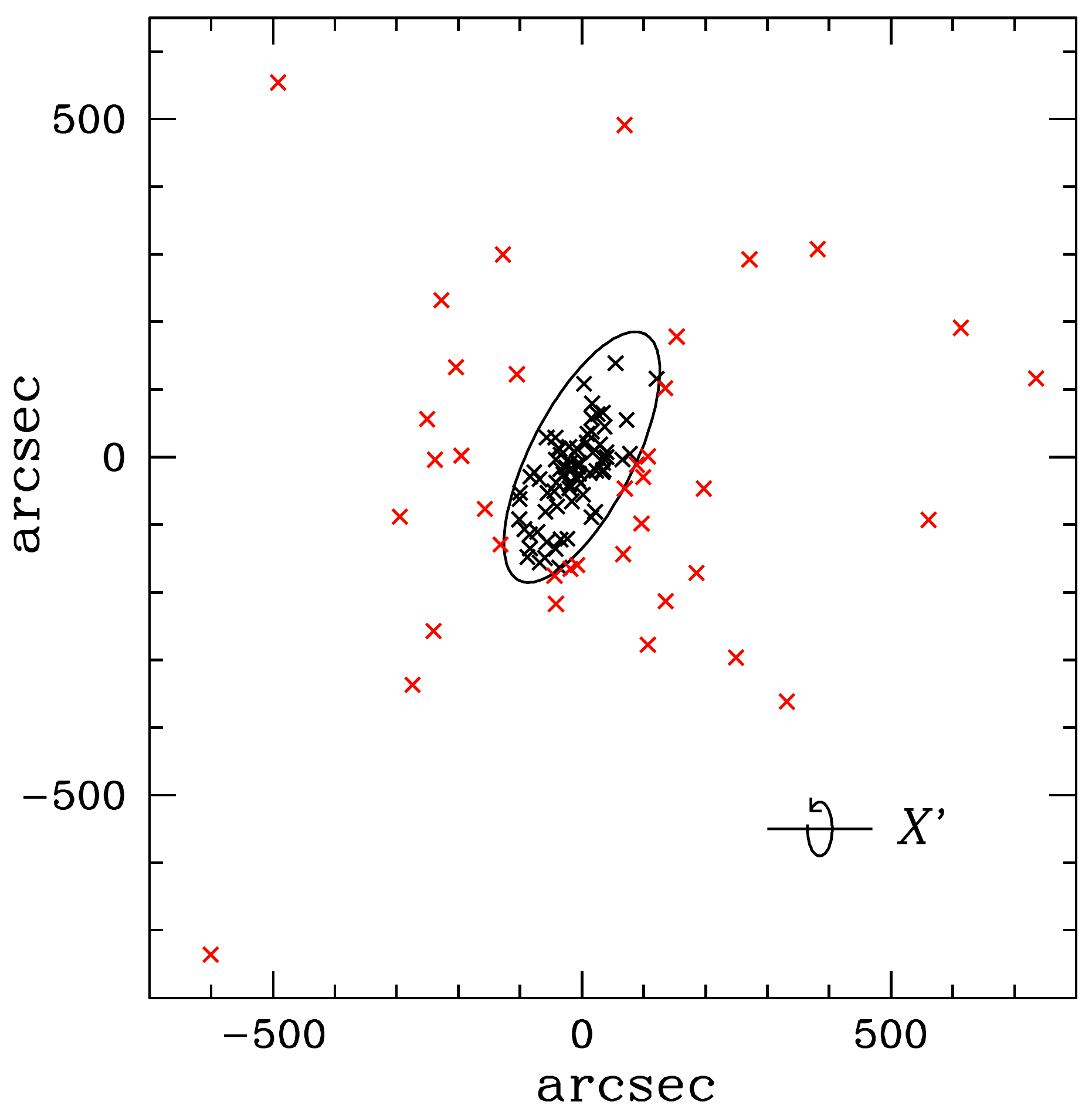}{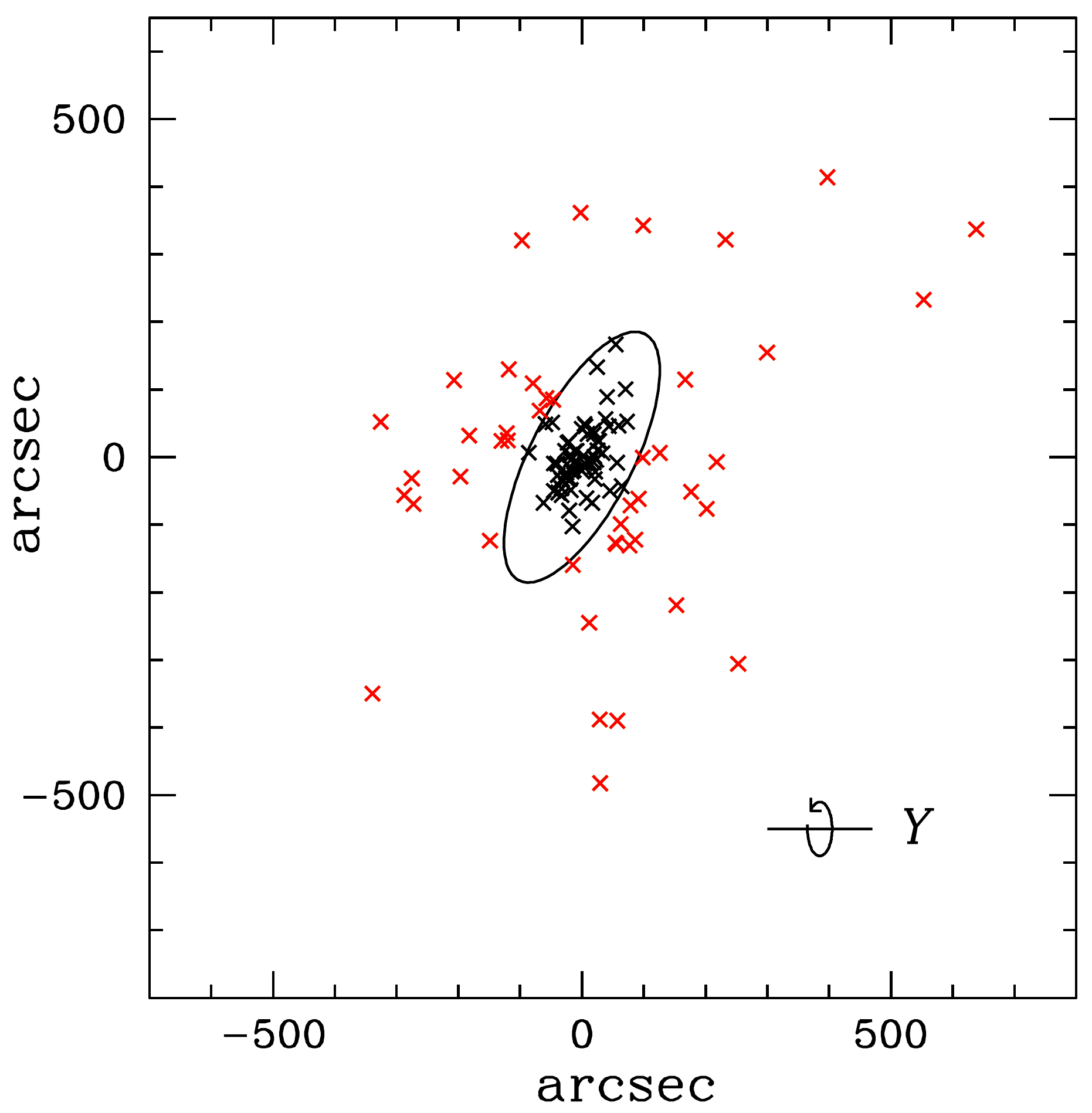}
\caption{
MW GC system, at the distance and orientation of NGC~4258. {\it Left:} 3-D rotation about Galactic $X^\prime$ axis 
before projection on the plane of the sky.
{\it Right:} 3-D rotation about Galactic $Y$ axis. {\it Solid black line:} default mask; {\it red crosses:} sources visible in the 
WIRCam FOV; {\it black crosses:} masked-out GCs. 
\label{fig:masks}}
\end{figure}

There are four possible orientations of the FOV and mask that preserve the alignment of
the galaxy; they can be seen as mirror reflections of the projected coordinates 
($+X_{\rm proj},+Y_{\rm proj}$;$+X_{\rm proj},-Y_{\rm proj}$; $-X_{\rm proj},+Y_{\rm proj}$; $-X_{\rm proj},-Y_{\rm proj}$, with two 
sets of four pairs each for rotation around the Galactic $X^\prime$ and $Y$ axes, respectively).  
We show in Figure~\ref{fig:masks} the results of the artificial observations of the MW for the rotation around the 
$X^\prime$ and $Y$ axes, respectively, and $+X_{\rm proj},+Y_{\rm proj}$; angular distances in arcsec are measured relative to
the center of the galaxy, with the horizontal axis increasing in the direction of decreasing RA.
The ellipses delineated with the black solid line mark the outer edge of the default mask; the red crosses are clusters visible in the
WIRCam FOV, whilst sources represented by black crosses have been masked out. 

For the 4 realizations rotating around the $X^\prime$ axis we detect, respectively, \edit1{38} 
($+X_{\rm proj},+Y_{\rm proj}$), \edit1{36} ($+X_{\rm proj},-Y_{\rm proj}$), \edit1{36}
($-X_{\rm proj},+Y_{\rm proj}$), and \edit1{36} ($-X_{\rm proj},-Y_{\rm proj}$)
sources; for rotating around the $Y$ axis, the numbers are \edit1{45, 46, 45, 46}.  
We repeated the exercise with a slightly different mask for the most crowded area, i.e., an ellipse with a center displaced 
horizontally from the center of the galaxy 
6$\farcs$45 to the east, semi-major axis = 3$\farcm$4 , axis ratio = cos(64$\fdg$5), P.A. = 153$\degr$ (alternative mask); 
the difference was one globular cluster less detected, on average. Changing the P.A.\ by $\pm$2$\degr$
or the inclination angle to the line of sight, also by $\pm$2$\degr$, had no effect.

With the numbers from the default mask, there would be 
\edit1{41}$\pm$5 simulated MW clusters visible in NGC~4258, and 
eq.~\ref{eq:kissler}
gives for NGC~4258 a total \edit1{$N_{\rm GC} = 144\pm31$}. This error is statistical only; 
includes errors in the assumed number of total GCs in the MW, 
Poisson errors in the observed number of GCCs in NGC~4258, and Poisson errors in the number
of simulated clusters in the WIRCam FOV.  

To this error, we add potential systematics. 
Errors in the distance to the galaxy result in uncertainties in the detection limiting magnitude, and in the 
effective areas of the galaxy covered by the FOV and excluded by the mask. 
For the given, rather small, uncertainties in the distance to NGC~4258 ($\pm$ 0.23 Mpc, if we add in quadrature the 
random and systematic errors),
$N_{\rm GC}$ would vary by \edit1{+12/-3}. Another important concern is the difference in the number of obscured clusters
between the MW and NGC~4258, particularly relevant, once again, because the latter is not edge-on. 

Albeit with small number statistics, we performed the following experiment to try to assess the effect 
of dust in the galactic disk on the detection of GC candidates. We have inspected the  
optical images of NGC 4258 and confirmed, by the sheer numbers of background galaxies seen behind the 
disk beyond $R_{25}$, that it is quite transparent there. If the effects of dust (or crowding, for that matter) 
were significantly worse
between 0.4 and 1 $R_{25}$, then we should derive a significantly larger total number of clusters from 
the region outside $R_{25}$ than from the whole field or from the area inside $R_{25}$. Masking out the
region inside the ellipse aligned with the projected galaxy and semi-major axis = $R_{25}$, there are
on average \edit1{15.5$\pm$0.8} simulated MW clusters, and (after correction for decontamination) \edit1{9.5} detected clusters  
in NGC 4258. Thus, from the outside region one would obtain \edit1{98$\pm$33} total clusters for NGC~4258, i.e., 
{\it fewer} than from the whole FOV. In view of this brief analysis, we estimate that the 
effect of obscuration should not be larger than a 25\% variation, and that the total number of 
globular clusters in NGC~4258 is \edit1{$N_{\rm GC}$ = 144$\pm$31$^{+38}_{-36}$}, with the first error statistical and
the second, systematic.  

We take the opportunity here to discuss the fact that the projected spatial distribution of the detected clusters
in NGC~4258 appears somewhat disky/flattened. At this point, it would be hard to derive 
the intrinsic shape of the system, since we have a sample of only \edit1{39} objects, all outside
the centermost region, and we lack kinematical
information. However, we have compared the distribution
of azimuthal angles $\phi$ of the GCCs in NGC~4258 with those of the MW projections, and also with randomly generated uniform distributions
of $\phi$ with \edit1{39} points. The Kolmogorov-Smirnov test cannot rule out with high significance, neither that
the NGC~4258 system has been drawn from a uniform distribution of azimuthal angles, nor that the MW projections and the NGC~4258 system
have been drawn from the same distribution.
Spectroscopic follow-up would be highly desirable to investigate whether we are looking at a large disk GC 
configuration. It has been reported that satellite galaxies of both the MW and M~31 may not 
be isotropically distributed and, instead, form coplanar groups 
\citep[e.g.,][]{metz07,metz08,metz09,ibat13}. Regarding in particular the MW globular clusters, 
two subgroups seem to have coplanar configurations \citep{pawl12}: the bulge/disk clusters, aligned with the plane of the Galaxy,  
and the young halo clusters \citep{mack05}, at Galactocentric distances larger than 20 kpc, and lying in the same polar plane as the
satellite galaxies.

\section{Specific frequency, and the $N_{\rm GC}$ vs.\ $M_\bullet$ and $M_{\rm GC}$ vs.\ $M_\bullet$ relations} 

From the distance to the galaxy, and the values of $B_{T,0} = 9.10 \pm 0.07$ mag and ($B\ - V$)$_{T,0} = 0.55$ mag 
given in \citet{rc3}, we find $M_V = -21.42 \pm 0.10$ mag, and a specific frequency $S_N = 
 N_{\rm GC} \times 10^{0.4 \times [M_V + 15]} = \edit1{0.39 \pm 0.09}$, if we consider random errors only, and
$S_N = \edit1{0.39 \pm 0.13}$ if we include the uncertainty in the number of obscured clusters.\footnote{The systematic error in the
distance basically cancels out, because the absolute magnitude of the galaxy also changes.}
This specific frequency is comparable to the Milky Way's, which has $S_N = \edit1{0.5}\pm0.1$ \citep{ashm98}. 

\begin{figure}[ht!]
\plottwo{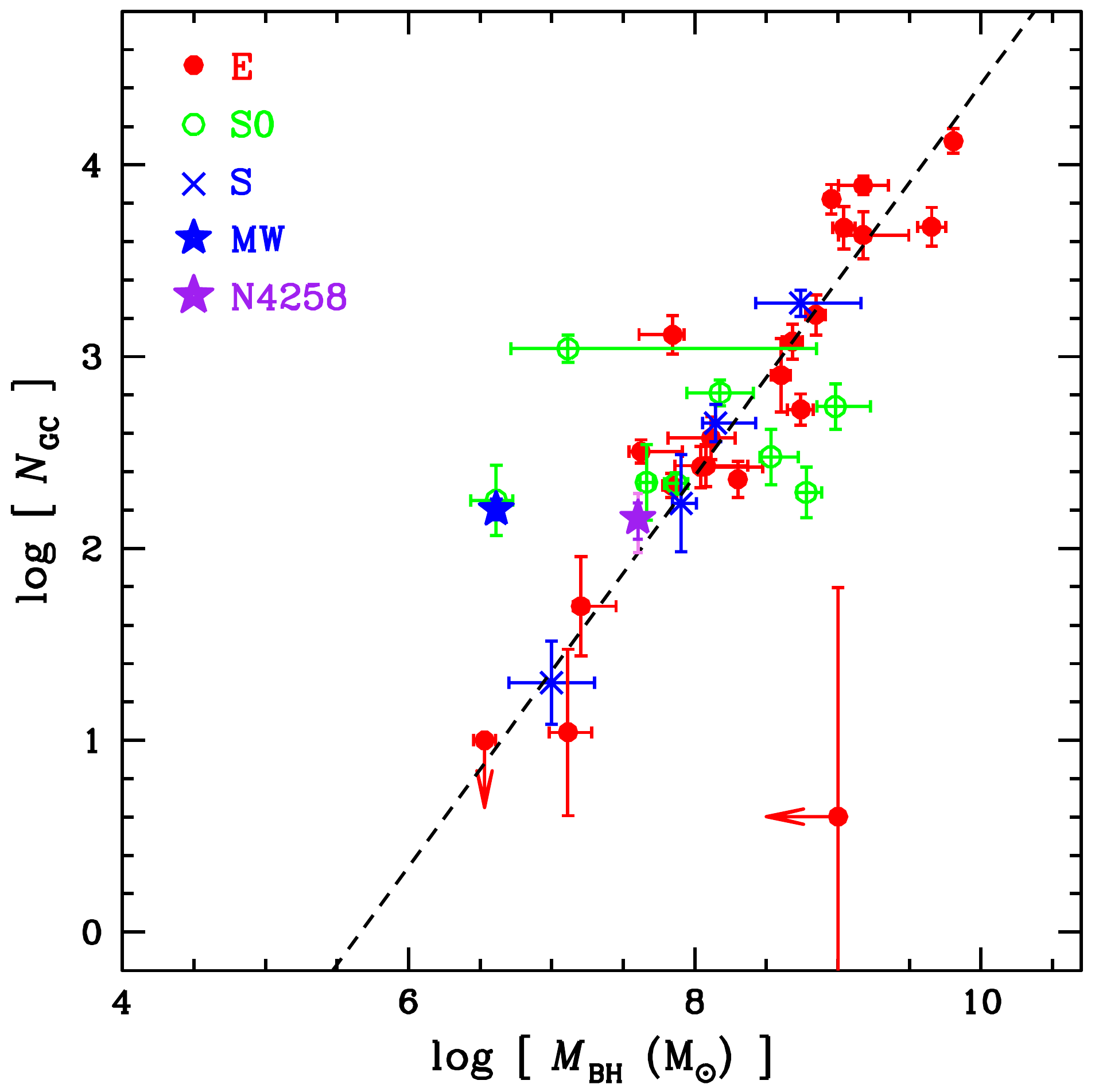}{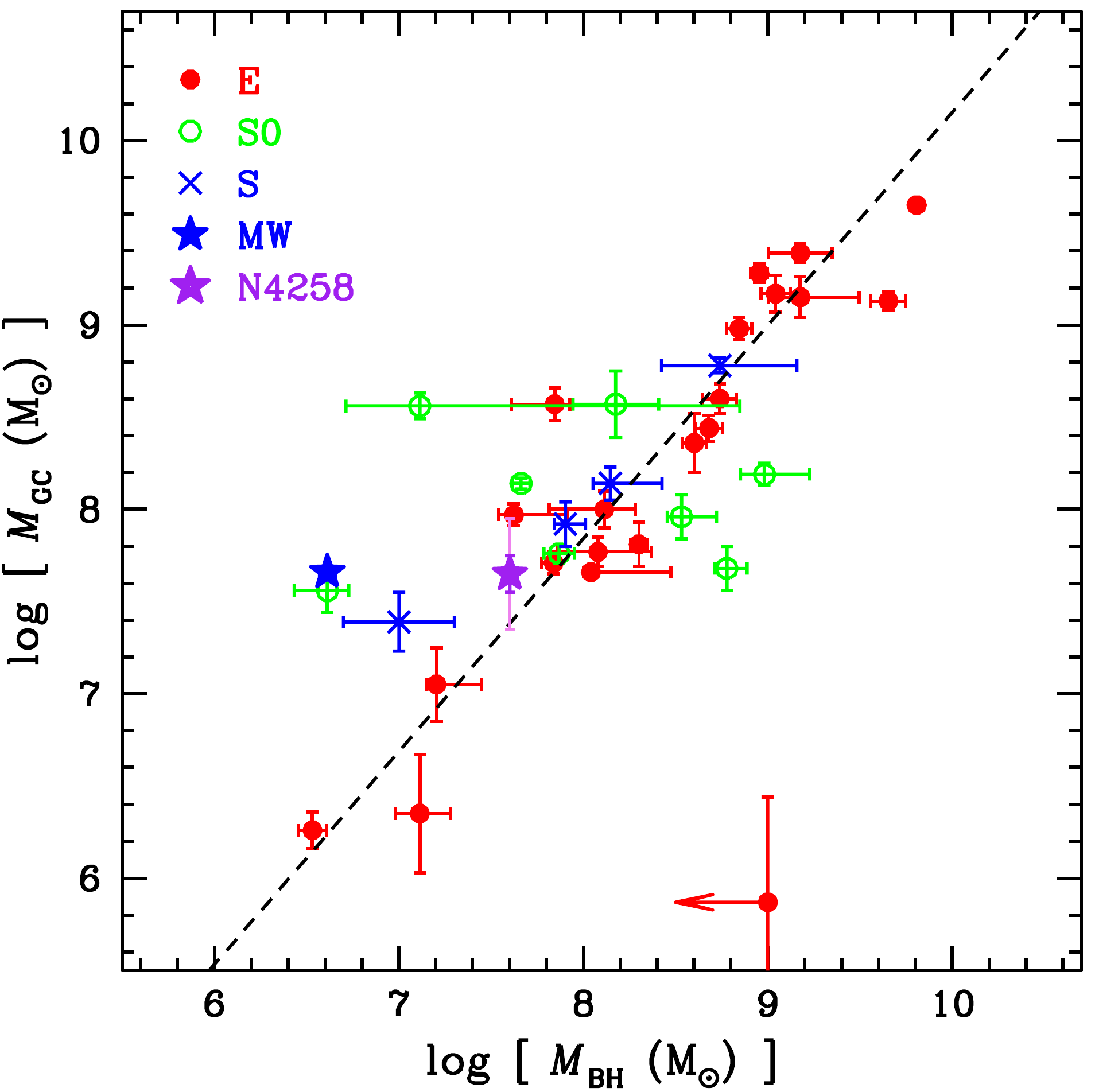}
\caption{Relation of central black hole mass with number of globular clusters
(log $N_{\rm GC}$ vs.\ log $M_\bullet$), {\it left}, and with total mass of
GC system (log $M_{\rm GC}$ vs.\ log $M_\bullet$), {\it right}.  
Solid ({\it red}) and open ({\it green}) circles, and ({\it blue}) crosses
represent, respectively, the elliptical,
lenticular, and spiral galaxies in the sample of \citet{harr11,harr14}. 
The {\it solid blue} star is the Milky Way, also in their sample;
the {\it solid purple} star is NGC~4258, for which we derive $N_{\rm GC}$ and $M_{\rm GC}$
in this work. The purple bars correspond 
to random errors only; systematic errors,
added in quadrature to the former, are shown in magenta. 
In the log $N_{\rm GC}$ vs.\ log $M_\bullet$ plot, 
the {\it dashed line} is Harris \& Harris's fit to the ellipticals only that are not
upper or lower limits in either $N_{\rm GC}$ or $M_\bullet$.
In the log $M_{\rm GC}$ vs.\ log $M_\bullet$ graph, 
the {\it dashed line} is our fit to the ellipticals only in the Harris \& Harris sample 
that are not upper or lower limits in either $M_{\rm GC}$ or $M_\bullet$.
\label{fig:ngcvsmbh}}
\end{figure}

Figure~\ref{fig:ngcvsmbh}, left,  displays the location 
of NGC~4258 (purple star) in the log $N_{\rm GC}$ vs.\ log $M_\bullet$ diagram, 
together with the \citet{harr11,harr14} sample (ellipticals represented by
solid red circles, lenticulars by open green ones).  We show the random errors for NGC~4258 with 
purple bars, and in magenta the systematic errors added in quadrature to the former.
The spiral galaxies M~31, M~81, M~104, and NGC~253 (blue crosses) fall on the elliptical
correlation, the dashed line, calculated by \citet{harr11} as log 
$N_{\rm GC} = (-5.78 \pm 0.85) + (1.02 \pm 0.10)$ log $M_\bullet/M_\odot$. 
The MW (solid blue star), on the other hand, deviates significantly from it.
For NGC~4258, the relation
predicts $N_{\rm GC,predicted} = 94 \pm 14$, which we must compare with our derived 
\edit1{$N_{\rm GC,derived}$ = 144$\pm$31$^{+38}_{-36}$}. If only random errors are considered, 
$N_{\rm GC,derived}$ is consistent with the prediction within 2 $\sigma$. With systematic errors added in
quadrature, for 1 $\sigma$ the minimum $N_{\rm GC,derived,min} = \edit1{96}$, while the maximum    
$N_{\rm GC,predicted,max} = 108$.

We take the opportunity to include an $M_{\rm GC}$ vs.\ log $M_\bullet$ diagram, where 
$M_{\rm GC}$ is the total mass of the GC system. \citet{burk10} had already noticed that the
total mass of the GC system is, to a good approximation, equal to the mass of the central black hole;
they assumed a mean GC mass of 2$\times 10^5 M_\odot$, and found $M_\bullet \propto M_{\rm GC}^{1.08 \pm 0.04}$ 
for their sample. \citet{harr11} speculate, however, that this equality could be coincidental and transitory, 
if GC systems have been continually losing mass, whereas black holes have been accreting it, since
their respective births at a similar redshift.

The $M_{\rm GC}$ vs.\ log $M_\bullet$ for the \citet{harr11,harr14} sample plus NGC~4258
is shown in Figure~\ref{fig:ngcvsmbh}, right panel.
$M_{\rm GC}$ values for objects in the Harris \& Harris sample have been taken from \citet{harr13}. 
Our fit to the ellipticals only (excluding NGC~4486B, whose SMBH mass is an upper limit)  
yields log $M_{\rm GC}/M_\odot = (-1.40 \pm 0.79) + (1.15 \pm 0.09)$ log $M_\bullet/M_\odot$. 

In order to place NGC~4258 in the plot, we derive the total mass of its GC system in two independent
ways, and obtain the same result. (1) We calculate the dynamical mass of the galaxy as 
$M_{\rm dyn} = 4 R_e \sigma_e^2 / G$ \citep{wolf10,harr13}; $R_e =$ 4.6 kpc and $\sigma_e =$ 115 km
s$^{-1}$ \citep{rc3}. We then infer the mean GC mass $<M_{\rm GC}>$ from Figure 13 in \citet{harr13}; 
we get $10^{5.4} M_\odot$,
and multiply it by the total number of GCs that we have estimated for NGC~4258 in Section~\ref{sec:totalnum}.
The result is log $M_{\rm GC} = 7.6 \pm 0.1 \pm 0.1$, random and systematic uncertainty,
respectively, taken directly from the  
uncertainty in $N_{\rm GC}$.
(2) On the other hand, 99\% of the mass of a GC system is contained in the clusters with 
M$_V <$ -6.5 mag, i.e., brighter than one mag below the GCLF turnover, which is the case 
for all the 38 candidates we have detected.
We notice (see Table~\ref{tab:colordist} and Figure~\ref{fig:colordist}) that 
the (\gp - \rp) color of the GCCs in NGC~4258, \edit1{0.49}$\pm$0.06 mag, is very narrow 
and perfectly consistent with an SSP with age 12 Gyr, $Z =$ 0.0004, and a Chabrier IMF \citep{bruz03}. By comparing 
the absolute \gp\ and \rp\ magnitudes of our clusters with the values 
given by the models for one solar mass of such population, 
we derive their individual masses.  We add them up and obtain, for both \gp\ and \rp, log $M_{\rm GC} = \edit1{7.7^{+0.3}_{-0.3}}$.
Here, we estimate the lower error bar assuming that the dispersion in the color distribution is due to age; an SSP with 
(\gp - \rp) $\sim$ \edit1{0.43} mag is about 5 Gyr old, and its mass-to-light ratio $M/L$ is roughly half as for 12 Gyr. The upper error bar 
is dominated by the [39$ \times (\frac{144}{65} -1)$] bright clusters that we could 
have missed in the crowded region inside $\sim 0.4 R_{25}$; this error would of course be smaller if the 
masked-out region had been more reduced, i.e., for elliptical or 
spiral edge-on galaxies. We show this very conservative systematic error bars in Figure~\ref{fig:ngcvsmbh}, right panel. 
\footnote{ 
Taking the $g$ and $r$ SDSS absolute magnitudes for the Sun from 
\url{http://www.ucolick.org/~cnaw/sun.html}, and transforming them to the MegaCam system, 
we derive $(M/L)_{g^\prime}$ = 3.9, and
$(M/L)_{r^\prime}$ = 3.6 for our clusters.}

Since it does not depend on detecting or accounting for faint clusters, the $M_{\rm GC}$ vs.\ log $M_\bullet$
correlation could be a very robust way to predict $M_\bullet$ in galaxies where other methods are unavailable. 
Conversely, $N_{\rm GC}$ could be biased when  
the GCLF turnover is not reached with at least 50\% completeness.  
The $M_{\rm GC}$ values in \citet{harr13} were not calculated from the brighter clusters but, rather,
by adding up all clusters and assuming a mean GC mass; hence, it is not possible here to 
assess the intrinsic scatter of the correlation when only clusters brighter than $M_V = -6.5$ are considered.
Clearly, however, this should be investigated. 

\section{Summary and conclusions} \label{sec:concl}

Up until now, the study of GC systems in spiral galaxies has been hampered by the sparsity of their globulars, compared to
ellipticals, and mostly limited to edge-on objects, given the difficulty of detecting individual GCs projected
on a spiral disk.   
Given our goal of comparing the number of GCs vs.\ the mass of the SMBH in spirals, we need to study the
GC systems of galaxies with measured $M_\bullet$, regardless of their orientation to the line of sight.

We have successfully applied the \uiks\ diagram GC selection technique \citep{muno14}, for the first time to 
a spiral galaxy, NGC~4258. 
GCCs occupy a region in the \uiks\ color-color plot that is virtually free
not only of foreground Galactic stars and background galaxies, but also of young stellar clusters in the 
disk of the spiral galaxy of interest. We have also demonstrated, specifically for the case of NGC~4258, how superior
the \uiks\ color-color diagram is as a GCC selection tool, compared to other more widely used diagrams, like  
(\rp\ - \ip) vs.\ (\gp\ - \ip),  (F438W - F606W) vs.\ (F606W - F814W),
and ($U\ - B$) vs.\ ($V\ - I$).
 
We complemented the \uiks\ plot with the structural parameters (i.e., mainly effective radius $r_e$, FWHM, and SExtractor's SPREAD\_MODEL) of the objects in the 
selection region to determine our final sample. 
We thus increased the sample of spirals with measurements of both the GC system and SMBH mass to 6 galaxies (by 20\%).

The combination of the \uiks\ diagram plus structural parameters stands as the most efficient 
photometric tool to investigate the GC systems in both
ellipticals and spirals. Even though deep $u^*/U$ and near-infrared images with good resolution are
more expensive than optical data between 4000 and 8500 $\AA$, they are much easier to obtain than the alternative spectroscopy. 

In spite of the extreme shallowness of the \ks-band data, we were able to detect 
\edit1{39} GCCs in NGC~4258. 
NGC~4258 is of special interest, since it is the archaetypical megamaser galaxy, 
and it harbors the central black hole with the best mass measurement 
outside the MW.  
The unimodal color distribution of the GCCs is consistent with those of the
globular clusters in the MW and M~31. After completeness, GCLF, and space coverage corrections, we derive \edit1{$N_{GC} = 144\pm 31$} and 
\edit1{$S_N = 0.4 \pm 0.1$} (random uncertainty only) for NGC~4258. 
The galaxy falls within 2 $\sigma$ on the $N_{\rm GC}$ vs.\ $M_\bullet$ relation determined by
elliptical galaxies.
A spectroscopic study will further validate our procedures of source detection and selection; 
confirm the membership of our candidates in the NGC~4258 GC system; 
allow us determine their kinematics, the shape of the system, and the dark matter content of the halo 
within $\sim$ 25 kpc; investigate whether the galaxy also follows the correlation, discussed by 
\citet{sado12}, between $M_\bullet$ and the velocity dispersion of the GC system. 

Regarding the $N_{\rm GC}$ vs.\ $M_\bullet$ relation for spirals, and its bearing
on galaxy formation and assembly, clearly a larger sample is required. 
At the same time that 
$M_\bullet$ scaling relations have been least explored for low mass galaxies,
this is precisely the mass range where different proposed scenarios could
be distinguished.
For example, if the ability to fuel a central black hole is correlated with  
a classical bulge \citep[e.g.,][]{hu08,gree08,gado09}, then galaxies with pseudobulges could  
show a  correlation sequence between $N_{\rm GC}$ and $M_\bullet$ that is
offset towards lower MBH masses relative to the relation for ellipticals. If, on the other hand,
scaling relations are determined by statistical convergence through merging \citep{peng07,jahn11},
their scatter should increase with decreasing galaxy/halo mass. Black hole masses and galaxy properties
determined for low mass galaxies of varied morphologies are urgently needed, and the method we have
presented here opens promising possibilities for the study of GC systems in these galaxies.

\acknowledgments

\edit1{We thank the anonymous referee for her/his constructive and insightful remarks.}
RAGL and LL acknowledge the support from DGAPA, UNAM, through project PAPIIT IG100913; 
RAGL also acknowledges support from CONACyT, Mexico, 
through project SEP-CONACyT I0017-151671.
GBA acknowledges support for this work from the National Autonomous University of M\'exico (UNAM), through grant PAPIIT IG100115. YO-B 
acknowledges financial support through CONICYT-Chile (grant CONICYT-PCHA/Doctorado Nacional/2014-21140651).
THP acknowledges support by the FONDECYT Regular Project Grant (No.\ 1161817) and the BASAL Center for 
Astrophysics and Associated Technologies (PFB-06).
Pascal Fouqu\'e aided us to recover the information about the strategy used to observe NGC~4258 in the \ks-band with CFHT.
John Blakeslee generously offered us his expertise every time we asked.

Based on observations obtained with MegaPrime/MegaCam, a joint project of CFHT and CEA/IRFU, at the Canada-France-Hawaii Telescope (CFHT) which is operated by the National Research Council (NRC) of Canada, the Institut National des Science de l'Univers of the Centre National de la Recherche Scientifique (CNRS) of France, and the University of Hawaii. This work is based in part on data products produced at Terapix available at the Canadian Astronomy Data Centre as part of the Canada-France-Hawaii Telescope Legacy Survey, a collaborative project of NRC and CNRS. 

Based on observations obtained with WIRCam, a joint project of CFHT, Taiwan, Korea, Canada, France, at the Canada-France-Hawaii Telescope (CFHT) which is operated by the National Research Council (NRC) of Canada, the Institute National des Sciences de l'Univers of the Centre National de la Recherche Scientifique of France, and the University of Hawaii. This work is based in part on data products produced at TERAPIX, the WIRDS (WIRcam Deep Survey) consortium, and the Canadian Astronomy Data Centre. This research was supported by a grant from the Agence Nationale de la Recherche ANR-07-BLAN-0228  

\vspace{5mm}
\facilities{CFHT}

\software{SExtractor, PSFEx, TopCat, IRAF, Kapteyn, \sc{ishape}}

\listofchanges


\begin{thebibliography}{}
\bibitem[Annunziatella et al.(2013)]{annu13} Annunziatella, M., Mercurio, A., Brescia, M., Cavuoti, S., \& Longo, G.\ 2013, \pasp, 125, 68 
\bibitem[Ashman \& Zepf(1998)]{ashm98} Ashman, K.~M., \& Zepf, S.~E.\ 1998, Cambridge Astrophysics Series, 30, 
\bibitem[Barmby et al.(2000)]{barm00} Barmby, P., Huchra, J.~P., Brodie, J.~P., et al.\ 2000, \aj, 119, 727
\bibitem[Bertin(2011)]{bert11} Bertin, E.\ 2011, Astronomical Data Analysis Software and Systems XX, 442, 435 
\bibitem[Bertin \& Arnouts(1996)]{bert96} Bertin, E., \& Arnouts, S.\ 1996, \aaps, 117, 393 
\bibitem[Bruzual \& Charlot(2003)]{bruz03} Bruzual, G., \& Charlot, S.\ 2003, \mnras, 344, 1000
\bibitem[Boulade et al.(2003)]{boul03} Boulade, O., Charlot, X., Abbon, P., et al.\ 2003, \procspie, 4841, 72 
\bibitem[Burkert \& Tremaine(2010)]{burk10} Burkert, A., \& Tremaine, S.\ 2010, \apj, 720, 516 
\bibitem[Capuzzo-Dolcetta \& Donnarumma(2001)]{capu01} Capuzzo-Dolcetta, R., \& Donnarumma, I.\ 2001, \mnras, 328, 645 
\bibitem[Capuzzo-Dolcetta \& Mastrobuono-Battisti(2009)]{capu09} Capuzzo-Dolcetta, R., \& Mastrobuono-Battisti, A.\ 2009, \aap, 507, 183 
\bibitem[Capuzzo-Dolcetta \& Vicari(2005)]{capu05} Capuzzo-Dolcetta, R., \& Vicari, A.\ 2005, \mnras, 356, 899 
\bibitem[Cardelli et al.(1989)]{card89} Cardelli, J.~A., Clayton, G.~C., \& Mathis, J.~S.\ 1989, \apj, 345, 245
\bibitem[Chabrier(2003)]{chab03} Chabrier, G.\ 2003, \pasp, 115, 763
\bibitem[Davis et al.(2007)]{davi07} Davis, M., Guhathakurta, P., Konidaris, N.~P., et al.\ 2007, \apjl, 660, L1 
\bibitem[Desai et al.(2012)]{desa12} Desai, S., Armstrong, R., Mohr, J.~J., et al.\ 2012, \apj, 757, 83
\edit1{\bibitem[de Souza et al.(2015)]{deso15} de Souza, R.~S., Hilbe, J.~M., Buelens, B., et al.\ 2015, \mnras, 453, 1928} 
\bibitem[de Vaucouleurs et al.(1991)]{rc3} de Vaucouleurs, G., de Vaucouleurs, A., Corwin, H.~G., Jr., et al.\ 1991, Third Reference Catalogue of Bright Galaxies.~Volume I: Explanations and references.~ Volume II: Data for galaxies between 0$^{h}$ and 12$^{h}$.~ Volume III: Data for galaxies between 12$^{h}$ and 24$^{h}$., by de Vaucouleurs, G.; de Vaucouleurs, A.; Corwin, H.~G., Jr.; Buta, R.~J.; Paturel, G.; Fouqu{\'e}, P..~Springer, New York, NY (USA), 1991, 2091 p.
\bibitem[Dressler(1989)]{dres89} Dressler, A.\ 1989,  in IAU Symp. 134, Active Galactic Nuclei, ed. D. E. Osterbrock \& J. S. Miller (Dordrecht: Kluwer), 217 
\bibitem[Fabian(2012)]{fabi12} Fabian, A.~C.\ 2012, \araa, 50, 455 
\bibitem[Fedotov et al.(2011)]{fedo11} Fedotov, K., Gallagher, S.~C., Konstantopoulos, I.~S., et al.\ 2011, \aj, 142, 42 
\bibitem[Ferrarese et al.(2012)]{ferra12} Ferrarese, L., C{\^o}t{\'e}, P., Cuillandre, J.-C., et al.\ 2012, \apjs, 200, 4 
\bibitem[Ferrarese \& Merritt(2000)]{ferr00} Ferrarese, L., \& Merritt, D.\ 2000, \apj, 539, L9 
\bibitem[Gadotti \& Kauffmann(2009)]{gado09} Gadotti, D.~A., \& Kauffmann, G.\ 2009, \mnras, 399, 621 
\bibitem[Gebhardt et al.(2000)]{gebh00} Gebhardt, K., et al.\ 2000, \apj, 539, L13 
\bibitem[Georgiev et al.(2006)]{geor06} Georgiev, I.~Y., Hilker, M., Puzia, T.~H., et al.\ 2006, \aap, 452, 141
\edit1{\bibitem[Georgiev \& B{\"o}ker(2014)]{geor14} Georgiev, I.~Y., \& B{\"o}ker, T.\ 2014, \mnras, 441, 3570} 
\bibitem[Gnedin et al.(2014)]{gned14} Gnedin, O.~Y., Ostriker, J.~P., \& Tremaine, S.\ 2014, \apj, 785, 71 
\bibitem[Goodwin et al.(1998)]{good98} Goodwin, S.~P., Gribbin, J., \& Hendry, M.~A.\ 1998, The Observatory, 118, 201 
\bibitem[Goudfrooij et al.(2003)]{goud03} Goudfrooij, P., Strader, J., Brenneman, L., et al.\ 2003, \mnras, 343, 665
\bibitem[Graham(2012a)]{grah12a} Graham, A.~W.\ 2012a, \apj, 746, 113 
\bibitem[Graham(2012b)]{grah12b} \samename\ 2012b, \mnras, 422, 1586 
\bibitem[Greene et al.(2008)]{gree08} Greene, J.~E., Ho, L.~C., \& Barth, A.~J.\ 2008, \apj, 688, 159 
\bibitem[Greene et al.(2010)]{gree10} Greene, J.~E., et al.\ 2010, \apj, 721, 26 
\bibitem[G{\"u}ltekin et al.(2009)]{gult09} G{\"u}ltekin, K., et al.\ 2009, \apj, 698, 198 
\bibitem[Gwyn(2008)]{gwyn08} Gwyn, S.~D.~J.\ 2008, \pasp, 120, 212 
\bibitem[Gwyn(2012)]{gwyn12} \samename\ 2012, \aj, 143, 38 
\bibitem[Gwyn(2014)]{gwyn14} \samename\ 2014, Astronomical Data Analysis Software and Systems XXIII, 485, 387 
\bibitem[Hammer et al.(2010)]{hamm10} Hammer, D., Verdoes Kleijn, G., Hoyos, C., et al.\ 2010, \apjs, 191, 143 
\bibitem[H{\"a}ring \& Rix(2004)]{hari04} H{\"a}ring, N., \& Rix, H.-W.\ 2004, \apj, 604, L89 
\bibitem[Harris 1996 (2010 edition) ]{harr96} Harris, W.~E.\ 1996, \aj, 112, 1487 
\bibitem[Harris \& Harris(2011)]{harr11} Harris, G.~L.~H., \& Harris, W.~E.\ 2011, \mnras, 410, 2347 
\edit1{\bibitem[Harris et al.(2009)]{harr09} Harris, W.~E., Kavelaars, J.~J., Hanes, D.~A., Pritchet, C.~J., \& Baum, W.~A.\ 2009, \aj, 137, 3314} 
\bibitem[Harris et al.(2014)]{harr14} Harris, G.~L.~H., Poole, G.~B., \& Harris, W.~E.\ 2014, \mnras, 438, 2117 
\bibitem[Harris et al.(2013)]{harr13} Harris, W.~E., Harris, G.~L.~H., \& Alessi, M.\ 2013, \apj, 772, 82
\bibitem[Herrnstein et al.(1999)]{herr99} Herrnstein, J.~R., Moran, J.~M., Greenhill, L.~J., et al.\ 1999, \nat, 400, 539 
\bibitem[Hu(2008)]{hu08} Hu, J.\ 2008, \mnras, 386, 2242 
\bibitem[Humphreys et al.(2013)]{hump13} Humphreys, E.~M.~L., Reid, M.~J., Moran, J.~M., Greenhill, L.~J., \& Argon, A.~L.\ 2013, \apj, 775, 13 
\bibitem[Ibata et al.(2013)]{ibat13} Ibata, R.~A., Lewis, G.~F., Conn, A.~R., et al.\ 2013, \nat, 493, 62 
\bibitem[Jahnke \& Macci{\`o}(2011)]{jahn11} Jahnke, K., \& Macci{\`o}, A.~V.\ 2011, \apj, 734, 92 
\bibitem[Jalali et al.(2012)]{jala12} Jalali, B., Baumgardt, H., Kissler-Patig, M., et al.\ 2012, \aap, 538, A19 
\bibitem[Jester et al.(2005)]{jest05} Jester, S., Schneider, D.~P., Richards, G.~T., et al.\ 2005, \aj, 130, 873
\bibitem[Jord{\'a}n et al.(2005)]{jord05} Jord{\'a}n, A., C{\^o}t{\'e}, P., Blakeslee, J.~P., et al.\ 2005, \apj, 634, 1002 
\edit1{\bibitem[Jord{\'a}n et al.(2007)]{jord07} Jord{\'a}n, A., McLaughlin, D.~E., C{\^o}t{\'e}, P., et al.\ 2007, \apjs, 171, 101} 
\edit1{\bibitem[King(1962)]{king62} King, I.\ 1962, \aj, 67, 471} 
\edit1{\bibitem[King(1966)]{king66} King, I.~R.\ 1966, \aj, 71, 64}
\bibitem[Kissler-Patig et al.(1999)]{kiss99} Kissler-Patig, M., Ashman, K.~M., Zepf, S.~E., \& Freeman, K.~C.\ 1999, \aj, 118, 197 
\bibitem[Kormendy(1993)]{korm93} Kormendy, J.\ 1993, The Nearest Active Galaxies, 197 
\bibitem[Kormendy \& Gebhardt(2001)]{korm01} Kormendy, J., \& Gebhardt, K.\ 2001, 
20th Texas Symposium on relativistic astrophysics, 586, 363
\bibitem[Kormendy \& Ho(2013)]{korm13} Kormendy, J., \& Ho, L.C.\ (2013), \araa, 51, 511
\bibitem[Kormendy \& Richstone(1995)]{korm95} Kormendy, J., \& Richstone, D.\ 1995, \araa, 33, 581 
\bibitem[Kron(1980)]{kron80} Kron, R.~G.\ 1980, \apjs, 43, 305 
\bibitem[Laor(2001)]{laor01} Laor, A.\ 2001, \apj, 553, 677
\edit1{\bibitem[Larsen(1999)]{lars99} Larsen, S.~S.\ 1999, \aaps, 139, 393}
\bibitem[L{\"a}sker et al.(2014)]{lask14} L{\"a}sker, R., Ferrarese, L., van de Ven, G., \& Shankar, F.\ 2014, \apj, 780, 70
\bibitem[L{\"a}sker et al.(2016)]{lask16} L{\"a}sker, R., Greene, J.~E., Seth, A., et al.\ 2016, \apj, 825, 3
\bibitem[Mackey \& van den Bergh(2005)]{mack05} Mackey, A.~D., \& van den Bergh, S.\ 2005, \mnras, 360, 631
\bibitem[Magnier \& Cuillandre(2004)]{magn04} Magnier, E.~A., \& Cuillandre, J.-C.\ 2004, \pasp, 116, 449
\bibitem[Magorrian et al.(1998)]{mago98} Magorrian, J., et al.\ 1998, \aj, 115, 2285 
\bibitem[Marconi \& Hunt(2003)]{marc03} Marconi, A., \& Hunt, L.~K.\ 2003, \apj, 589, L21 
\bibitem[McLaughlin et al.(1994)]{mcla94} McLaughlin, D.~E., Harris, W.~E., \& Hanes, D.~A.\ 1994, \apj, 422, 486 
\bibitem[McLure \& Dunlop(2002)]{mclu02} McLure, R.~J., \& Dunlop, J.~S.\ 2002, \mnras, 331, 795 
\bibitem[McConnell et al.(2011)]{mcco11} McConnell, N.~J., Ma, C.-P., Gebhardt, K., et al.\ 2011, \nat, 480, 215 
\bibitem[Metz et al.(2007)]{metz07} Metz, M., Kroupa, P., \& Jerjen, H.\ 2007, \mnras, 374, 1125 
\bibitem[Metz et al.(2009)]{metz09} \samename\ 2009, \mnras, 394, 2223
\bibitem[Metz et al.(2008)]{metz08} Metz, M., Kroupa, P., \& Libeskind, N.~I.\ 2008, \apj, 680, 287-294
\bibitem[Mu{\~n}oz et al.(2014)]{muno14} Mu{\~n}oz, R.~P., Puzia, T.~H., Lan{\c c}on, A., et al.\ 2014, \apjs, 210, 4 
\edit1{\bibitem[Muratov \& Gnedin(2010)]{mura10} Muratov, A.~L., \& Gnedin, O.~Y.\ 2010, \apj, 718, 1266}
\bibitem[O'Donnell(1994)]{odon94} O'Donnell, J.~E.\ 1994, \apj, 422, 158
\bibitem[Oke(1974)]{oke74} Oke, J.~B.\ 1974, \apjs, 27, 21 
\bibitem[Pawlowski et al.(2012)]{pawl12} Pawlowski, M.~S., Pflamm-Altenburg, J., \& Kroupa, P.\ 2012, \mnras, 423, 1109
\bibitem[Peacock et al.(2010)]{peac10} Peacock, M.~B., Maccarone, T.~J., Knigge, C., et al.\ 2010, \mnras, 402, 803 
\bibitem[Peng(2007)]{peng07} Peng, C.~Y.\ 2007, \apj, 671, 1098
\bibitem[Pota et al.(2015)]{pota15} Pota, V., Brodie, J.~P., Bridges, T., et al.\ 2015, \mnras, 450, 1962 
\edit1{\bibitem[Pota et al.(2013)]{pota13} Pota, V., Graham, A.~W., Forbes, D.~A., et al.\ 2013, \mnras, 433, 235} 
\bibitem[Powalka et al.(2016)]{powa16} Powalka, M., Puzia, T.~H., Lan{\c c}on, A., et al.\ 2016, arXiv:1608.08628 
\bibitem[Puget et al.(2004)]{puge04} Puget, P., Stadler, E., Doyon, R., et al.\ 2004, \procspie, 5492, 978 
\bibitem[Puzia et al.(2014)]{puzi14} Puzia, T.~H., Paolillo, M., Goudfrooij, P., et al.\ 2014, \apj, 786, 78
\bibitem[Reid et al.(2014)]{reid14} Reid, M.~J., Menten, K.~M., Brunthaler, A., et al.\ 2014, \apj, 783, 130 
\bibitem[Rhode(2012)]{rhod12} Rhode, K.~L.\ 2012, \aj, 144, 154 
\bibitem[Robin et al.(2003)]{robi03} Robin, A.~C., Reyl{\'e}, C., Derri{\`e}re, S., \& Picaud, S.\ 2003, \aap, 409, 523
\bibitem[Sadoun \& Colin(2012)]{sado12} Sadoun, R., \& Colin, J.\ 2012, \mnras, 426, L51
\bibitem[Sani et al.(2011)]{sani11} Sani, E., Marconi, A., Hunt, L.~K., \& Risaliti, G.\ 2011, \mnras, 413, 1479 
\bibitem[Schlafly \& Finkbeiner(2011)]{schlaf11} Schlafly, E.~F., \& Finkbeiner, D.~P.\ 2011, \apj, 737, 103 
\bibitem[Schlegel et al.(1998)]{schleg98} Schlegel, D.~J., Finkbeiner, D.~P., \& Davis, M.\ 1998, \apj, 500, 525 
\bibitem[Silk \& Rees(1998)]{silk98} Silk, J., \& Rees, M.~J.\ 1998, \aap, 331, L1 
\bibitem[Skrutskie et al.(2006)]{skru06} Skrutskie, M.~F., Cutri, R.~M., Stiening, R., et al.\ 2006, \aj, 131, 1163
\bibitem[Snyder et al.(2011)]{snyd11} Snyder, G.~F., Hopkins, P.~F., \& Hernquist, L.\ 2011, \apjl, 728, L24 
\bibitem[Spitler \& Forbes(2009)]{spit09} Spitler, L.~R., \& Forbes, D.~A.\ 2009, \mnras, 392, L1 
\bibitem[Terlouw \& Vogelaar(2015)]{terl15} Terlouw, J.~P., \& Vogelaar, M.~G.~R.\ 2015, 
        Kapteyn Package, version 2.3, Kapteyn Astronomical Institute, Groningen, 
        available from \url{http://www.astro.rug.nl/software/kapteyn/} 
\bibitem[Tremaine et al.(2002)]{trem02} Tremaine, S., et al.\  2002, \apj, 574, 740 
\bibitem[Wang et al.(2014)]{wang14} Wang, S., Ma, J., Wu, Z., \& Zhou, X.\ 2014, \aj, 148, 4
\bibitem[Wolf et al.(2010)]{wolf10} Wolf, J., Martinez, G.~D., Bullock, J.~S., et al.\ 2010, \mnras, 406, 1220 
\bibitem[van den Bergh(1995)]{vand95} van den Bergh, S.\ 1995, \aj, 110, 1171
\bibitem[Williams et al.(1996)]{will96} Williams, R.~E., Blacker, B., Dickinson, M., et al.\ 1996, \aj, 112, 1335 


\end{thebibliography}
\end{document}